\documentclass[11pt,a4paper]{report}
\usepackage{graphicx,bbm,slashed,amssymb}
\usepackage{palatino}

\begin{document}

\bibliographystyle{h-physrev3}

\newcommand {\intsum}{\ \int\!\!\!\!\!\!\!\!\!\sum}
\newcommand {\real}{\mathrm{Re}}
\newcommand {\imag}{\mathrm{Im}}
\newcommand {\Ds} {\slashed{D}}
\newcommand {\cS}{\mathcal{S}}
\newcommand {\bM}{{\bf M}}
\newcommand {\cN}{\mathcal{N}}
\newcommand {\cA}{\mathcal{A}}
\newcommand {\cB}{\mathcal{B}}
\newcommand {\cC}{\mathcal{C}}
\newcommand {\cD}{\mathcal{D}}
\newcommand {\cE}{\mathcal{E}}
\newcommand {\cF}{\mathcal{F}}
\newcommand {\cG}{\mathcal{G}}
\newcommand {\cO}{\mathcal{O}}
\newcommand {\bA}{{\bf A}}
\newcommand {\bB}{{\bf B}}
\newcommand {\bC}{{\bf C}}
\newcommand {\bD}{{\bf D}}
\newcommand {\bE}{{\bf E}}
\newcommand{\Res}[1]{\!\!\!\!\begin{array}{c}\raisebox{-8.5pt}{Res}
  \\{\scriptstyle \ #1}\end{array}\!\!\!\!}
\newcommand {\Tr}{\mathrm{Tr}}
\newcommand {\gf}{g_{\rm eff}}
\newcommand {\Qs}{\slashed{Q}}
\newcommand{\Li}{\mathrm{Li}}
\newcommand{\emu}{\varepsilon }
\newcommand{\bfs}{\sffamily\bfseries}

\begin{titlepage}
\begin{figure}
  \centering
  \includegraphics[width=2.5cm]{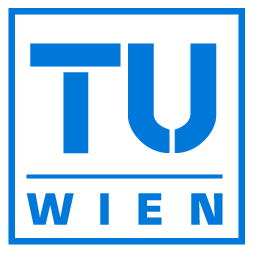}  
\end{figure}
\begin{center}
\vfill\vfill
\Large{DISSERTATION} \\ 
\vfill
\Huge{Aspects of Cold Dense \\Quark Matter}\\
\vfill
\large{ausgef\"uhrt zum Zwecke der Erlangung des akademischen Grades
eines Doktors der technischen Wissenschaften unter der Leitung von\\}
\vfill
\large{Ao. Univ.-Prof. Dr. Anton Rebhan}\\
\large{Institutsnummer: E 136}\\
\large{Institut f\"ur Theoretische Physik} \\
\vfill
\large{eingereicht an der Technischen Universit\"at Wien} \\
\large{Fakult\"at f\"ur Physik}\\
\vfill
\large{von}\\
\vfill
\large{\bf Dipl.-Ing. Andreas Gerhold}\\
\large{Matrikelnummer: 9555935}\\
\large{Gallgasse 53/6, A-1130 Wien}\\
\large{Austria}\\
\vfill

\end{center}
\end{titlepage}

\thispagestyle{empty}
\quad\newpage

\newpage
\renewcommand{\thepage}{\roman{page}}
\setcounter{page}{2}

\section*{Abstract}
\addtocontents{toc}{\protect\contentsline {chapter}{\numberline {}Abstract}{\thepage}}
This thesis is devoted to properties of quark matter at high density and (comparatively) low temperature. In nature 
matter under these conditions can possibly be found in the interior of (some) neutron stars. Under ``normal''
conditions quarks are confined in the hadrons. Under the extreme conditions in the core of a neutron star, however, 
the strong interaction between the quarks becomes weaker because of asymptotic freedom. Then the
hadrons will break up, and for the remaining interaction between the quarks a (semi-)perturbative
treatment should be sufficient, at least at asymptotic densities.

It turns out that cold dense quark matter does not behave like a Fermi liquid as a consequence of long-range chromomagnetic
interactions between the quarks. While the specific heat of a Fermi liquid is a linear function
of the temperature ($T$) at low temperature, the specific heat of normal (non-supercon\-duc\-ting) quark matter is 
proportional to $T\log T$ at low temperature. In this thesis the specific heat and the quark self energy in normal quark matter 
are discussed in detail. A discrepancy between earlier papers on the specific heat is resolved by showing
that in one of these papers gluonic contributions to the energy density were overlooked.
Moreover higher order corrections to the known leading order terms are computed in this thesis
in a systematic manner.

The quark-quark interaction mediated by one-gluon exchange is attractive in the color antitriplet channel. Therefore
quarks form Cooper pairs at sufficiently small temperatures, which leads to the phenomenon of color superconductivity.
During the last years the issue of gauge (in)dependence of the color superconductivity gap has been discussed in the
literature. Here a formal proof is given that the fermionic quasiparticle dispersion relations in a color
superconductor are gauge independent. In a color superconductor the gluon field acquires a non-vanishing expectation value in general, 
which acts as an effective chemical potential for the color charge. This expectation value 
is computed for two different color superconducting phases (2SC and CFL). It is shown that at leading order this 
expectation value is determined from a tadpole diagram with a quark loop. Tadpole diagrams with gluon or
Nambu-Goldstone loops turn out to be negligible at our order of accuracy.

\newpage
\thispagestyle{empty}
\quad
\vskip5cm
\begin{center}
  {\it\large meinen Eltern gewidmet}
\end{center}

\newpage

\quad
\vskip2cm
\section*{\center{Acknowledgements}}
\addtocontents{toc}{\protect\contentsline {chapter}{\numberline {}Acknowledgements}{\thepage}}
I am deeply indebted to my supervisor Anton Rebhan, who introduced me to quantum field theory at
finite temperature and density, and who guided my research activities during the last two and a half years.
I greatly benefited from innumerable discussions with him.

Furthermore I would like to thank Andreas Ipp for a 
very fruitful and enjoyable collaboration on non-Fermi-liquid physics.
I would also like to thank Jean-Paul Blaizot, Urko Reinosa and Paul Romatschke for extensive discussions on 
quantum field theory  at finite temperature and density.
I also gratefully acknowledge very helpful comments from M. Alford, C. Manuel, K. Rajagopal, D. Rischke, T. Sch\"afer and K. Schwenzer.

I want to express my gratitude to all the members of the Institute for Theoretical Physics 
at the Vienna University of Technology, in particular to
Sebastian Guttenberg, Robert Sch\"ofbeck and Robert Wimmer, for providing
a very pleasant working atmosphere.

I also want to thank my family, in particular my parents who made my studies possible with
continuous moral and financial support.
Finally I want to thank Veronika for her love and her patience, and for trying to teach me Ukrainian grammar.
 \raisebox{-3.5pt}{\includegraphics{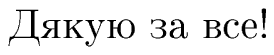}}

\tableofcontents

\newpage

\quad\vskip2cm
``Die Astronomen m\"ussen aus dem Strahlen den Stern erkennen 
- (und tappen doch im Dunkeln herum, denn sie k\"onnen durch ihr
Verfahren, man nennt es Spectralanalyse, doch nur die erdverwandten 
Stoffe auffinden - das dem Stern Ureigene bleibt ewig
unerforschlich) \ldots''

\vspace{.5cm}
\qquad\qquad\qquad\quad\quad\qquad\qquad--\ \,Gustav Mahler in a letter to Alma Schindler 

\newpage\quad\thispagestyle{empty}

\chapter{Introduction}
\renewcommand{\thepage}{\arabic{page}}
\setcounter{page}{1}

We know four fundamental interactions in nature, namely gravitational, electromagnetic, weak and strong interactions.
Since this thesis is devoted to properties of quark matter, we shall mainly be concerned with the strong interactions.
Today the accepted theory of strong interactions is Quantum Chromo\-dynamics (QCD). It is a gauge theory with
gauge group 
$SU(3)_c$, where the index $c$ stands for \emph{color}.
The quarks belong to the fundamental representation of this group,
and the gauge bosons (gluons) belong  to the adjoint representation. 
Denoting the quark field with $\psi$ and the gluon field with $A$, one writes the Lagrangian of QCD as 
\cite{Fritzsch:1973pi,Weinberg:1973un}
\begin{equation}
  \mathcal{L}_{QCD}=-{1\over4}F_{\mu\nu}^a F^{a\mu\nu}+\sum_{f=1}^{N_f} \bar\psi_{i,f} 
  \left(i \slashed{D}_{ij}-m_f\delta_{ij}\right)\psi_{i,f}, \label{qcd}
\end{equation}
with $\slashed{D}=\gamma^\mu D_\mu$,
$F_{\mu\nu}^a=\partial_\mu A_\nu^a-\partial_\nu A^a_\mu-gf^{abc}A_\mu^b A_\nu^c,$ and
$D_{\mu,ij} =\partial_\mu\delta_{ij}+i g A_\mu^a T^a_{ij}.$
In Eq. (\ref{qcd}) we have written explicitly the sum over quark flavors.
Apart from the invariance under $SU(3)_c$
gauge transformations, the Lagrangian (\ref{qcd}) is also invariant under global unitary transformations in flavor space
as long as the differences of the quark masses can be neglected.
In the chiral limit ($m_f\to0$) the Lagrangian is even
invariant under separate flavor space transformations of left and right handed quarks, with the corresponding
symmetry groups $SU(N_f)_L$ and $SU(N_f)_R$. Moreover the Lagrangian (\ref{qcd}) is invariant under global 
$U(1)$ phase transformations of the quark fields, which is related to quark number conservation\footnote{At the
classical level the Lagrangian is also invariant under axial phase transformations of the quark fields, which constitute the
group $U(1)_A$. In the quantum theory, however,
this symmetry is broken by an anomaly, see e.g. \cite{Bertlmann:1996xk}.}.

The quantum theory based on the Lagrangian (\ref{qcd}) is renormalizable \cite{'tHooft:1971fh,'tHooft:1972fi,Becchi:1975nq}.
Neglecting the quark masses for the moment, one finds for the one-loop beta function\footnote{By now the beta function
is known to four-loop order \cite{vanRitbergen:1997va}.}
\begin{equation}
   \beta(g)=-{g^3\over16\pi^2}\left(11-{\textstyle{2\over3}}N_f\right).\label{beta}
\end{equation}
Nature chooses $N_f=6$. (For many physical situations the number of ``active'' flavors is smaller than six.
E.g. for the astrophysical systems, on which we eventually focus, we need to take 
into account only the three lightest flavors.) Therefore
the beta function is negative, which shows that QCD possesses the all-important property of asymptotic freedom
 \cite{Gross:1973id,Politzer:1973fx}. 

Experimental confirmations of QCD come from deep inelastic scattering experiments, and other
high energy processes (see \cite{Eidelman:2004wy} and references therein).
These experiments give a current world average of \cite{Eidelman:2004wy}
\begin{equation}
  \alpha_s(M_Z)=0.1187\pm0.002
\end{equation}
with $\alpha_s={g^2\over4\pi}$, and $M_Z=91.1876\pm0.0021\,\mathrm{GeV}$ is the mass of the $Z$ boson.

At small energies quarks and gluons are confined into hadrons. In the confined phase chiral symmetry is spontaneously broken by 
a non-vanishing quark condensate $\langle\bar qq\rangle$, with $\pi$, $K$ and $\eta$ 
playing the role of (pseudo-)Nambu-Goldstone bosons. The
low-energy dynamics of the hadrons can be described with chiral perturbation theory \cite{Gasser:1983yg,Gasser:1984gg}. 

From asymptotic freedom one expects at sufficiently high temperatures a phase transition to a 
deconfined phase \cite{Collins:1974ky}. 
The resulting ``soup'' of quarks and gluons is called quark gluon plasma (QGP).
This idea is corroborated by lattice simulations, which predict a phase transition at
 $T_c\sim 170\, \mathrm{MeV}$ for $ N_f=2$ \cite{AliKhan:2000iz,Karsch:2000kv},
and at $T_c\sim 150\, \mathrm{MeV}$ for $ N_f=3$ \cite{Karsch:2000kv}.
It is assumed that the QGP existed in the early universe. Experimentalists try 
to reproduce the QGP in heavy ion collisions at SPS (CERN), at RHIC (Brookhaven), and  from 2007 onwards at LHC (CERN).

The equation of state (the pressure as a function of the temperature) of the high temperature QGP 
has been computed perturbatively
up to order $g^6\log g$ (see \cite{Kajantie:2002wa} and references therein\footnote{The coefficient of the $g^6$-term cannot
be computed within perturbation theory, since Feynman diagrams of arbitrarily high order contribute to this
coefficient \cite{Linde:1980ts,Gross:1980br}, and there is no known way of how to resum them.}). The expansion in the coupling constant 
turns out to be only poorly convergent for non-asymptotic temperatures. It is however possible to improve the
situation by performing systematic resummations of the perturbative series, such as  resummations within dimensional reduction 
\cite {Kajantie:2002wa,Blaizot:2003iq}, HTL-screened perturbation theory
\cite{Andersen:1999fw,Andersen:1999sf}, or 
approximately selfconsistent resummations based on the 2PI effective action,
\cite{Blaizot:1999ip,Blaizot:1999ap,Blaizot:2000fc}
which show good agreement with the lattice data for $T\gtrsim 2T_c$. For reviews on the high temperature QGP see
e.~g. \cite{Blaizot:2001nr,Karsch:2001cy,Blaizot:2003tw,Laermann:2003cv,Kraemmer:2003gd,Adams:2005dq}.

It is of great importance to understand the properties of the quark gluon plasma also at finite chemical 
potential\footnote{The chemical potential can be viewed as a measure for the density at a given temperature.
E.g. for an ideal relativistic gas of free fermions the particle number density is proportional to $\mu^3$ for $T\to0$.
Thus a high chemical potential is equivalent to
high particle number density in this case.}. This region of the QCD phase diagram is relevant for the interior 
of compact stars \cite{Glendenning:1997wn,Prakash:2000jr,Yakovlev:2004iq}\footnote{
As far as the macroscopic structure of compact stars is concerned, the curvature of
spacetime is in general not negligible. However, the change in the metric is tiny for the typical length scales of particle 
physics \cite{Glendenning:1997wn}.
Therefore it is sufficient to consider the field equations of matter in flat spacetime.}.
From asymptotic freedom  one expects a phase transition to a deconfined phase to occur not only at high temperature, but also at
high density. 
Unfortunately one does not know an efficient algorithm for 
lattice simulations at finite $\mu$.
The reason is that the fermion determinant in the path integral is complex at finite $\mu$, which precludes
Monte Carlo importance sampling.
So far this ``sign problem'' can only be circumvented for
$\mu<T$ (see \cite{Laermann:2003cv} and references therein).
But for the region $\mu\gg T$, which is relevant for compact stars, no method of implementing lattice
simulations is known to date. Therefore simplified models, such as Nambu-Jona-Lasinio models 
\cite{Nambu:1961tp,Nambu:1961fr,Buballa:2003qv}, and \mbox{(semi-)}perturbative methods are
the only viable approaches at the moment. In compact stars the chemical potential might be as large as
$500\,\mathrm{MeV}$. At this energy scale $\alpha_s(\mu)$ is of the order one,
therefore the applicability of \mbox{(semi-)}perturbative methods is rather questionable. 
Nevertheless one can try to extrapolate results which have been obtained for small values of the coupling constant
to larger values of the coupling constant, 
hoping that the qualitative features remain valid.


Compared to the chemical potential, the temperature in the interior of 
a compact star will be rather small (of the order of tens of $\mathrm{keV}$ \cite{Prakash:2000jr}).
In this region of the QCD phase diagram quark matter is expected to be in a color superconducting phase,
which is characterized by a non-vanishing diquark condensate. 
Such diquark condensates can arise from the fact that one gluon exchange is attractive in the color antitriplet channel,
leading to the formation of Cooper pairs at sufficiently small temperatures.
Color superconductivity has been discussed already in the late 1970's \cite{Barrois:1977xd,Bailin:1983bm},
but at that time it was believed
that the color superconductivity gap would be of the order $1\,\mathrm{MeV}$, 
and thus almost negligible.
Only some years ago \cite{Alford:1997zt,Rapp:1997zu} it was realized that the gap may be as large as $\sim10-100\,\mathrm{MeV}$,
with corresponding critical temperatures of the order $\sim5-50\,\mathrm{MeV}$. This discovery stimulated extensive research in
this field during the last few years, see e.g. \cite{Rajagopal:2000wf,Alford:2001dt,Nardulli:2002ma,Schafer:2003vz,
Rischke:2003mt,Ren:2004nn,Huang:2004ik,Shovkovy:2004me} for reviews.

Due to the interplay of finite quark masses and the constraints from  color and electric neutrality, there may be
various color superconducting phases in the QCD phase diagram, 
which are distinguished mainly by the particular form of the diquark condensate.
These matters will be reviewed in more detail in  chapter \ref{ccsc} of this thesis.


In nature color superconducting phases could in principle be discovered from analyses of neutron star data.
There exist indeed some proposals for signatures that could indicate the presence of color superconductivity.
For instance, the cooling behavior of a neutron star depends on the specific heat and the neutrino emissivity, which are both sensitive
to the phase structure (see e.g. \cite{Blaschke:1999qx,Page:2000wt,Reddy:2002xc,Jaikumar:2002vg,Kundu:2004mz}).
Other possible signatures include $r$-mode instabilities \cite{Madsen:1999ci,Manuel:2004iv}, and pulsar glitches due to crystalline structures in
inhomogeneous color superconducting phases \cite{Alford:2000ze}.

\vspace{.5cm}
This thesis is organized as follows.
In chapter \ref{csym} we discuss some formal properties of gauge theories, which will be used in chapter \ref{ccsc}.
In chapters \ref{csigma} and \ref{cspecific} we compute the quark self energy and the specific heat of
normal degenerate quark matter. These results may be relevant for the cooling properties of (proto-) neutron stars \cite{Schafer:2004jp}.
Chapter \ref{ccsc} is devoted to color superconductivity. We give a
general proof that the fermionic quasiparticle dispersion relations in a color superconductor are gauge independent.
Furthermore we compute gluon tadpole diagrams, which are related to color neutrality.
Chapter \ref{cconc} finally contains our conclusions.

\vspace{.5cm}
Throughout this thesis we shall use the following conventions.
We use natural units, $\hbar=c=k_B=1$. The Minkowski metric is $g^{\mu\nu}=\mathrm{diag}(1,-1,-1,-1)$. Four momenta
are denoted as $K^\mu=(k_0,{\bf k})$. The absolute value of the three momentum is denoted as $k:=|{\bf k}|$. A unit three
vector is denoted as ${\bf\hat k}:={\bf k}/k$.

\chapter{Symmetries in quantum field theory \label{csym}}

\section{General gauge dependence identities}
In this section we will recapitulate the derivation of a general gauge dependence identity, following
\cite{Lee:1973fn,Fukuda:1975di, Kobes:1990dc}.

Let us consider an arbitrary gauge theory which is defined by an action functional $S_{inv}[\varphi]$ that is 
invariant under some gauge transformations,
\begin{equation}
  \delta\varphi^i=D^i_\alpha[\varphi]\delta\xi^\alpha.
\end{equation}
Here we use the DeWitt notation \cite{Dewitt:1967ub}, where an index $i$ comprises all discrete and continuous
field labels, and a Greek index ($\alpha$) comprises group and space-time indices.
 E.g. for QCD, as defined by the Lagrangian (\ref{qcd}), one has explicitly [with $\alpha\leftrightarrow (x,a)$]
\begin{equation}
  \begin{array}{rll}
    D^i_\alpha\to& (\delta^{ab}\partial_\mu^{y}-g f^{abc}A_\mu^c(y))\delta(x-y),&\quad i\leftrightarrow A_\mu^b(y),\vspace{1.5mm}\\
    &-igT^a\psi(y)\,\delta(x-y),&\quad i\leftrightarrow\psi(y),\vspace{1.5mm} \\
    &ig\bar\psi(y) T^a\,\delta(x-y),&\quad i\leftrightarrow\bar\psi(y).  
  \end{array}
\end{equation}
 
In general we assume that the gauge generators form an off-shell closed algebra, which means that the
commutator of two gauge transformations is again a gauge transformation,
\begin{equation}
  D^i_{\alpha,j}[\varphi]D^j_\beta[\varphi]-D^i_{\beta,j}[\varphi] D^j_\alpha[\varphi]
  =c^\gamma_{\alpha\beta}[\varphi]D^i_\gamma[\varphi], \label{alg}
\end{equation}
where the ``structure constants'' $c^\gamma_{\alpha\beta}$ can in principle be field dependent.

In order to quantize the theory we have to fix the gauge freedom, e.g. with a quadratic gauge fixing term,
\begin{equation}
  S_{inv}[\varphi]\to S_{gf}[\varphi]= S_{inv}[\varphi]+{1\over2}F^\alpha[\varphi]F_\alpha[\varphi].
\end{equation}
The quantum theory is defined via the path integral representation of the generating functional of
connected Green's functions\footnote{We assume for the moment 
that $\varphi^i$ are bosonic fields in order to keep the notation simple. It is easy to check, however,
that the final identity Eq. (\ref{gauge}) is also valid for the fermions in QCD, provided one 
chooses suitable conventions for the fermionic derivatives.},
\begin{equation}
  \exp(iW[J])=\int {\mathcal D}\varphi \det\left(F^\alpha_{,i}[\varphi]D^i_\beta[\varphi]\right)
  \exp\left(i\left(S_{gf}[\varphi]+J_i\varphi^i\right)\right).
\end{equation}
With the help of a Legendre transformation
\begin{equation}
  W[J]=\Gamma[\bar\varphi]+J_i\bar\varphi^i, \qquad \bar\varphi^i:={\delta W[J]\over\delta J_i},
\end{equation}
one obtains the effective action,
\begin{equation}
  \exp(i\Gamma[\bar\varphi])=\int {\mathcal D}\varphi \det\left(F^\alpha_{,i}[\varphi]D^i_\beta[\varphi]\right)
  \exp\left(i\left(S_{gf}[\varphi]-\Gamma_{,i}[\bar\varphi](\varphi-\bar\varphi)^i\right)\right). \label{ea}
\end{equation}

If a change $\delta F^\alpha[\varphi]$ in the gauge fixing condition is accompanied by the
following (non-local) gauge transformation,
\begin{equation}
  \delta\varphi^i=D^i_\alpha[\varphi]\delta\Xi^\alpha[\varphi], \quad
  \delta\Xi^\alpha[\varphi]=\cG ^\alpha_{\ \beta}[\varphi]\delta F^\beta[\varphi], \label{gt}
\end{equation}
the gauge fixed action $S_{gf}$ will remain invariant.
Here $\cG ^\alpha_{\ \beta}$ is the ghost propagator in a background field $\varphi$, which is defined via
\begin{equation}
  \left(F^\alpha_{,i}D^i_\beta\right)[\varphi]\cG ^\beta_{\ \gamma}[\varphi]=-\delta^\alpha_\gamma. \label{gh}
\end{equation}
Let us examine in which way the measure and the Faddeev Popov determinant change under the gauge
transformation (\ref{gt}) and $\delta F^\alpha$. From the measure we get a Jacobian, which can be evaluated for small
$\delta\varphi^i$ using the formula 
\begin{equation}
  \det({\bf 1}+\mathcal{X})\simeq 1+\Tr\mathcal{X},
\end{equation}
which is valid if $|\mathcal{X}_{ij}|\ll1$. The variation of the Faddeev Popov determinant can be calculated
using Eq. (\ref{c1}). In this way we find 
\begin{eqnarray}
  &&\!\!\!\!\!\!\!\!\!\!\!\!\!
  \delta\left(\mathcal{D}\varphi\det\left(F^\alpha_{,i}D^i_\beta\right)[\varphi]\right)=
   \mathcal{D}\varphi\det\left(F^\alpha_{,i}D^i_\beta\right)
  \left[\Tr {\delta(\delta\varphi)^i\over\delta\varphi^j}+\delta\Tr\log\left(F^\alpha_{,i}D^i_\beta\right)
  \right]\nonumber\\
  &&\!\!\!\!\!=\mathcal{D}\varphi\det\left(F^\alpha_{,i}D^i_\beta\right)
  \left[D^i_{\alpha,i}\cG ^\alpha_{\ \beta}\delta F^\beta
  +\cG ^\alpha_{\ \beta}F^\beta_{,i}\left(D^j_\alpha D^i_{\gamma,j}- D^j_\gamma D^i_{\alpha,j}\right) 
  \cG ^\gamma_{\ \delta}\delta F^\delta\right].\qquad
\end{eqnarray}
The first term between the brackets in the second line vanishes if $D^i_{\alpha,i}=0$. Using Eqs. (\ref{alg}) and (\ref{gh}) one
finds that the remaining part vanishes provided that $c^\beta_{\beta\alpha}=0$. These two conditions are fulfilled
for QCD\footnote{In order to prove this one uses the fact that the structure constants are totally antisymmetric
and that the Gell-Mann matrices are traceless.}, and also for the Higgs models which we will discuss in this chapter.

The only term in Eq. (\ref{ea}) which is not invariant is the source term $\Gamma_{,i}[\bar\varphi]\varphi^i$.
Therefore we find the following gauge dependence identity,
\begin{equation}
  \delta\Gamma[\bar\varphi]=-\Gamma_{,i}[\bar\varphi]\langle D^i_\alpha[\varphi]\cG ^\alpha_{\ \beta}[\varphi]
  \delta F^\beta[\varphi]\rangle[\bar\varphi]=:\Gamma_{,i}[\bar\varphi]\delta X^i[\bar\varphi], \label{gauge}
\end{equation}
where we use the notation
\begin{eqnarray}
  &&\!\!\!\!\!\!\!\!\!\!\!\!\!\!\!\langle K[\varphi]\rangle[\bar\varphi]\nonumber\\
  &&\!\!\!\!\!\!\!\!\!\!\!\!=e^{-i\Gamma[\bar\varphi]}\int \mathcal{D}\varphi\,K[\varphi]
  \det\left(F^\alpha_{,i}[\varphi]D^i_\beta[\varphi]\right)
  \exp\left(i\left(S_{gf}[\varphi]-\Gamma_{,i}[\bar\varphi](\varphi-\bar\varphi)^i\right)\right)\qquad
\end{eqnarray}
for any $K[\varphi]$.

As an application let us consider the effective potential $V_{eff}$ for a translationally invariant system, 
\begin{equation}
  \Gamma[\bar\varphi]|_{\bar\varphi=\mathrm{const.}}=-V_{eff}(\bar\varphi)\int d^4x.
\end{equation}
In general $V_{eff}$ will be a gauge dependent quantity. However, the identity
(\ref{gauge}) ensures that
\begin{equation}
  \delta V_{eff}(\bar\varphi)+\delta X^i[\bar\varphi] {\partial V_{eff}(\bar\varphi)\over\partial\bar\varphi^i}=0.
\end{equation} 
This is the Nielsen identity for the effective potential \cite{Fukuda:1975di,Nielsen:1975fs}, 
which states that a change in the gauge fixing function
(the first term on the left hand side) can be compensated by a change in $\bar\varphi$ (the second term on the left hand side). This identity
implies that the value of the effective potential at its minimum (where $\partial V_{eff}/\partial\bar\varphi^i=0$) is
gauge independent.

A further consequence of the identity (\ref{gauge}) is the gauge independence of the position of the
propagator singularities \cite{Kobes:1990dc,Nielsen:1975fs,Aitchison:1983ns,Kobes:1990xf}, which we will discuss in more detail in 
the sections \ref{sha} and \ref{shna} and in chapter \ref{ccsc}. 

\section{Implications of global symmetries \label{sglob}}
The aim of this section is to examine the consequences of global symmetries for
the one-point and two-point functions in quantum field theory. 

Let us consider an arbitrary quantum field theory with field content $\Phi^i(x)$, $\bar\Phi^i(x)$. 
We assume that the theory is invariant with respect to a group $G$ of continuous global symmetry
transformations, which act on the fields as
\begin{equation}
  \Phi_i^\prime(x)=g_{ij}\Phi_j(x),\quad \bar\Phi_i^\prime(x)=\bar\Phi_j(x)g_{ji}^\dag,
\end{equation}
where the $g_{ij}$ form some representation of $G$.
The effective action up to second order can be written as
\begin{eqnarray}
  &&\Gamma[\Phi]\simeq\int d^4x\,\Gamma_i(x)\Phi_i(x)+\int d^4x\,\bar\Gamma_i(x)\bar\Phi_i(x)\nonumber\\
  &&\quad\qquad+\int d^4x\,d^4y\,\Gamma_{ij}(x,y)\Phi_i(x)\bar\Phi_j(y)+\mathcal{O}(\Phi^3).
\end{eqnarray}
We assume that the symmetry is not (spontaneously)
broken. Then we have
$
  \Gamma[\Phi]=\Gamma[\Phi^\prime],
$
which implies
\begin{eqnarray}
  \Gamma_i(x)=g_{ji}\Gamma_j(x),\label{g4} \\
  \bar\Gamma_i(x)=\bar\Gamma_j(x)g_{ij}^\dag,\label{g5} \\
  \Gamma_{ij}(x,y)=g_{ki}\Gamma_{kl}(x,y)g_{jl}^\dag, \label{g6}
\end{eqnarray} 
for all $g$. First let us consider Eqs. (\ref{g4}) and (\ref{g5}). 
With the additional assumption that the representation formed by the set of all $g$'s is non-trivial and 
irreducible, these equations imply that
\begin{equation}
   \Gamma_i(x)=0,\quad \bar\Gamma_i(x)=0.
\end{equation}
which means that the tadpole diagrams vanish (in other words, the expectation values of the 
field operators are zero). 

If we assume that the representation is unitary, we can write Eq. (\ref{g6}) as
\begin{equation}
  \Gamma^T_{ji}(x,y)=g_{jl}^{-1}\Gamma^T_{lk}(x,y)g_{ki}.
\end{equation}
If we assume furthermore that the representation is irreducible, we can now invoke Schur's lemma,
which gives the result
\begin{equation}
  \Gamma_{ij}\propto\delta_{ij},
\end{equation}
i.e. the (inverse) propagator is proportional to the unit matrix with respect to the group indices.

Next we would like to examine in which way parity ($P$) and time reversal symmetry ($T$) constrain 
the propagator. For simplicity we work in Euclidean space and 
assume that our fields $\phi^i$ transform under $PT$ as\footnote{No 
summation over the index in parentheses.}
\begin{equation}
  \phi^i(x)\rightarrow c^{(i)}\phi^i(-x),
\end{equation}
with $c^{(i)}=\pm1$. [This comprises for instance vector fields and (pseudo-)scalar fields.]
Assuming translational invariance, the bilinear part of the effective action can be written as
\begin{equation}
  \Gamma_2[\phi]={1\over2}\int {d^4K\over(2\pi)^4}\phi^i(-K)\Gamma_{ij}(K)\phi^j(K), \label{f13}
\end{equation}
where $\Gamma_{ij}$ is the full inverse propagator. Invariance under $PT$ implies
\begin{equation}
  \Gamma_2[\phi]={1\over2}c^{(i)}c^{(j)}\int {d^4K\over(2\pi)^4}\phi^i(K)\Gamma_{ij}(K)\phi^j(-K).
\end{equation}
Performing the substitution $K\rightarrow-K$ in the last equation and comparing it with
Eq. (\ref{f13}) we find
$ 
  \Gamma_{ij}(K)=c^{(i)}c^{(j)}\Gamma_{ij}(-K).
$ 
The mere definition of the inverse propagator implies $\Gamma_{ij}(K)=\Gamma_{ji}(-K)$, 
and therefore we have
\begin{equation}
  \Gamma_{ij}(K)=c^{(i)}c^{(j)}\Gamma_{ji}(K).
\end{equation}
Using Dyson's equation 
\begin{equation}
  \Delta^{ij}\Gamma_{jk}=-\delta^i_k \label{dy}
\end{equation}
we arrive at the following identity for the full propagator,
\begin{equation}
   S^{ij}(K)=c^{(i)}c^{(j)}S^{ji}(K). \label{f18}
\end{equation}
For $c^{(i)}c^{(j)}=1$ the propagator is therefore symmetric in $i$ and $j$.

\section{Abelian Higgs model \label{sha}} 
This and the next section serve as an illustration of the application of the gauge dependence identities for
systems with spontaneous symmetry breaking at finite temperature.
We will discuss the  gauge independence of the locations of propagator singularities, both for the Abelian and a non-Abelian
Higgs model. This is actually more than a simple exercise, since the methods that we develop here will 
be useful also in the more complicated context of color superconductivity, which will be the subject of the last
chapter of this thesis.

Let us also mention that the static limit of the gauged Higgs model yields the Landau-Ginzburg Lagrangian, which
can be used for instance for the description of superconducting systems in the vicinity of the transition 
temperature \cite{Bailin:1983bm}. In this thesis, however, we shall not go into the details of the Landau-Ginzburg
description of superconducting systems.

There exists a vast literature on various aspects of Higgs models,  let us just mention Refs. \cite{Kirzhnits:1972ut,Dolan:1973qd,Weinberg:1974hy,Arnold:1992rz,Buchmuller:1992rs,Buchmuller:1993bq,Buchmuller:1994qy,Rebhan:1994ie}, in which the finite temperature behavior of
Higgs models is discussed.

\subsection{Definition of the model}
The Abelian Higgs is defined by the Lagrangian
\begin{equation}
  \mathcal L_{inv} = -{1\over4}F_{\mu\nu}F^{\mu\nu}+(D_\mu\Phi)(D^\mu\Phi)^*-V(\Phi) \label{a1}
\end{equation}
with $F_{\mu\nu}=\partial_\mu A_\nu-\partial_\nu A_\mu$, $D_\mu=\partial_\mu+ieA_\mu$,
$\Phi={1\over\sqrt{2}}(\varphi_1+i\varphi_2)$, and
\begin{equation} 
  V(\Phi)=m^2\Phi^*\Phi+{\lambda\over3!}(\Phi^*\Phi)^2.
\end{equation} 
The Lagrangian (\ref{a1}) is invariant under the gauge transformation
\begin{equation}
 \delta\Phi=i\alpha\Phi,\quad \delta A_\mu=-{1\over e}\partial_\mu\alpha.
\end{equation}
We assume $m^2<0$ so that the gauge symmetry is spontaneously broken. The expectation value
of $\Phi$ is taken to be real so that $\varphi_1$ is the Higgs boson and $\varphi_2$ is the would-be
Nambu-Goldstone boson.
$\mathcal L_{inv}$ is invariant with respect to a ${\mathbbm Z}_2$ symmetry \cite{Fukuda:1975di}:
\begin{equation}
  A_\mu \rightarrow -A_\mu,\qquad \varphi_2 \rightarrow -\varphi_2 \label{e4}
\end{equation}
Following \cite{Fukuda:1975di} we choose a gauge fixing which does not break the ${\mathbbm Z}_2$ symmetry:
\begin{equation} 
  \mathcal L_{gf}={1\over2\alpha}(f\cdot A+\kappa \varphi_2)^2 \label{e8}
\end{equation}
with some constant $\kappa$ and $f_\mu=\partial_\mu$, $f_\mu=(0,\partial_i)$ or $f_\mu=u_\mu$, 
where $u_\mu=(1,0,0,0)$.
In momentum space we write $\tilde{f}_\mu=\beta(K)K_\mu+\gamma(K)\tilde{n}_\mu$ 
with $\tilde{n}^\mu=(g^{\mu\nu}-K^\mu K^\nu/K^2)u_\nu$.
If $f_\mu=u_\mu$ we will assume $\kappa=0$ because otherwise time reversal
symmetry would be violated.
The corresponding ghost Lagrangian is given by
\begin{equation}
  \mathcal L_{ghost}=\bar c\left(-{1\over e}f\cdot\partial+2\alpha\kappa\varphi_1\right)c.
\end{equation}
The total Lagrangian 
\begin{equation}
  \mathcal L= \mathcal{L}_{inv}+\mathcal{L}_{gf}+\mathcal{L}_{ghost}
\end{equation}
 is still invariant under (\ref{e4}), which implies that Green functions 
(derivatives of the effective action, 
$\Gamma_{,ijk\ldots}|_{A=\varphi_2=0,\varphi_1=\bar\varphi}$) with an odd total number of
external $A$- and $\varphi_2$-legs will vanish. Using Dyson's equation (\ref{dy})
is is easy to see that this is also true for general Green functions.



\subsection{ Free propagators}
The free propagators can be obtained easily by evaluating the path integral for the free theory.
One finds that the free propagators are simply given by inverting the ``coefficients'' of the
kinetic (bilinear) terms in the Lagrangian. If $\phi_i(x)$ is the solution of the free field equation,
\begin{equation}
  {\delta S_{bil.}\over\delta\phi_i(x)}=J_i(x), \label{pp1}
\end{equation}
one finds for the free propagator
\begin{equation}
  G_{ij}(x,y)={\delta\phi_j(y)\over\delta J_i(x)}. \label{pp3}
\end{equation}
In this way one obtains the following tree level propagators for the Abelian Higgs model
(with the notation ``5''=$\varphi_2$):
\begin{eqnarray}
  \Delta_{(0)}^{\mu\nu}&=&{1\over K^2-e^2\bar\varphi_{(0)}^2}\Big(\cA^{\mu\nu}
  +\cB^{\mu\nu}-{\gamma  k \over\beta K^2-e \kappa\bar\varphi_{(0)}^2}\cC^{\mu\nu}\nonumber\\
  &&-{\gamma^2 k ^2+(\alpha K^2+\kappa^2)(K^2-e^2\bar\varphi_{(0)}^2)\over(\beta K^2
  -e\kappa\bar\varphi_{(0)})^2}\cD^{\mu\nu}\Big),\\
  \Delta_{(0)}^{\mu5}&=&{i k \over K^2(K^2-e^2\bar\varphi_{(0)}^2)(\beta K^2
  -e\kappa\bar\varphi_{(0)})^2}\nonumber\\
  &&\times\Big[\left(-e\gamma^2 k ^2\bar\varphi_{(0)}-K^2(\beta\kappa
  +\alpha e\bar\varphi_{(0)})(K^2-e^2\bar\varphi_{(0)}^2)\right)\cE^\mu\qquad\nonumber\\
  &&+{\gamma k \over\alpha}\left(2\beta^2K^2\kappa+\alpha\beta e K^2
  \bar\varphi_{(0)}+\alpha e^2\kappa\bar\varphi_{(0)}\right)\cF^\mu\Big],\\
  \Delta_{(0)}^{55}&=&-{e^2\gamma^2 k ^2\bar\varphi_{(0)}^2
  +K^2(K^2-e^2\bar\varphi_{(0)}^2)(\beta^2K^2+\alpha e^2\bar\varphi_{(0)}^2)
  \over K^2(K^2-e^2\bar\varphi_{(0)}^2)(\beta K^2-e\kappa\bar\varphi_{(0)})^2},\\
  \Delta_{(0)}^{\varphi_1\varphi_1}&=&-{1\over K^2-m^2-{\lambda\over2}
  \bar\varphi_{(0)}^2},
\end{eqnarray}
where
\begin{eqnarray}
  &&\cA^{\mu\nu}=g^{\mu\nu}-{K^\mu K^\nu\over K^2}-{\tilde n^\mu \tilde n^\nu\over\tilde n^2},
  \quad \cB^{\mu\nu}={\tilde n^\mu \tilde n^\nu\over\tilde n^2},\label{proj1}\\
  &&\cC^{\mu\nu}={1\over k}(\tilde n^\mu K^\nu+K^\mu\tilde n^\nu),\label{proj2}\quad
  \cD^{\mu\nu}={K^\mu K^\nu\over K^2},\\
  &&\cE^\mu={K^\mu\over k},\quad \cF^\mu={\tilde n^\mu\over\tilde n^2},
\end{eqnarray}
and $\bar\varphi_{(0)}$ is the position of the minimum of the tree level potential, i.e.
\begin{equation}
  \bar\varphi_{(0)}=\sqrt{-6m^2\over\lambda}.
\end{equation}
We notice that all the structure functions are gauge dependent at tree level 
apart from $\Delta_{A(0)}$, $\Delta_{B(0)}$ and 
$\Delta^{\varphi_1\varphi_1}_{(0)}$.
 
The (full) propagators have the following symmetries: 
the definition of propagators implies $\Delta^{ij}=\Delta^{ji}$, and therefore in 
the non-condensed notation
\begin{equation}
  \Delta^{\mu\nu}(K)=\Delta^{\nu\mu}(-K),\quad 
  \Delta^{\mu5}(K)=\Delta^{5\mu}(-K). \label{e1}
\end{equation}
Invariance with respect to parity and time reversal implies (see Sec. \ref{sglob})
\begin{equation}
  \Delta^{\mu\nu}(K)=\Delta^{\mu\nu}(-K),\quad 
  \Delta^{\mu5}(K)=-\Delta^{\mu5}(-K). \label{e2}
\end{equation}
Eqs. (\ref{e1}) and (\ref{e2}) yield 
\begin{equation}
  \Delta^{\mu\nu}(K)=\Delta^{\nu\mu}(K),\quad 
  \Delta^{\mu5}(K)=-\Delta^{5\mu}(K). \label{e5}
\end{equation}
If one used the gauge condition (\ref{e8}) with $f_\mu=u_\mu$ and $\kappa\neq0$,
(\ref{e2}) and (\ref{e5}) would not be valid, and an explicit calculation shows that a term which is
antisymmetric in $\mu\nu$ would indeed appear already in the tree level propagator 
$\Delta_{(0)}^{\mu\nu}$.\\
We may parametrize the full propagators as follows:
\begin{eqnarray}
  \Delta^{\mu\nu}&=&\Delta_A \cA^{\mu\nu}+\Delta_B \cB^{\mu\nu}+
  \Delta_C \cC^{\mu\nu}+\Delta_D \cD^{\mu\nu},\\
  \Delta^{\mu5}&=&\Delta_1 \cE^\mu+\Delta_2 \cF^\mu,  
\end{eqnarray}
We perform an analogous decomposition for the two-point functions $\Gamma_{\mu\nu}$
and $\Gamma_{\mu5}$
which are obtained from the effective action. These are one-particle irreducible
apart from possible tadpole insertions. We also define self energies in the usual
manner as $\Pi_{ij}=\Delta_{(0)ij}^{-1}-\Delta_{ij}^{-1}$.

With the help of Dyson's equation (\ref{dy}) one can express the
structure functions of the 
full propagators in terms of the self energies in the following way:  
\begin{eqnarray}
  \Delta_A&=&-{1\over\Gamma_A},\\
  \Delta_B&=&{\Gamma_1^2-\Gamma_{55}\Gamma_D\tilde{n}^2\over-\Gamma_1^2\Gamma_B
  +2\Gamma_1\Gamma_2\Gamma_C+\Gamma_2^2\Gamma_D+\Gamma_{55}(\Gamma_C^2
  +\Gamma_B\Gamma_D)\tilde{n}^2},\\
  \Delta_C&=&{\Gamma_1\Gamma_2+\Gamma_{55}\Gamma_C\tilde{n}^2\over\Gamma_1^2
  -\Gamma_{55}\Gamma_D\tilde{n}^2}\Delta_B,\\
  \Delta_D&=&-{\Gamma_2^2+\Gamma_{55}\Gamma_B\tilde{n}^2\over\Gamma_1^2
  -\Gamma_{55}\Gamma_D\tilde{n}^2}\Delta_B,\\
  \Delta_1&=&{(\Gamma_1\Gamma_B-\Gamma_2\Gamma_C)\tilde{n}^2\over\Gamma_1^2
  -\Gamma_{55}\Gamma_D\tilde{n}^2}\Delta_B,\\
  \Delta_2&=&{(\Gamma_1\Gamma_C-\Gamma_2\Gamma_D)\tilde{n}^2\over\Gamma_1^2
  -\Gamma_{55}\Gamma_D\tilde{n}^2}\Delta_B,\\
  \Delta_{55}&=&-{(\Gamma_C^2+\Gamma_B\Gamma_D)\tilde{n}^2\over\Gamma_1^2
  -\Gamma_{55}\Gamma_D\tilde{n}^2}\Delta_B,
\end{eqnarray}
with
\begin{eqnarray}
  \Gamma_A&=&-(K^2-e^2\bar\varphi_{(0)}^2)+\Pi_A,\\
  \Gamma_B&=&{\gamma^2\tilde{n}^2\over\alpha}-(K^2-e^2\bar\varphi_{(0)}^2)+\Pi_B,\\
  \Gamma_C&=&{\beta\gamma k \over\alpha}+\Pi_C,\\
  \Gamma_D&=&{\beta^2K^2\over\alpha}+e^2\bar\varphi_{(0)}^2+\Pi_D,\\
  \Gamma_1&=&-i\left(e\bar\varphi_{(0)}
   +{\kappa\beta\over\alpha}\right) k +\Pi_1,\\
  \Gamma_2&=&-{i\kappa\gamma\over\alpha}\tilde{n}^2+\Pi_2,\\
  \Gamma_{55}&=&K^2+{\kappa^2\over\alpha}+\Pi_{55}.
\end{eqnarray}

\subsection{Gauge independence of propagator singularities \label{sechig}}
Taking the second derivative of the gauge dependence identity (\ref{gauge}) one obtains
\begin{equation}
  \delta\Gamma_{,ij}=\Gamma_{,kij}\delta X^k+\Gamma_{,ki}\delta X^k_{,j} +
  \Gamma_{,kj}\delta X^k_{,i} +\Gamma_{,k}\delta X^k_{,ij}. \label{e3}
\end{equation}
We evaluate (\ref{e3}) at $A=\varphi_2=0,\varphi_1=\bar\varphi$. Then the last
term vanishes. For the inverse Higgs propagator we obtain in configuration space
\begin{eqnarray}
  &&\!\!\!\!\!\!\!\!\delta{\delta^2\Gamma\over\delta\varphi_1(x)\delta\varphi_1(y)}\Big|_{\bar\varphi}
  =\int dz {\delta^3\Gamma\over\delta\varphi_1(z)\delta\varphi_1(x)\delta\varphi_1(y)}
  \Big|_{\bar\varphi}\delta X^{\varphi_1}(z) \Big|_{\bar\varphi}\nonumber\\
  &&\qquad\quad\qquad+\int dz {\delta^2\Gamma\over\delta\varphi_1(z)\delta\varphi_1(x)}
  \Big|_{\bar\varphi}{\delta\over\delta{\varphi_1}(y)}\delta X^{\varphi_1}(z)
  \Big|_{\bar\varphi}+(x\leftrightarrow y).\qquad
\end{eqnarray}
Using translation invariance the first term on the right hand side can be rewritten as
\begin{equation}
  \int dz {\delta^3\Gamma\over\delta\varphi_1(z)\delta\varphi_1(x)\delta\varphi_1(y)}
  \Big|_{\bar\varphi}\delta X^{\varphi_1}(0) \Big|_{\bar\varphi}=
  {\partial\over\partial\bar\varphi}{\delta^2\Gamma\over\delta\varphi_1(x)\delta\varphi_1(y)}
  \Big|_{\bar\varphi}\delta X^{\varphi_1}(0) \Big|_{\bar\varphi}.
\end{equation}
In momentum space we obtain therefore
\begin{equation}
  \delta\Gamma_{\varphi_1\varphi_1}+
  \delta\bar\varphi {\partial\over\partial\bar\varphi}\Gamma_{\varphi_1\varphi_1}
  =2\Gamma_{\varphi_1\varphi_1}
  \delta X^{\varphi_1}_{,\varphi_1}
\end{equation}
or
\begin{equation}
  \delta\Delta_{\varphi_1\varphi_1}^{-1}+
  \delta\bar\varphi {\partial\over\partial\bar\varphi}\Delta_{\varphi_1\varphi_1}^{-1}
  =2\Delta_{\varphi_1\varphi_1}^{-1}
  \delta X^{\varphi_1}_{,\varphi_1}, \label{e43}
\end{equation}
with $\delta\bar\varphi: = -\delta X^{\varphi_1}(z=0)$. 
In a similar way one obtains for the gauge field propagator
\begin{equation}
  \delta\Delta^{\mu\nu}+\delta\bar\varphi {\partial\over\partial\bar\varphi}
  \Delta^{\mu\nu}=
  -\Delta^{\mu\rho}\delta X^\nu_{,\rho}
  -\Delta^{\nu\rho}\delta X^\mu_{,\rho}
  -\Delta^{\mu5}\delta X^\nu_{,5}
  -\Delta^{\nu5}\delta X^\mu_{,5}, \label{ee44}
\end{equation}
which implies
\begin{eqnarray}
  \delta\Delta_A^{-1}+\delta\bar\varphi {\partial\over\partial\bar\varphi}
  \Delta_A^{-1}&=&
  \Delta_A^{-1}\cA_\mu^\rho\delta X^\mu_{,\rho},\label{g7}\\
  \delta\Delta_B^{-1}+\delta\bar\varphi {\partial\over\partial\bar\varphi}
  \Delta_B^{-1}&=&
  2\Delta_B^{-1}\Bigg[\left({\tilde{n}_\nu \tilde{n}^\rho\over \tilde{n}^2}+{\Gamma_1\Gamma_2+
  \Gamma_{55}\Gamma_C\over\Gamma_1^2-\Gamma_{55}\Gamma_D\tilde{n}^2} 
  {\tilde{n}_\nu K^\rho\over k }\right)\delta X^\nu_{,\rho}\nonumber\\
  &&+{\Gamma_1\Gamma_C-\Gamma_2\Gamma_D\over\Gamma_1^2-\Gamma_{55}\Gamma_D\tilde{n}^2}    \tilde{n}_\nu\delta X^\nu_{,5}\Bigg].\label{e7}
\end{eqnarray}

Eqs. (\ref{e43}), (\ref{e7}) and (\ref{g7}) imply that the locations of 
the poles of the Higgs propagator and of the transverse and longitudinal components of the
gauge field propagator are gauge independent \cite{Nielsen:1975fs}, provided that the
singularities of the $\delta X$'s do not coincide with 
those of $\Delta_{\varphi_1\varphi_1}$, $\Delta_A$ or $\Delta_B$,
respectively. In the case of $\Delta_B$ one also has to take into account
the various $\Gamma$'s and $\tilde n$'s on the right hand side of
(\ref{e7}) \cite{Kobes:1990xf, Kobes:1990dc}.
As in \cite{Kobes:1990xf, Kobes:1990dc} one may argue
that $\delta X$ is 1PI up to a full ghost propagator and
tadpole insertions. The singularities of the ghost propagator will be
different from the physical dispersion laws since there is
no HTL ghost self energy. The tadpoles are 1PI up to a full Higgs 
propagator evaluated at zero momentum, and at this point the Higgs 
propagator is non-singular.
In general, the 1PI parts of $\delta X$ are expected to have no singularities, 
apart from possible mass-shell singularities which can be 
avoided by introducing an infrared cut-off 
to be lifted only at the very end of the calculation \cite{Baier:1991dy,Rebhan:1992ak}.

On the right hand side of (\ref{e7}) additional 
singularities arise from the 
factor $1/K^2$ contained in the $\tilde n$'s and from 
$1/(\Gamma_1^2-\Gamma_{55}\Gamma_D\tilde{n}^2)$. These kinematical 
singularities have to be excluded from the gauge independence proof. 
The expression $\Gamma_1^2-\Gamma_{55}\Gamma_D\tilde{n}^2$ 
is obviously gauge dependent already at tree level,
and in general according to
\begin{eqnarray}
  &&\!\!\!\!\!\!\!\!\!\!\delta[\Gamma_1^2-\Gamma_{55}\Gamma_D\tilde{n}^2]
  +\delta\bar\varphi{\partial\over\partial\bar\varphi}
  [\Gamma_1^2-\Gamma_{55}\Gamma_D\tilde{n}^2]=\nonumber\\
  &&\!\!\!\!\!=2\Big[\Big((\Gamma_1^2
  -\Gamma_{55}\Gamma_D\tilde n^2){K^\mu K_\rho\over K^2}-(\tilde n^2\Gamma_{55}
  \Gamma_C+\Gamma_1\Gamma_2){K^\mu \tilde n_\rho\over k}\Big)
  \delta X^\rho_{,\mu}\nonumber\\
  &&\!\!\!\!+\Big((\Gamma_1\Gamma_C-\Gamma_D\Gamma_2)\tilde n_\rho
  -2\Gamma_1\Gamma_D{\tilde n^2 K_\rho\over k}\Big)\delta X^\rho_{,5}
  -(\Gamma_1^2+\tilde n^2\Gamma_{55}\Gamma_D)\delta X^5_{,5}\Big].\qquad
\end{eqnarray}

\section{A non-Abelian Higgs model \label{shna}}
Let us consider a non-Abelian Higgs model with the gauge field and
the scalar field both belonging to the adjoint representation of $SU(2)$.
The Lagrangian is given by
\begin{equation}
  \mathcal L_{inv} = -{1\over4}F_{\mu\nu a}F^{\mu\nu}_a+(D_\mu\varphi)_a
  (D^\mu\varphi)_{a}-V(\varphi),
\end{equation}
with $F^{\mu\nu}_a=\partial^\mu A^\nu_a-\partial^\nu A^\mu_a+g 
\varepsilon_{abc}A^\mu_bA^\nu_c$, $(D^\mu \varphi)_a=\partial^\mu\varphi_a
+g\varepsilon_{abc}A^\mu_b\varphi_c$, and
\begin{equation} 
  V(\varphi)={m^2\over2}\varphi_a\varphi_a+{\lambda\over4!}
  (\varphi_a\varphi_a)^2,
\end{equation} 
with $m^2<0$. 
We assume that the expectation value of $\varphi_a$ is 
proportional to $\delta_{a3}$. Using the discrete symmetries of the
Lagrangian \cite{Fukuda:1975di}, it is easy to see that the gauge field propagator
is diagonal in color space. It should be noted, however, that the gauge field propagator in the Higgs phase is
not proportional to the unit matrix, which can be expected from the 
discussion in Sec. \ref{sglob}.

It turns out that the Higgs propagator
$\Delta_{\varphi_3\varphi_3}$ fulfills the same gauge dependence identity as 
in the Abelian case
[Eq. (\ref{e43})]. 
For the gauge field propagator one finds that
$\Delta^{\mu\nu}_{11}$ and $\Delta^{\mu\nu}_{22}$ both fulfill Eqs. 
(\ref{ee44})-(\ref{e7}),
and that $\Delta^{\mu\nu}_{33}$ obeys the same gauge dependence identity as in
QED \cite{Kobes:1990dc}. The fact that the gauge dependence identities are 
identical with those of Abelian models is of course only due to the particular
simplicity of the $SU(2)$ model.  

 The $SU(2)$ Higgs model is an interesting toy model, since
we will see in chapter~\ref{ccsc} that also in the case of color superconductivity the gluon propagator is
in general not proportional to the unit matrix in color space.

\chapter{Quark self energy in ultradegenerate QCD \label{csigma}}

\section{General remarks}
In this chapter we will compute the quark self energy in normal ultradegenerate QCD, both at zero and (small) finite temperature.
As discussed in the introduction, and in more detail in chapter \ref{ccsc},  quark matter at high density and
sufficiently small temperature is actually in a color superconducting phase.
Nevertheless, it is interesting to study also the properties of normal, i.e. non-supercon\-ducting quark matter.
First, a good understanding of normal
quark matter is certainly helpful before tackling the more complicated case of color superconductivity.
Moreover, it is conceivable that quark matter in young (proto-)neutron 
stars is already dense enough for deconfinement,
but still not cold enough for color superconductivity. If this is the case, the results of our computations will be relevant 
in particular for the cooling behavior of proto-neutron stars.

The quark self energy will be used in the next chapter to compute the specific heat of normal cold dense quark matter.
We will evaluate the quark self energy only on the light cone, since this is the quantity which enters
the formula for the specific heat at leading order [see Eq. (\ref{x9}) below]. Furthermore it is sufficient for the
computation of the specific heat to consider the fermionic 
quasiparticles in the vicinity of the Fermi surface.  Thus the quark momentum is hard, while the momentum of 
the gluon in the one-loop self energy (see Fig. \ref{figquark}) may be arbitrarily soft. 
Therefore \cite{Braaten:1989mz} we have to use a resummed
gluon propagator in order to avoid an IR divergence in the quark self energy.
For this reason we will review the computation of the HDL resummed gluon propagator in the 
next section.

\section{Gluon self energy \label{secgluon}}

\begin{figure}
 \begin{center}
  \includegraphics{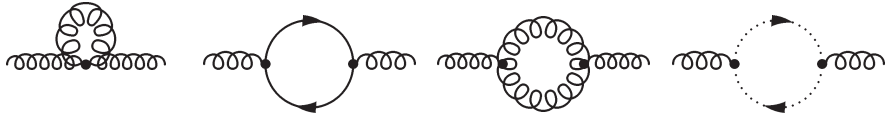}
  \vspace{-2mm}
  {\it \caption{One-loop gluon self energy.} \label{figgluon}}
 \end{center}
\end{figure}

The gluon self energy at the one-loop level is given by the
diagrams of Fig. \ref{figgluon}. The second one of the
four diagrams depends on the chemical potential {\it and} the
temperature, while the three
remaining ones give only finite temperature corrections to the
gluon self energy. Below we will only need the gluon self energy
at zero temperature, therefore we evaluate here only the
diagram with a quark loop in Fig. \ref{figgluon}. 
For QCD the diagram contains an additional factor of $\Tr[T^aT^b]={1\over2}\delta^{ab}$
compared to the photon self energy in QED.
In the following we shall drop the Kronecker delta, and define
\begin{equation}
\gf^2 = \left\{
        \begin{array}{cc} 
        \displaystyle \frac{g^2 N_f}{2} \, , & {\rm QCD} \, , \\ & \\
        g^2 N_f \, , & {\rm QED} \, . \\ \end{array} \right.\label{geffdef}
\end{equation}
Then our final results will be applicable both for QCD and QED.

In the imaginary time 
formalism \cite{LeB:TFT} we get for the second diagram of Fig. \ref{figgluon}
\begin{equation}
  \Pi_{\mu\nu}(i\omega,q)=-\gf^2T\sum_{\omega_n}\int{d^3k\over(2\pi)^3}
  \Tr\left[\gamma_\mu S_0(K-Q)\gamma_\nu S_0(K)\right], \label{gl1}
\end{equation}
where $Q^\mu=(i\omega,{\bf q})$ and $K^\mu=(i\omega_n,{\bf k})$,
and the free quark propagator is given by  
\begin{equation}
  S_0(i\omega_n,k)=\int_{-\infty}^\infty{dk_0\over2\pi}{\slashed{K}\rho_0(K)
  \over k_0-i\omega_n-\mu} \label{e63}
\end{equation}
with
\begin{equation}
  \rho_0(K)=2\pi\mathrm{sgn}(k_0)\,\delta\left(k_0^2-k^2\right)=
  {\pi\over k}\left[\delta(k_0-k)-\delta(k_0+k)\right].
\end{equation}
Using the methods outlined in appendix \ref{appa} it is straightforward to evaluate
the Matsubara sum in Eq. (\ref{gl1}), 
\begin{equation}
  T\sum_{\omega_n}{1\over k_0-i\omega_n-\mu}{1\over p_0-i\omega_n+i\omega-\mu}=
  {n_f(k_0-\mu)-n_f(p_0-\mu)\over k_0-p_0-i\omega},
\end{equation}
where $n_f$ is the Fermi-Dirac distribution defined in Eq. (\ref{nf}).
After performing the Dirac trace in Eq. (\ref{gl1}) we get \cite{Ipp:2003qt},
\begin{eqnarray}
  &&
  \Pi_{\mu\nu}(i\omega,q)=-4\gf^2\int {d^3k\over(2\pi)^3}\int_{-\infty}^\infty
  {dk_0\over2\pi}\int_{-\infty}^\infty{dp_0\over2\pi}\rho_0(k_0,k)
  \rho_0(p_0,p)\nonumber\\
  &&\quad\times{n_f(k_0-\mu)-n_f(p_0-\mu)\over k_0-p_0-i\omega}\left(k_\mu p_\nu
  +p_\mu k_\nu-g_{\mu\nu}k\cdot p\right),
\end{eqnarray}
where ${\bf p}={\bf k}-{\bf q}$. Now we can perform the analytic 
continuation $i\omega\to q_0+i\eta$ to get the retarded self energy.  Using
\begin{equation}
  {1\over x+i\eta}={\cal P}{1\over x}-\delta(x)
\end{equation}
we can separate the real and imaginary parts of the self energy. 
It is convenient to define
\begin{equation}
  \Pi_G:=g^{\mu\nu}\Pi_{\mu\nu},\quad \Pi_H:=u^\mu u^\nu\Pi_{\mu\nu},
\end{equation}
with $u^\mu=(1,0,0,0)$. These two structure functions are related
to the transverse and longitudinal self energies via
\begin{equation}
  \Pi_L=-{q_0^2-q^2\over q^2}\Pi_H, \quad
  \Pi_T={1\over2}\left(-\Pi_L+\Pi_G\right). \label{g10}
\end{equation}
After subtraction of the vacuum contribution one finds 
for the real parts \cite{Weldon:1982aq}
\begin{eqnarray}
  &&\!\!\!\!\!\!\!\!\!\!\!\!\!\!\!\!\!\!\!\!\!\!\!\!\!\!\!\!
  \real\Pi_G(q_0,q)={\gf^2\over\pi^2}\int_0^\infty dk\, n(k)\nonumber\\
  &&\times\left(4k+{q_0^2-q^2\over2q}\log\bigg|{(2k+q_0+q)(2k-q_0+q)
  \over(2k+q_0-q)(2k-q_0-q)}\bigg|\right), \label{repig}\\
  &&\!\!\!\!\!\!\!\!\!\!\!\!\!\!\!\!\!\!\!\!\!\!\!\!\!\!\!\!
  \real\Pi_H(q_0,q)={\gf^2\over\pi^2}\int_0^\infty dk\, n(k)
  \bigg[2k\left(1-{q_0\over q}\log\bigg|{q_0+q\over q_0-q}\bigg|\right)\nonumber\\
  &&+{(2k+q_0+q)(2k+q_0-q)\over4q}\log\bigg|{2k+q_0+q\over2k+q_0-q}\bigg|\nonumber\\
  &&-{(2k-q_0-q)(2k-q_0+q)\over4q}\log\bigg|{2k-q_0-q\over2k-q_0+q}\bigg|\bigg],
\end{eqnarray}
where we use the abbreviation
\begin{equation}
  n(k)={1\over2}
  \left(n_f(k-\mu)+n_f(k+\mu)\right).
\end{equation}
For the imaginary parts one finds \cite{Ipp:2003qt}
\begin{equation}
  \imag\Pi_{G,H}(q_0,q)=-{1\over2}(\pi^+_{G,H}(q_0,q)-\pi^+_{G,H}(-p_0,p)), \label{imp1}
\end{equation}
with
\begin{eqnarray}
  &&\!\!\!\!\!\pi^+_G(q_0,q)={\gf^2\over\pi p}\int_0^\infty dk\,n(k)(q_0^2-q^2)F(k,q_0,q),\\
  &&\!\!\!\!\!
  \pi^+_H(q_0,q)={\gf^2\over2\pi p}\int_0^\infty dk\,n(k)(2k+q-q_0)(2k-q-q_0)F(k,q_0,q),\qquad
\end{eqnarray}
where
\begin{equation}
  F(k,q_0,q):=\mathrm{sgn}(k-q_0)\Theta(|k-q|\le|k-q_0|\le|k+q|), \label{imp4}
\end{equation}
where we use the notation $\Theta({\rm true\ expression})=1$
and $\Theta({\rm false\ expression})=0$. 

In general the remaining 
integrations over $k$ can be performed analytically only for
the imaginary parts \cite{Ipp:2003qt}. In the zero temperature limit, 
however, a closed
result can be obtained also for the real parts. 
In this limit one finds for for $\Pi_L$ and $\Pi_T$ for $q_0\ge0$ \cite{Ipp:2003qt}
\begin{eqnarray}
  &&\!\!\!\!\!\!\!\!\!\!\!\!\!\!\!\!\!\!\!\!\!\!\!
  \real\Pi_L(q_0,q)\big|_{T=0}={\gf^2\left(q^2-q_0^2\right)\over48\pi^2q^2}
  \Big[32\mu^2+{1\over q}N(q_0,q)\Big],\\
  &&\!\!\!\!\!\!\!\!\!\!\!\!\!\!\!\!\!\!\!\!\!\!\!
  \real\Pi_T(q_0,q)\big|_{T=0}={\gf^2\over96\pi^2q^3}\Big[16\mu^2q\left(q^2+2q_0^2\right)\nonumber\\
  &&+\left(q^2-q_0^2\right)\left(6q^2M(q_0,q)-N(q_0,q)\right)\Big],\\
  &&\!\!\!\!\!\!\!\!\!\!\!\!\!\!\!\!\!\!\!\!\!\!\!
  \imag\Pi_L(q_0,q)\big|_{T=0}=-{\gf^2\left(q^2-q_0^2\right)\over48\pi q^3}\nonumber\\
  &&\times\Big[W(q_0,q)U(q_0,q)-W(-q_0,q)U(-q_0,q)\Big],\\
  &&\!\!\!\!\!\!\!\!\!\!\!\!\!\!\!\!\!\!\!\!\!\!\!
  \imag\Pi_T(q_0,q)\big|_{T=0}={\gf^2\left(q^2-q_0^2\right)\over96\pi q^3}\nonumber\\
  &&\!\!\!\!\!\times\Big[\left(W(q_0,q)+6q^2\right)U(q_0,q)
  -\left(W(-q_0,q)+6q^2\right)U(-q_0,q)\Big],
\end{eqnarray}
where we have used the abbreviations \cite{Ipp:2003qt}
\begin{eqnarray}
  U(q_0,q)&=&(2\mu-|q_0+q|)\Theta(2\mu-|q_0+q|),\\
  V(q_0,q)&=&(2\mu-q-q_0)(2\mu+2q-q_0),\\
  W(q_0,q)&=&\Theta(q_0+q)V(q_0,q)+\Theta(-q_0-q)V(-q_0,-q),\\
  R(q_0,q)&=&(q_0+q)\log|q_0+q|-(2\mu+q_0+q)\log|2\mu+q_0+q|,\\
  S(q_0,q)&=&\left(-12\mu^2q_0+(2q-q_0)(q+q_0)^2\right)\log|q_0+q|\nonumber\\
  &&\quad+(2\mu-2q+q_0)(2\mu+q+q_0)^2\log|2\mu+q+q_0|,\\
  M(q_0,q)&=&R(q_0,q)-R(q_0,-q)+R(-q_0,q)-R(-q_0,-q),\\
  N(q_0,q)&=&S(q_0,q)-S(q_0,-q)+S(-q_0,q)-S(-q_0,-q).
\end{eqnarray}

If $q,q_0\ll\mu$ the leading contribution to the self energy is quadratic in $\mu$. 
This contribution comes from hard momenta in the quark loop ($k\sim\mu$), therefore
the corresponding self energy ist called Hard Dense Loop (HDL) self energy 
\cite{Braaten:1989mz,Altherr:1992mf,Vija:1994is,Manuel:1995td}. In the
following we shall denote the HDL approximation with a tilde ($\tilde{\ }$).
The explicit expressions are considerably simpler than the general ones given above \cite{LeB:TFT},
\begin{eqnarray}
  \real\tilde\Pi_L(q_0,q)&=&2m^2{q^2-q_0^2\over q^2}\left(1-{q_0\over2q}
  \log\bigg|{q_0+q\over q_0-q}\bigg|\right),\\
  \real\tilde\Pi_T(q_0,q)&=&m^2{q_0\over q}\left({q_0\over q}+{q^2-q_0^2\over2q^2}
  \log\bigg|{q_0+q\over q_0-q}\bigg|\right),\label{repith}\\
  \imag\tilde\Pi_L(q_0,q)&=&m^2{q_0\left(q^2-q_0^2\right)\over q^3}\pi\Theta(q^2-q_0^2),\\
  \imag\tilde\Pi_T(q_0,q)&=&-m^2{q_0\left(q^2-q_0^2\right)\over2q^3}\pi\Theta(q^2-q_0^2),\label{impith}
\end{eqnarray} 
where $m^2=\gf^2\mu^2/(2\pi^2)$. The leading finite temperature corrections have the same
functional dependence on $q_0$ and $q$, one only has to include a finite temperature correction into 
$m^2$. For QCD one finds
\begin{equation}
   m^2=N_f{g^2\mu^2\over4\pi^2}+\left(N_c+{N_f\over2}\right){g^2 T^2\over6}, \label{g32}
\end{equation}
where the contribution proportional to $N_c$ comes from the gluon and ghost loops
in Fig. \ref{figgluon}.
The part of the gluon self energy which is quadratic in the temperature is called
Hard Thermal Loop (HTL) self energy.

Using Dyson's equation $\Pi_{\mu\nu}=D^{-1}_{\mu\nu}-D^{-1}_{0,\mu\nu}$ 
one can obtain the one-loop resummed gluon propagator.
In the HDL approximation one finds in covariant gauge \cite{LeB:TFT}, using the 
tensors defined in Eqs. (\ref{proj1}) and (\ref{proj2}),
\begin{equation}
  \tilde D_{\mu\nu}(q_0,q)=-\tilde D_T\cA_{\mu\nu}-
  \tilde D_L\cB_{\mu\nu}-{\xi\over q^2-q_0^2}\cD_{\mu\nu}
\end{equation}
and in Coulomb gauge (see e.g. \cite{Manuel:2000mk})
\begin{equation}
  \tilde D_{\mu\nu}(q_0,q)=-\tilde D_T\cA_{\mu\nu}+
  \tilde D_L {q_0^2-q^2\over q^2}g_{\mu0}g_{\nu0}+\xi_C {Q_\mu Q_\nu\over q^4} \label{e634} 
\end{equation}
where
\begin{eqnarray}
  \tilde D_L&=&{1\over q^2-q_0^2+\tilde\Pi_L}, \label{g35}\\
  \tilde D_T&=&{1\over q^2-q_0^2+\tilde\Pi_T}. \label{g36}
\end{eqnarray}
The HDL spectral density for the transverse part is given by
\begin{equation}
  \rho_{T}(q_0,q)=2\,\imag \tilde D_{T}(q_0+i\eta,q). \label{g37}
\end{equation}
In covariant gauge the longitudinal propagator
contains an unphysical factor of $1/(q^2-q_0^2)$ \cite{Kobes:1990dc,Kobes:1990xf}, 
therefore one defines the longitudinal spectral density as
\begin{equation}
  \rho_{L}(q_0,q)=2{q_0^2-q^2\over q^2}\imag \tilde D_{L}(q_0+i\eta,q). \label{g38}
\end{equation}
Both $\rho_T$ and $\rho_L$ contain a pole and a cut contribution.
The explicit expressions can be found in \cite{LeB:TFT}, and are also given in appendix \ref{appc} for convenience.
In terms of the spectral densities, the structure functions of the propagator can be
written as \cite{LeB:TFT}
\begin{eqnarray}
  \tilde D_T(q_0,q)&=&\int_{-\infty}^\infty {dq_0^\prime\over2\pi}{\rho_T(q_0^\prime,q)\over 
  q_0^\prime-q_0}, \label{e639} \\
  {q_0^2-q^2\over q^2}
  \tilde D_L(q_0,q)&=&\int_{-\infty}^\infty {dq_0^\prime\over2\pi}{\rho_L(q_0^\prime,q)\over 
  q_0^\prime-q_0}-{1\over q^2}. \label{e640}
\end{eqnarray}

\section{Fermion self energy on the light cone \label{secq}}

\begin{figure}
\begin{center}
 \parbox{3cm}{
 \includegraphics{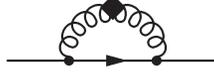}}
 \vspace{-.5cm}
 {\it\caption{One-loop quark self energy with resummed gluon propagator.} \label{figquark}}
\end{center}
\end{figure}

The fermion self energy is defined through
\begin{equation}
  S^{-1}(P)=S_0^{-1}(P)+\Sigma(P),
\end{equation}
where $S_0(P)=-(P\!\!\!\!/)^{-1}$ is the free fermion propagator,
and $P^\mu=(p^0,{\bf p})$ with $p^0=i\omega_n+\mu$,
$\omega_n=(2n+1)\pi T$ in the imaginary time formalism,
and $P^\mu=(E,{\bf p})$ after analytic continuation to Minkowski space.

With the energy projection operators 
\begin{equation}
  \Lambda_{\bf p}^\pm={1\over2}(1\pm\gamma_0\gamma^i\hat p^i) \label{ep}
\end{equation}
we decompose $\Sigma(P)$ in the quasiparticle and antiquasiparticle self energy,
\begin{equation}
  \Sigma(P)=\gamma_0\Lambda_{\bf p}^+\Sigma_+(P)-\gamma_0\Lambda_{\bf p}^-\Sigma_-(P),
\end{equation}
and
\begin{equation}
  \gamma_0 S^{-1}=S_+^{-1}\Lambda_{\bf p}^+ + 
  S_-^{-1}\Lambda_{\bf p}^-
\end{equation}
so that $S_\pm^{-1}=-[p^0\mp(|\mathbf p|+\Sigma_\pm)]$.

The one-loop fermion self energy is given by
\begin{equation}
  \Sigma(P)=-g^2 C_f T\sum_{\omega}\int {d^3q\over(2\pi)^3}\,\gamma^\mu S_0(P-Q)\gamma^\nu D_{\mu\nu}(Q),
\end{equation}
where $ D_{\mu\nu}$ is the gauge boson propagator.

Following \cite{Braaten:1991dd} we introduce an intermediate scale $q^*$, such that $m\ll q^*\ll\mu$, 
and we divide the $q$-integration into a soft part ($q<q^*$) and a hard part ($q>q^*$),
\begin{equation}
  \Sigma_+=\Sigma_+^\mathrm{(s)}+\Sigma_+^\mathrm{(h)}.
\end{equation}
For the hard part we can use the free gluon propagator, whereas for the 
soft part we shall use the HDL resummed gluon propagator.

The hard part of the fermion self energy on the light cone 
in a general Cou\-lomb gauge is given by the gauge independent
expression
\begin{eqnarray} 
  &&\!\!\!\!\!\!\!
  \Sigma_+^\mathrm{(h)}(E)=-{g^2C_f\over 8\pi^2}\int_{q^*}^\infty dq\,q^2\int_{-1}^1 dt\int_{-\infty}^\infty dr_0
  \left[\delta(r_0-r)-\delta(r_0+r)\right]\nonumber\\
  &&\times\Bigg\{2\left(\mathrm{sgn}(r_0)-\hat{\bf p}\!\cdot\!\hat{\bf q}\:\hat{\bf r}\!\cdot\!\hat{\bf q}\right)
  \int_{-\infty}^\infty dq_0{1\over2q}\left[\delta(q_0-q)-\delta(q_0+q)\right] \nonumber\\
  &&\quad\times{1+n_b(q_0)-n_f(r_0-\mu)-\Theta(q_0)+\Theta(-r_0)\over r_0+q_0-E-i\eta}\nonumber\\
  &&\quad+\left(\mathrm{sgn}(r_0)+\hat{\bf r}\!\cdot\!\hat{\bf p}\right)
  {1\over q^2}\left[n_f(r_0-\mu)-\Theta(-r_0)\right]\Bigg\}, \label{sh}
\end{eqnarray}
where ${\bf r}={\bf p}-{\bf q}$ and $E= p$. The distribution functions $n_b$ and $n_f$
are given by Eqs. (\ref{nb}) and (\ref{nf}).
In Eq. (\ref{sh}) we have subtracted the vacuum parts of the distribution functions,
since we know anyway that the vacuum part of the fermion self energy on the light
cone vanishes because of gauge and Lorentz invariance. This subtraction is in fact necessary, because otherwise
we would get a spurious vacuum contribution coming
from the fact the we do not use an $O(4)$ invariant cutoff for the energy-momentum integration.
The finite-$T$ and finite-$\mu$ parts, however, are not affected by these
subtleties in the UV region.

After performing the angular integration we find
\begin{eqnarray}
  &&\Sigma_+^\mathrm{(h)}(E)={g^2C_f\over8\pi^2p^2}\int_{q^*}^\infty dq\,q\,\bigg\{
  2\left(q-|p-q|+p\coth\left({q\over2T}\right)\right)\nonumber\\
  &&\quad\qquad\qquad+T\log\left[{\left(1+e^{(|p-q|-\mu)/ T}\right)\left(1+e^{(|p-q|+\mu)/ T}\right)\over
  \left(1+e^{(p+q-\mu)/ T}\right)\left(1+e^{(p+q+\mu)/ T}\right)}\right]\bigg\}.
\end{eqnarray}
The integral is finite in the limit $q^*\to0$, with the result
\begin{equation}
  \Sigma_+^\mathrm{(h)}={M_\infty^2\over 2p} (1+\cO(T^2/\mu^2)),
\end{equation}
with $M_\infty^2=g^2C_f\mu^2/(4\pi^2)$. 
Here $q^*$ enters as a correction proportional to $(q^*/\mu)^2$, so that
we can send $q^*$ to zero. Conversely we expect that in the soft contribution
we should be able to send $q^*$ to infinity without encountering
divergences, as will indeed be the case, but only after all
soft contributions are added together.
 
For the soft part one finds on the light cone in a general Coulomb gauge
the gauge independent expression
\cite{Manuel:2000mk} 
\begin{eqnarray}
  &&\!\!\!\!\!\!\!\!\!\!\!\!\!\!\!\!
  \Sigma_\pm^\mathrm{(s)}(E)=-{g^2C_f\over 8\pi^2}\int_0^{q^*} dq\,q^2\int_{-1}^1 dt\int_{-\infty}^\infty dr_0
  \left[\delta(r_0-r)-\delta(r_0+r)\right]\nonumber\\
  &&\times\Bigg\{2\left(\pm\mathrm{sgn}(r_0)-\hat{\bf p}\!\cdot\!\hat{\bf q}\:\hat{\bf r}\!\cdot\!\hat{\bf q}\right)
  \int_{-\infty}^\infty{dq_0\over2\pi}\rho_T(q_0,q){1+n_b(q_0)-n_f(r_0-\mu)\over r_0+q_0\mp|E|-i\eta}\nonumber\\
  &&\quad+\left(\pm\mathrm{sgn}(r_0)+\hat{\bf r}\!\cdot\!\hat{\bf p}\right)
  \bigg[\int_{-\infty}^\infty{dq_0\over2\pi}\rho_L(q_0,q){1+n_b(q_0)-n_f(r_0-\mu)\over r_0+q_0\mp|E|-i\eta}\nonumber\\
  &&\quad-{1\over q^2}\left({1\over2}-n_f(r_0-\mu)\right)\bigg]\Bigg\}, \label{r1}
\end{eqnarray}
where $E=\pm p$, and $\rho_T$ and $\rho_L$ are the spectral densities of transverse and longitudinal
gauge bosons, 
as given in Eqs. (\ref{g37}) and (\ref{g38}).
We may use $q\ll |E|,r$ because of $q<q^*$ and $|E|\sim\mu$.
Depending on the sign of $E$,
we can drop the term $\delta(r_0+r)$ 
or the term $\delta(r_0-r)$
in Eq. (\ref{r1}), since its contribution is suppressed with $\sim q/E$ 
compared to the remaining contribution. 
Then we find for the soft contribution to the real part of $\Sigma_+$
\begin{eqnarray}
  &&\!\!\!\!\!\!\!\!\!\!
  \real\Sigma_+^\mathrm{(s)}=-{g^2C_f\over 8\pi^2}\int_0^{q^*} dq\,q^2\int_{-1}^1 dt
  \bigg[\int_{-\infty}^\infty{dq_0\over\pi}\left[(1-t^2)\rho_T(q_0,q)+\rho_L(q_0,q)\right]\nonumber\\
  &&\times\:\mathcal{P}{1+n_b(q_0)-n_f(E-\mu-qt)\over q_0-qt}-{1\over q^2}\left(1-2n_f(E-\mu-qt)\right)\bigg].
  \label{si4}
\end{eqnarray}
This quantity vanishes for $E=\mu$ by symmetric integration. After
performing the $q_0$-integration we therefore have
\begin{eqnarray}
  &&\real\Sigma_+^\mathrm{(s)}=
  {g^2C_f\over4\pi^2}\int_0^{q^*} dq\,q^2\int_{-1}^1dt
  \left(n_f(E-\mu-qt)-n_f(-qt)\right)\nonumber\\
  &&\qquad\qquad\qquad\times(1-t^2)\left[\real \tilde D_T(qt,q)-\real \tilde D_L(qt,q)\right], \label{rsi0}
\end{eqnarray}
where $\tilde D_{T}$ and  $\tilde D_{L}$ are given in Eqs. (\ref{g35}) and (\ref{g36}), respectively.
For $\imag\Sigma_+$ (which receives no hard contribution) we find
in an analogous way
\begin{eqnarray}
  &&\!\!\!\!\!\!\!\!\!\!
  \imag\Sigma_+=-{g^2C_f\over 8\pi^2}\int_0^{q^*} dq\,q^2\int_{-1}^1 dt
  \left((1-t^2)\rho_T(qt,q)+\rho_L(qt,q)\right)\nonumber\\
  &&\qquad\times\:\left[1+n_b(qt)-n_f(E-\mu-qt)\right].
\end{eqnarray}

The antiquasiparticle self energy $\Sigma_-^\mathrm{(s)}$ is obtained by
inserting negative values of $E$ in the above expressions for $\Sigma_+^\mathrm{(s)}$ and
including an overall factor $(-1)$.
 With $\mu>0$ we can
then replace $n_f(E-\mu-qt)$ by 1.

\section{Expansion 
for small $|E-\mu|$ and small $T$}
In this section we will perform an expansion of  $\Sigma_+$ in the region
\begin{equation}
  T\sim |E-\mu|\ll g\mu\ll\mu. \label{ineq1}
\end{equation}
This region is relevant for the computation of the low temperature specific heat, see chapter \ref{cspecific}.
We will use the expansion parameter $a:=T/m$,
and we define
$\lambda:=(E-\mu)/T$. From (\ref{ineq1}) we have $a\ll1$ and $\lambda\sim\mathcal{O}(1)$. 

\subsection{The first few terms in the series}
In the part with
the transverse gluon propagator we substitute 
\begin{equation}
  q=ma^{1/3}z, \quad t=a^{2/3}v/z. \label{su1}
\end{equation}
After expanding the
integrand with respect to $a$ we find for the transverse contribution
\begin{eqnarray}
  &&\!\!\!\!\!\!\!\!\!\!
  \real\Sigma^\mathrm{(s)}_{+(T)}=-{g^2C_fma\over\pi^2}\int_{-{q^*\over am}}^{q^*\over am}dv 
  \int_{a^{2/3}|v|}^{q^*\over a^{1/3}m}dz\,{e^\lambda-1\over(1+e^v)(1+e^{\lambda-v})}\nonumber\\
  &&\!\!\!\!\times\bigg[{z^5\over v^2\pi^2+4z^6}+{2v^2z(v^2\pi^2-4z^6)\over(v^2\pi^2+4z^6)^2}a^{2/3}
  -{16v^4z^3(3v^2\pi^2-4z^6)\over(v^2\pi^2+4z^6)^3}a^{4/3}+\ldots\bigg].\nonumber\\ \label{si7}
\end{eqnarray}
The $z$-integrations are straightforward. In the $v$-integrals we may send the integration limits to
$\pm\infty$. Using the formulae\footnote{$\Gamma(\alpha)$ is the Euler Gamma function, defined as
$
  \Gamma(\alpha):=\int_0^\infty dt\, t^{\alpha-1}e^{-t}.
$
The Gamma function satisfies $\Gamma(n)=(n-1)!$ for $n\in\mathbbm N$ and $\Gamma'(1)=-\gamma_E$, 
where $\gamma_E$ is Euler's constant,
$
  \gamma_E=\lim_{m\to\infty}\left(\sum_{k=1}^m{1\over k}-\log m\right)\simeq0.5772156\ldots
$
Furthermore, $\Li_\alpha(s)$ is the polylogarithm, defined as $\Li_\alpha(s)=\sum_{k=1}^\infty {s^k k^{-\alpha}}$.
}
\begin{eqnarray}
  &&\int_{-\infty}^\infty dv{e^\lambda-1\over(1+e^v)(1+e^{\lambda-v})}|v|^\alpha\nonumber\\
  &&\qquad=\Gamma(\alpha+1)\left[\Li_{\alpha+1}(-e^{-\lambda})-\Li_{\alpha+1}(-e^{\lambda})\right]
  \quad\forall \alpha\ge0,\quad\\
  &&\int_{-\infty}^\infty dv{e^\lambda-1\over(1+e^v)(1+e^{\lambda-v})}\log|v|\nonumber\\
  &&\qquad=-\gamma_E\lambda+{\partial\over\partial \alpha}
  \left(\Li_{\alpha+1}(-e^{-\lambda})-\Li_{\alpha+1}(-e^{\lambda})\right)\Big|_{\alpha=0},
\end{eqnarray}
we find, neglecting terms which are suppressed at least with $(m/q^*)^4$,
\begin{eqnarray}
  &&\!\!\!\!\!\!\!\!\!\!\!\! \real\Sigma^\mathrm{(s)}_{+(T)}=-g^2C_fm\nonumber\\
  &&\!\!\!\!\!\times\Bigg\{{a\over12\pi^2}\left[\lambda\log\left({2(q^*)^3\over a m^3\pi}\right)+\gamma_E\lambda
  -{\partial\over\partial \alpha}\left(\Li_{\alpha+1}(-e^{-\lambda})-\Li_{\alpha+1}(-e^{\lambda})\right)\Big|_{\alpha=0}
  \right]\nonumber\\
  &&+{2^{1/3}a^{5/3}\over9\sqrt{3}\pi^{7/3}}\Gamma\left(\textstyle{5\over3}\right)
  \left(\Li_{5/3}(-e^{-\lambda})-\Li_{5/3}(-e^\lambda)\right)\nonumber\\
  &&-20{2^{2/3}a^{7/3}\over27\sqrt{3}\pi^{11/3}}\Gamma\left(\textstyle{7\over3}\right)
  \left(\Li_{7/3}(-e^{-\lambda})-\Li_{7/3}(-e^\lambda)\right)\nonumber\\
  &&+{8(24-\pi^2)a^3\log a\over27\pi^6}\lambda(\lambda^2+\pi^2)+\mathcal{O}(a^3)\Bigg\}. 
\end{eqnarray}

In the longitudinal part we substitute $q=mx$ and $t=au/x$. In  a similar way as for the transverse part
we find
\begin{equation}
  \real\Sigma^\mathrm{(s)}_{+(L)}=-g^2C_fm\left[ {a\lambda\over8\pi^2}\log\left({2m^2\over(q^*)^2}\right)
  -{(\pi^2-4)a^3\log a\over96\pi^2}\lambda(\lambda^2+\pi^2)+\mathcal{O}(a^3) \right].
\end{equation}

Turning now to $\imag\Sigma_+$ we notice that it vanishes at $E=\mu$
only in the case of $T=0$. For finite temperature, however small,
there is an IR divergent contribution in the transverse sector
\cite{LeB:TFT},
\begin{equation}
\imag\Sigma_{+(T)}\big|_{E=\mu}=
-{g^2 C_f T\over4\pi} \log{m\over\Lambda_{\rm IR}}
\end{equation}
where the infrared cutoff may be provided at finite temperature
by the nonperturbative magnetic screening mass of QCD. In QED,
where no magnetostatic screening is possible, a resummation
of these singularities leads to nonexponential damping behavior
\cite{Blaizot:1996az}.

After subtraction of the energy independent part we have
\begin{eqnarray}
  &&\imag\Sigma_+
-\imag\Sigma_+\big|_{E=\mu}
={g^2C_f\over8\pi^2}\int_0^{q^*} dq\,q^2\int_{-1}^1dt
  \left(n_f(E-\mu-qt)-n_f(-qt)\right)\nonumber\\
  &&\qquad\qquad\qquad\qquad\qquad\qquad
\times\left[(1-t^2)\rho_T(qt,q)+\rho_L(qt,q)\right]. \label{imsi0}
\end{eqnarray}
Following the
steps which led to Eq. (\ref{si7}), we find for the transverse contribution
\begin{eqnarray}
  &&\!\!\!\!\!\!\!\!\!\!\!
  \imag\Sigma_{+(T)}-\imag\Sigma_{+(T)}\big|_{E=\mu}
  ={g^2C_fma\over2\pi}\int_{-{q^*\over am}}^{q^*\over am}dv 
  \int_{a^{2/3}|v|}^{q^*\over a^{1/3}m}dz\,{e^\lambda-1\over(1+e^v)(1+e^{\lambda-v})}\nonumber\\
  &&\qquad\times\bigg[-{z^2v\over v^2\pi^2+4z^6}+{16v^3z^4\over(v^2\pi^2+4z^6)^2}a^{2/3}
  +{16v^5(v^2\pi^2-12z^6)\over(v^2\pi^2+4z^6)^3}a^{4/3}+\ldots\bigg].\nonumber\\
\end{eqnarray}
Using the formula\footnote{$\zeta(\alpha)$ is the Riemann zeta function, which is defined as
$
  \zeta(\alpha):=\sum_{k=1}^\infty k^{-\alpha}\equiv\Li_\alpha(1).
$}
\begin{eqnarray}
  &&\!\!\!\!\!\!\!\!\!
  \int_{-\infty}^\infty dv{e^\lambda-1\over(1+e^v)(1+e^{\lambda-v})}
  |v|^\alpha\mathrm{sgn}(\alpha)\nonumber\\
  &&\!\!\!\!=-\Gamma(\alpha+1)\left[\Li_{\alpha+1}(-e^{-\lambda})+\Li_{\alpha+1}(-e^{\lambda})
  +2\left(1-2^{-\alpha}\right)\zeta(\alpha+1)\right]
  \quad\forall \alpha\ge0 \nonumber\\
\end{eqnarray}
we find in a similar way as above
\begin{eqnarray}
  &&\!\!\!\!\!\!\!\!\!\!\!\!\!\!\!\!
  \imag\Sigma_{+(T)}-\imag\Sigma_{+(T)}^\mathrm{(s)}\big|_{E=\mu}
  ={g^2C_fm}\Bigg\{-{a\over12\pi}\log\cosh\left({\lambda\over2}\right)\nonumber\\
  &&-{2^{1/3}a^{5/3}\over9\pi^{7/3}}\Gamma\left(\textstyle{5\over3}\right)
  \left(\Li_{5/3}(-e^{-\lambda})+\Li_{5/3}(-e^\lambda)+2\left(1-2^{-2/3}\right)
  \zeta\left(\textstyle{5\over3}\right)\right)\nonumber\\
  &&-20{2^{2/3}a^{7/3}\over27\pi^{11/3}}\Gamma\left(\textstyle{7\over3}\right)
  \left(\Li_{7/3}(-e^{-\lambda})+\Li_{7/3}(-e^\lambda)+2\left(1-2^{-4/3}\right)
  \zeta\left(\textstyle{7\over3}\right)\right)\nonumber\\
  &&+\mathcal{O}(a^3)\Bigg\}.
\end{eqnarray}
For the longitudinal part we obtain
\begin{equation}
  \imag\Sigma_{+(L)}-\imag\Sigma_{+(L)}^\mathrm{(s)}\big|_{E=\mu}
  =-{g^2C_fm}\left[{a^2\lambda^2\over64\sqrt{2}}+\mathcal{O}(a^3)\right]. \label{imsl}
\end{equation}


Putting the pieces together, and using the abbreviation $\emu=E-\mu$, 
we obtain for the real part
\begin{eqnarray}
  &&\!\!\!\!\!\!\!\!\!\!\!\!
  \real\Sigma_+={M_\infty^2\over 2E} - g^2C_fm\, \mathrm{sgn}(\emu)
  \Bigg\{{|\emu|\over12\pi^2m}\left[\log\left({4\sqrt{2}m\over\pi T f_1(\emu/T)}\right)+1\right]\nonumber\\
  &&+{2^{1/3}\sqrt{3}\over45\pi^{7/3}}\left({T\over m}f_2\left({\emu\over T}\right)\right)^{5/3}
  -20{2^{2/3}\sqrt{3}\over189\pi^{11/3}}\left({T\over m}f_3\left({\emu\over T}\right)\right)^{7/3}\nonumber\\
  &&-{6144-256\pi^2+36\pi^4-9\pi^6\over864\pi^6}\left({T\over m} 
  f_4\left({\emu\over T}\right)\right)^3\log\left({m\over T}\right)
  +\left({T\over m}\right)^3c_3\left({\emu\over T}\right)\nonumber\\
  &&+\cO\left({T\over m}\right)^{11/3}\Bigg\}, \label{resit}
\end{eqnarray}
where
\begin{eqnarray}
  f_1(\lambda)&=& \exp\left[1-\gamma_E+{1\over\lambda}{\partial\over\partial\alpha}  \label{fs1}
  \left(\Li_{\alpha+1}(-e^{-\lambda})-\Li_{\alpha+1}(-e^{\lambda})\right)\Big|_{\alpha=0}\right],\\
  f_2(\lambda)&=& \left|\Gamma\left(\textstyle{8\over3}\right)
  \left(\Li_{5/3}(-e^{-\lambda})-\Li_{5/3}(-e^{\lambda})\right)\right|^{3/5} ,\\
  f_3(\lambda)&=& \left|\Gamma\left(\textstyle{10\over3}\right)
  \left(\Li_{7/3}(-e^{-\lambda})-\Li_{7/3}(-e^{\lambda})\right)\right|^{3/7} ,\\
  f_4(\lambda)&=& \left|\lambda(\lambda^2+\pi^2)\right|^{1/3}. \label{fs4}
\end{eqnarray}
The determination of the function $c_3(\lambda)$ requires resummation of IR enhanced contributions, 
which will be discussed in the next subsection.
We note furthermore that the dependence on $q^*$ indeed
drops out in the sum of the transverse and longitudinal parts.

The functions $f_i(\lambda)$ show a simple asymptotic behavior.
In the zero temperature limit ($|\lambda|\to\infty$)\footnote{This limit can be taken by replacing the
distribution functions in Eq. (\ref{rsi0}) with their zero temperature counterparts, or by using in Eqs.
(\ref{fs1})-(\ref{fs4}) the asymptotic expansion of the polylogarithm given e.g. in \cite{Griguolo:2004jp}.}
we have $f_i(\lambda)\to|\lambda|$. If the temperature
is much higher than $|E-\mu|$ (i.e. $\lambda\to0$) we have $f_1(\lambda)\to c_0:= {\pi\over2}\exp(1-\gamma_E)
=2.397357\ldots$ and $f_{2,3,4}(\lambda)\to 0$.
For $|\lambda|\gg c_0$ or $|\lambda|\ll c_0$ we may approximate $f_1(\lambda)$ with $\mathrm{max}(c_0,|\lambda|)$,
which is qualitatively the result quoted in \cite{Brown:1999aq}. 
It should be noted, however, that the calculation of
Ref. \cite{Brown:1999aq} only took into account transverse gauge bosons, 
and therefore the scale under the logarithm
and its parametric dependence on the coupling was not correctly rendered.

For the imaginary part we find
\begin{eqnarray}
  && \imag\Sigma_{+}-\imag\Sigma_{+}\big|_{E=\mu}
  =g^2C_fm\bigg[-{T\over24\pi m}g_1\left({\emu\over T}\right)
  +3{2^{1/3}\over45\pi^{7/3}}\left({T\over m}g_2\left({\emu\over T}\right)\right)^{5/3}\nonumber\\
  &&\quad\qquad\qquad
- {1\over64 \sqrt{2}}\left({T\over m}g_3\left({\emu\over T}\right)\right)^{2}
  +20{2^{2/3}\over63\pi^{11/3}}\left({T\over m}g_4\left({\emu\over T}\right)\right)^{7/3}\nonumber\\
  &&\qquad\qquad\qquad+\left({T\over m}\right)^3d_3\left({\emu\over T}\right)\bigg)+\cO\left({T\over m}\right)^{11/3}
  \bigg], \label{imsit}
\end{eqnarray}
where
\begin{eqnarray}
  \!\!\!\!\!\!\!\!\!\!\!\!\!\!g_1(\lambda)&\!\!=\!\!&2\log\cosh\left({\lambda\over2}\right),\\
  \!\!\!\!\!\!\!\!\!\!\!\!\!\!g_2(\lambda)&\!\!=\!\!&\left[-\Gamma\left(\textstyle{8\over3}\right)
  \left(\Li_{5/3}(-e^{-\lambda})+\Li_{5/3}(-e^\lambda)+2\left(1-2^{-2/3}\right)
  \zeta\left(\textstyle{5\over3}\right)\right)\right]^{3/5},\\
  \!\!\!\!\!\!\!\!\!\!\!\!\!\!g_3(\lambda)&\!\!=\!\!&|\lambda|,\\
  \!\!\!\!\!\!\!\!\!\!\!\!\!\!g_4(\lambda)&\!\!=\!\!&\left[-\Gamma\left(\textstyle{10\over3}\right)
  \left(\Li_{7/3}(-e^{-\lambda})+\Li_{7/3}(-e^\lambda)+2\left(1-2^{-4/3}\right)
  \zeta\left(\textstyle{7\over3}\right)\right)\right]^{3/7}.
\end{eqnarray}
In the zero temperature limit we have $g_i(\lambda)\to|\lambda|$. If the temperature
is much higher than $|E-\mu|$  we have $g_i(\lambda)\to0$.
Again the determination of the function $d_3(\lambda)$ requires resummation of IR enhanced contributions, 
which will be discussed in the next subsection.

Explicitly, our $T=0$ result reads
\begin{eqnarray}
  \Sigma_+\big|_{T=0}&=&{M_\infty^2\over 2E}-g^2C_fm\,
  \Bigg\{{\emu\over12\pi^2m}\left[\log\left({4\sqrt{2}m\over\pi|\emu|}\right)+1\right]+{i|\emu|\over24\pi m}\nonumber\\
  &&\!\!\!\!\!\!\!+{2^{1/3}\sqrt{3}\over45\pi^{7/3}}\left({|\emu|\over m}\right)^{5/3}(\mathrm{sgn}(\emu)-\sqrt{3}i)\nonumber\\
  &&\!\!\!\!\!\!\!
  + {i\over64 \sqrt{2}}\left({\emu\over m}\right)^2
  -20{2^{2/3}\sqrt{3}\over189\pi^{11/3}}\left({|\emu|\over m}\right)^{7/3}(\mathrm{sgn}(\emu)+\sqrt{3}i)\nonumber\\
  &&\!\!\!\!\!\!\!-{6144-256\pi^2+36\pi^4-9\pi^6\over864\pi^6}\left({\emu\over m} 
  \right)^3 \left[\log\left({m\,0.927\ldots\over |\emu|}\right) 
 -{i\pi\mathrm{sgn}(\emu)\over 2}  \right]\nonumber\\
  &&\!\!\!\!\!\!\!
  +\mathcal{O}\bigg(\left({|\emu|\over m}\right)^{11/3}\bigg) \Bigg\}, \label{sigma0}
\end{eqnarray}
where we have included the zero temperature limits of $c_3$ and $d_3$, which will be derived in the next subsection.
We observe that the self energy at $T=0$ is at leading order proportional to $\varepsilon\log|\varepsilon|$,
which is nonanalytic at $\varepsilon=0$. This behavior is a consequence of
quasistatic chromomagnetic fields, which are only dynamically screened. On the other hand chromoelectric fields,
which are screened by the Debye mass, give an analytic term ($\propto\varepsilon$) in the self energy at leading
order, and nonanalytic contributions ($\propto\varepsilon^3\log|\varepsilon|$) only at higher orders. 

Apart from the first logarithmic term, the leading imaginary parts
contri\-buted by the transverse and longitudinal gauge bosons were
known previously \cite{Manuel:2000mk,LeBellac:1996kr,Vanderheyden:1996bw}.
As our results show,
the damping rate obtained by adding these two leading terms 
\cite{Manuel:2000mk,LeBellac:1996kr} is actually incomplete 
beyond the leading term, 
because the subleading transverse term of order $|\emu|^{5/3}$
is larger than the leading contribution from $\Sigma_{L}$. 

\subsection{Evaluation of the $a^3$-coefficient \label{st3}}
We write the function $c_3(\lambda)$ in Eq. (\ref{resit})
as the sum of the transverse and the longitudinal contribution, $c_3 = c_{3T}+c_{3L}$.
The terms through order $a^3$ in the integrand of Eq. (\ref{si7}) give the following contribution to $c_{3T}(\lambda)$,
\begin{eqnarray}
  &&\!\!\!\!\!\!\!\!\!\!\!\!\!\!\!\!
  c_{3T}^{(1)}(\lambda)={1\over3}\lambda(\lambda^2+\pi^2)\nonumber\\
  &&\times\left[{1\over\pi^4}
  +{2\over\pi^6}\left(-8\left(11-6\log\left({\textstyle{\pi\over2}}\right)\right)
  +2\pi^2\left(1-\log\left({\textstyle{\pi\over2}}\right)\right)\right)\right]\nonumber\\
  &&\!\!\!\!-{8(24-\pi^2)\over 9\pi^6}\bigg[{1\over6}(3-2\gamma_E)\lambda (\lambda^2+\pi^2)\nonumber\\
  &&\quad+2{\partial\over\partial \alpha}
  \left(\Li_{\alpha+3}(-e^{-\lambda})-\Li_{\alpha+3}(-e^{\lambda})\right)\Big|_{\alpha=0}
  \bigg], \label{s42}
\end{eqnarray}
Evaluating explicitly the next few terms in the expansion of the integrand, 
one finds additional contributions of order $a^{3}$.
They arise from the fact that the $z$-integrations would be IR divergent, being screened only
by $z_{min}= a^{2/3}|v|$. Therefore also terms which are formally of higher order in the integrand 
contribute to the order $a^{3}$ in the self energy. In the following we will perform a systematic
summation of all these terms.

With the substitution (\ref{su1})  the transverse gluon self energy in the HDL approximation can be written as
\begin{equation}
  \tilde\Pi_T(qt,q)=\gf^2\mu^2H_T\left({a^{2/3} v\over z}\right),
\end{equation}
with some function $H_T$. We may
neglect the term $q^2-q_0^2$ from the free propagator, because this term does not become singular for 
small $x$.\footnote{However,  $q^2$ and $q_0^2$ would have to be taken into account when summing
up the IR contributions to the coefficient of $a^5$, since for this coefficient also less IR singular contributions are
important.} 
After expansion of the integrand with respect to $a$ as in  Eq. (\ref{si7}) we
get then integrals of the following type contributing to the self energy,
\begin{equation}
  ma^{5/3}\int_{z_{min}}dz\,z \left({a^{2/3}v\over z}\right)^n \sim {ma^3 v^n |v|^{2-n}\over{n-2}}
  ={ma^3 v^2\over{n-2}}. \label{s44}
\end{equation}
[In the last step we have used the fact that $\real\tilde D_T(qt,q)$ is an even function of~$t$.] 
Now we see clearly that from arbitrary powers of $a$ in the integrand we get contributions to the order
$a^{3}$ in $\real\Sigma_+$. The case $n=2$ corresponds to the term of order $a^{3}\log a$, which we have 
evaluated already in the previous section.
As we are interested only in contributions from the IR region, we may take $\infty$ as upper integration limit in 
Eq. (\ref{s44}), since for $n>2$ we get then no 
contribution from the upper integration limit. [The cases $n<2$ have been evaluated explicitly in Eq. (\ref{s42}).]
Furthermore we see from Eq. (\ref{s44}) that from the $v$-integration we always get a factor
\begin{equation}
  \int_{-\infty}^\infty dv\,{e^\lambda-1\over(1+e^v)(1+e^{\lambda-v})}v^2={1\over3}\lambda(\lambda^2+\pi^2).
\end{equation}
Now we can write the function $c_{3T}(\lambda)$ as
\begin{eqnarray}
  &&c_{3T}(\lambda)= c_{3T}^{(1)}(\lambda)-{\lambda(\lambda^2+\pi^2)m^2\over12\pi^2\tilde b^9}
  \int_{\tilde b^2}^\infty dz\,z\nonumber\\
  &&\qquad\times\sum_{n=11}^\infty {\tilde b^n\over n!}\left(\left[{\partial^n\over\partial \tilde b^n}\tilde b^5
  \left(1-{\tilde b^2\over z}\right)
  \real\tilde D_T^\prime(m\tilde b^3, m\tilde bz)\right]\Bigg|_{\tilde b=0}\right), \label{cls}
\end{eqnarray}
where $\tilde b=a^3$, and the prime denotes the omission of the tree level propagator.
Eq. (\ref{cls}) is in fact independent of $\tilde b$, therefore we may simply set $\tilde b=1$.
Summing up the (Taylor) series we find
\begin{equation}
  c_{3T}(\lambda)= c_{3T}^{(1)}+{\lambda(\lambda^2+\pi^2)\over3}J_1
\end{equation}
with
\begin{equation}
  J_1=-{1\over4\pi^2}\int_1^\infty dz \left[-{8z\over\pi^2}+{16(24-\pi^2)\over3\pi^4z}
  +{z^2-1\over z} m^2 \real  {\tilde D_T^\prime}|_{q\to zq_0}\right].
\end{equation}
This integral can probably not be done analytically. Numerically one finds readily $J_1=0.000233964448\ldots$.

For the longitudinal part a completely analogous calculation gives
\begin{eqnarray}
  &&\!\!\!\!\!\!\!\!\!\!\!\
  c_{3L}={1\over3}\lambda(\lambda^2+\pi^2)\left[-{1\over16\pi^2}
  -{1\over128\pi^2}\left(8-3\pi^2+2(\pi^2-4)\log2\right)+J_2\right]\nonumber\\
  &&+{4(\pi^2-4)\over128\pi^2}\bigg[{1\over6}(3-2\gamma_E)\lambda (\lambda^2+\pi^2)\nonumber\\
  &&\quad+2{\partial\over\partial \alpha}
  \left(\Li_{\alpha+3}(-e^{-\lambda})-\Li_{\alpha+3}(-e^{\lambda})\right)\Big|_{\alpha=0}
  \bigg],
\end{eqnarray}
with
\begin{equation}
  J_2=-{1\over4\pi^2}\int_1^\infty dz \left[{z\over2}-{\pi^2-4\over8z}
  +z m^2 \real \tilde D_L^\prime|_{q\to zq_0}\right].
\end{equation}
Numerically one has $J_2=0.000250187944\ldots$

The constant under the logarithm in the fourth line of Eq. (\ref{sigma0}) can be determined from the zero temperature
limit of $c_3(\lambda)$, with the result
\begin{eqnarray}
  &&\exp\Bigg\{{1\over12( 6144-256\pi^2+36\pi^4-9\pi^6)}\bigg[-9\pi^6(-5+6\log2)\nonumber\\
  &&+72\pi^4(-4+3\log2)+128\pi^2(31-12\log({\textstyle{\pi\over2}}))-6144(7-6\log({\textstyle{\pi\over2}}))\nonumber\\
  &&+3456\pi^6(J_1+J_2)\bigg]\Bigg\} = 0.92789731088\ldots
\end{eqnarray}

\begin{figure}
  \begin{center}
    \includegraphics[width=8cm]{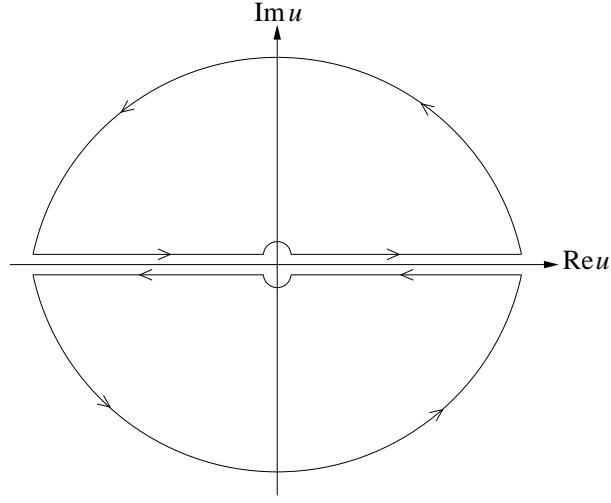}
  \end{center}
  \vspace{-0.5cm}
  {\it \caption{Integration contours for $A_3$. \label{figim}}}
\end{figure}

Let us turn now to the imaginary part.
The function $d_3(\lambda)$ in Eq. (\ref{imsit}) can be determined by summing up IR enhanced contributions 
in a similar way as in the computation of $c_3(\lambda)$. We find
\begin{equation}
  d_3(\lambda)=\left({1920-56\pi^2-9\pi^4\over144\pi^5}+J_3+J_4\right)
  \left[-2(\Li_{3}(-e^{-\lambda})+\Li_{3}(-e^{\lambda}))-3\zeta(3)\right], \label{di}
\end{equation}
with
\begin{eqnarray}
  J_3&=&{1\over8\pi^2}\int_1^\infty \left[-{4z^2\over\pi}+{64\over\pi^3}+{z^2-1\over z}m^2
  \bar\rho_T^\prime|_{q\to zq_0}\right],\\
  J_4&=&{1\over8\pi^2}\int_1^\infty \left[-{\pi\over2}+zm^2\rho_L^\prime|_{q\to zq_0}\right],
\end{eqnarray}
where the prime denotes the omission of the tree level propagators in the spectral densities.
In contrast to the real part of $\Sigma_+$, these integrals can be evaluated analytically rather easily,
as we will demonstrate in the following.
After the substitution $z=1/u$ we can write $J_3$ as
\begin{equation}
  J_3=A_1+A_2+A_3  \label{j3}
\end{equation}
with
\begin{eqnarray}
  &&A_1={1\over8\pi^2}\int_{-\epsilon}^\epsilon du\Bigg[-{2\over\pi u^4}+{32\over\pi^3u^2}\nonumber\\
  &&\qquad\qquad+{1-u^2\over u^3}
  {{\pi u\over2}(1-u^2)\over\left(u^2+{u\over2}(1-u^2)\log\left({1+u\over 1-u}\right)\right)^2+
  \left({\pi u\over2}(1-u^2)\right)^2}\Bigg],\\
  &&A_2={1\over 8\pi^2}\int_{\epsilon<|u|<1}du\left(-{2\over\pi u^4}+{32\over\pi^3u^2}\right),\\
  &&A_3={1\over 8\pi^2}\int_{\epsilon<|u|<\infty}du\ {1-u^2\over u^3}
  \imag {1\over u^2+{u\over2}(1-u^2)\log\left({1+u+i\eta\over -1+u+i\eta}\right)},
\end{eqnarray}
where $0<\epsilon\ll1$. In $A_3$ we have used the fact the the imaginary part in the integrand vanishes
for $|u|>1$, which allows us to extend the outer limits of the integration domain  from $\pm1$ to $\pm\infty$ without changing the
value of the integral.
It is easy to check that $A_1$ vanishes for $\epsilon\to0$. For $A_2$ we obtain
\begin{equation}
  A_2=-{1\over6\pi^3\epsilon^3}-{8\over\pi^5\epsilon}+{48-\pi^2\over6\pi^5}.
\end{equation}
$A_3$ can be written as
\begin{equation}
  A_3=-{i\over 16\pi^2}\int_\mathcal{C}du\ {1-u^2\over u^4}
  {1\over u+{1\over2}(1-u^2)\log\left({u+1\over u-1}\right)},
\end{equation}
where the contour $\mathcal{C}$ consists of the four straight lines pinching the real axis in 
Fig. \ref{figim}. The radius of the two small arcs in Fig. \ref{figim} is equal to $\epsilon$, 
whereas the radius of the two large arcs will be sent to infinity eventually. Since the
integral along a closed contour vanishes if the integrand is analytic in the domain enclosed by the
contour, we may replace the integral along the straight lines
with minus the integral along the arcs. After the substitution $u=r e^{i\varphi}$ we expand the
integrand for small $r$ to obtain the contribution from the small arcs, and for large
$r$ to obtain the contribution from the large arcs.
From the small arcs we get
\begin{equation}
   A_3^{(1)}={1\over6\pi^3\epsilon^3}-{8\over\pi^5\epsilon}-{2(24-\pi^2)\over3\pi^5}+\mathcal{O}(\epsilon),
\end{equation}
and from the large arcs we get
\begin{equation}
  A_3^{(2)}={3\over16\pi}.
\end{equation}
Adding up the $A$'s we find for $J_3$ after taking the limit $\epsilon\to0$
\begin{equation}
  J_3={-1152+40\pi^2+9\pi^4\over48\pi^5}.
\end{equation}
In a completely analogous way we find for the longitudinal contribution
\begin{equation}
  J_4=-{12-\pi^2\over64\pi}.
\end{equation}
Inserting these results for $J_3$ and $J_4$ into Eq. (\ref{di}), and taking the zero temperature limit, we find
\begin{equation}
  \lim_{\lambda\to\infty}\left({1\over|\lambda|^3}d_3(\lambda)\right)=-{6144-256\pi^2+36\pi^4-9\pi^6\over864\pi^6}\times {\pi\over2},
\end{equation}
as stated in the Eq. (\ref{sigma0}).

\begin{figure}
  \begin{center}
    \includegraphics[width=8.5cm]{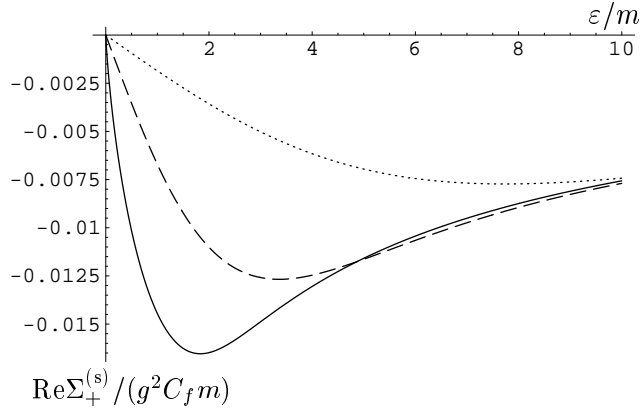}
  \end{center}
  \vspace{-0.5cm}
  {\it \caption{$\real\Sigma_+^\mathrm{(s)}$ at $T=0$ (continuous curve), $T=m$ (dashed curve), and $T=3m$ (dotted curve).
    \label{resiplot}}}
\end{figure}
\begin{figure}
  \begin{center}
    \includegraphics[width=9cm]{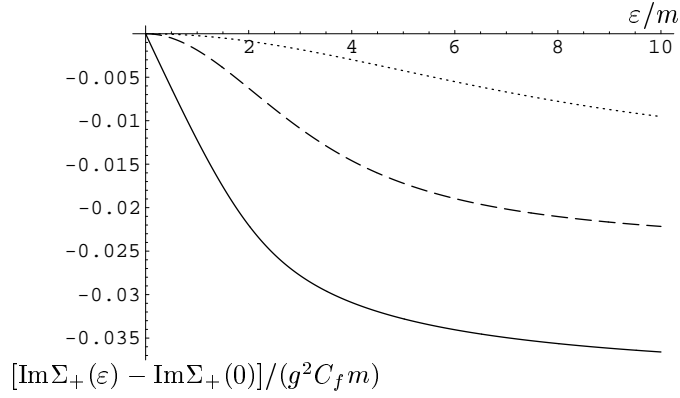}
  \end{center}
  \vspace{-0.5cm}
  {\it \caption{$\imag\Sigma_+(\varepsilon)-\imag\Sigma_+(0)$ 
  at $T=0$ (continuous curve), $T=m$ (dashed curve), and $T=3m$ (dotted curve).\label{imsiplot} }}
\end{figure}
\section{Numerical results}

For intermediate energies $|E-\mu|\gtrsim m$ the integrals in Eqs. (\ref{rsi0}) and (\ref{imsi0}) have to
be evaluated numerically. Figs. \ref{resiplot} and \ref{imsiplot} show the real and imaginary parts
of the self energy as a function of the energy at $T=0$, $T=m$ and $T=3m$. In the imaginary part we have
subtracted the quantity $\imag\Sigma_+(\varepsilon=0)$ which is divergent at finite $T$ [see the discussion
before Eq. (\ref{imsi0})]. The resulting function $\imag\Sigma_+(\varepsilon)-\imag\Sigma_+(0)$ is even 
in $\varepsilon$, which leads to a cusp for $\varepsilon\to0$ at $T=0$, while at finite $T$ it vanishes quadratically
for $\varepsilon\to0$.
The real part is an odd function with respect to $\varepsilon$. Again, the function is nonanalytic at $\varepsilon=0$ for $T=0$, 
but at finite $T$ it is analytic at this point. 
At $T=0$ we find 
\begin{equation}
  \lim_{\varepsilon\to\infty}\imag\Sigma_+(\varepsilon)\Big|_{T=0}=-g^2 C_f m \times 0.04053\ldots
\end{equation}

\begin{figure}
  \begin{center}
    \includegraphics[width=9cm]{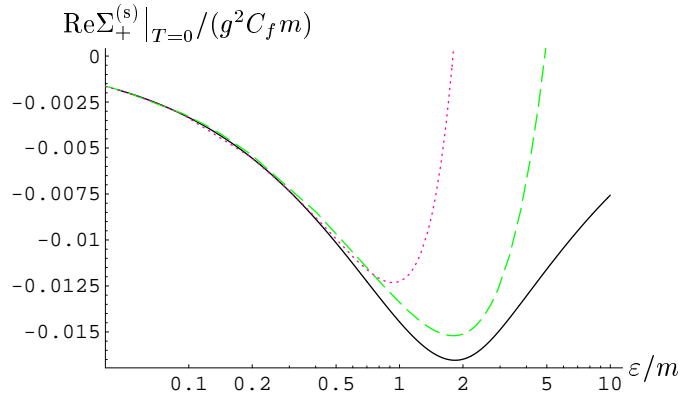}
  \end{center}
  \vspace{-0.5cm}
  {\it \caption{Exact expression for $\real\Sigma_+^\mathrm{(s)}\big|_{T=0}$ (continuous curve) compared to two perturbative approximations
  (dashed curve: $\cO(\varepsilon\log\varepsilon)$, dotted curve: $\cO(\varepsilon^3\log\varepsilon)$).\label{resipert}}}
\end{figure}

\begin{figure}
  \begin{center}
    \includegraphics[width=9cm]{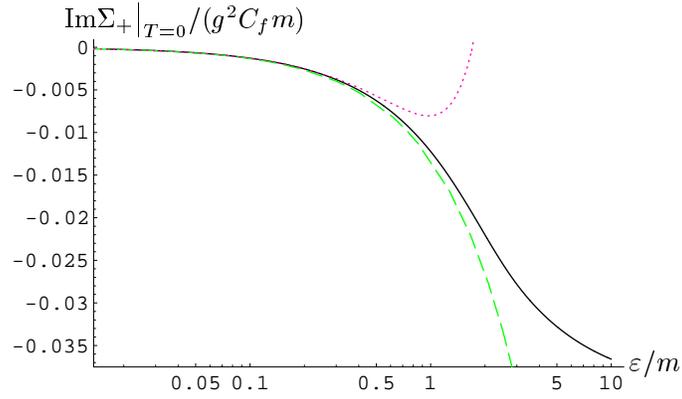}
  \end{center}
  \vspace{-0.5cm}
  {\it \caption{Exact expression for $\imag\Sigma_+\big|_{T=0}$ (continuous curve) compared to two perturbative approximations
  (dashed curve: $\cO(\varepsilon)$, dotted curve: $\cO(\varepsilon^3)$).\label{imsipert}}}
\end{figure}

In Figs. \ref{resipert} and \ref{imsipert} we show a comparison of the exact (numerical) results at $T=0$ with
perturbative results as obtained from Eq. (\ref{sigma0}). We observe that the leading terms ($\cO(\epsilon\log|\epsilon|)$ for
the real part and $\cO(|\epsilon|)$ for the imaginary part) are already quite good approximations for $\epsilon/m\lesssim1$. The
higher orders in the perturbative expansion do not lead to a considerable improvement of the approximation, which indicates
that the series converges well only for $\epsilon/m\ll1$. 

\begin{figure}
  \centerline{\includegraphics[width=9cm]{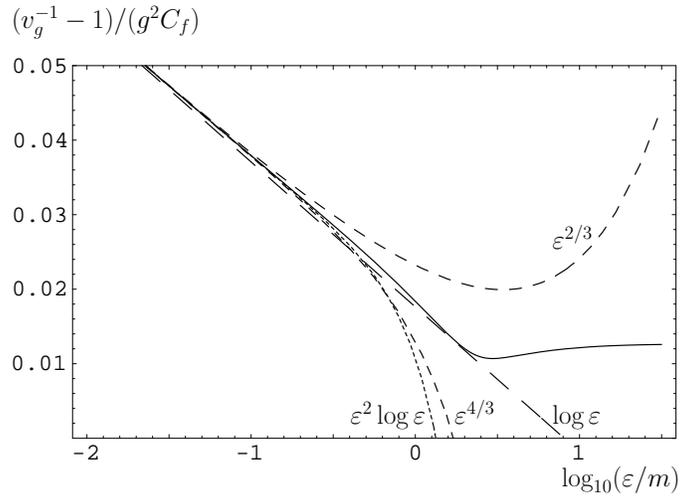}}
  \caption{\it{$(v_g^{-1}-1)/(g^2C_f)$ as a function of $\log_{10}((E-\mu)/m)$
   at zero temperature (from \cite{Gerhold:2005uu}).
  \label{finvv}}}
\end{figure}

Fig. \ref{finvv} shows a plot for the inverse group velocity \cite{Gerhold:2005uu},
\begin{equation}
  v_g^{-1}=1-{\partial\real\Sigma_+\over\partial E}. \label{vg}
\end{equation}
Again one can see that the leading logarithmic approximation is already quite close to the exact result for small
$\varepsilon$, while 
the hard contribution $M_\infty^2/(2E^2)$ [$\simeq~g^2C_f/(8\pi^2)$ for $\varepsilon\ll\mu$] dominates for larger
values of $\varepsilon$.


\section{Fermi surface properties}

The quasiparticle distribution function at zero temperature is given by
\begin{equation}
  n(p)={1\over\pi}\int_{-\infty}^\mu dp_0\,\imag{1\over{-p_0+p+\Sigma_+(p_0,p)-i\eta}}.
\end{equation}
We are interested in the behavior of the distribution function for $p\sim\mu$. A possible non-regular
behavior is expected from the singularity of the integrand. Thus we consider the quantity \cite{Luttinger119}
\begin{equation}
  n^\prime(p):={1\over\pi}\int_{\mu-\delta}^\mu dp_0\,\imag{1\over{-p_0+p+\Sigma_+(p_0,p)-i\eta}},
\end{equation}
with $\delta\ll\mu$. We will neglect the imaginary part of the self energy in the following, since the
imaginary part does not show a logarithmic enhancement in the vicinity of the Fermi surface.
Thus
we may write
\begin{equation}
  n^\prime(p)\simeq\int_{\mu-\delta}^\mu dp_0\,\delta(p_0-p-\real\Sigma_+(p_0,p))
  =\int_{\mu-\delta}^\mu dp_0\,\delta(p_0-\bar\omega(p))
  {1\over1-{\partial\real\Sigma_+\over\partial p_0}},
\end{equation}
where $\bar\omega(p)$ is the solution of the approximate dispersion law 
$p_0-p-\real\Sigma_+=0$. We obtain
\begin{equation}
  n^\prime(p)\simeq\Theta(\mu-\bar\omega(p)){1\over1-{\partial\real\Sigma_+\over\partial p_0}\big|_{p_0=
  \bar\omega(p)}}
\end{equation}
The second factor on the right hand side is the group velocity. In contrast to a normal Fermi liquid
it vanishes (like $1/\log|\bar\omega-\mu|$) for $\bar\omega\to\mu$.
Therefore also the discontinuity
at the Fermi surface vanishes. We remark that a similar behavior can also be found in condensed matter physics, in the context
of high-$T_c$ superconductors. These systems have been termed ``marginal Fermi liquids'' \cite{Varma1989}.

\chapter{Anomalous specific heat at low temperature \label{cspecific}}

\section{General remarks}

In this chapter we will continue our analysis of normal degenerate quark matter.
We will compute the specific heat, which may be relevant for cooling of (proto-)neutron stars at temperatures
above the color superconductivity  phase transition.
For a normal Fermi liquid (see Sec. \ref{sfermi}) the specific heat is linear in the temperature for small
temperatures. We will see, however, that the specific heat of cold dense quark matter contains
an anomalous leading term of the order $T\log T$. As for the quark self energy, this logarithmic enhancement
comes from long-range chromomagnetic fields. In this chapter we will also correct an error
in a recent computation of the specific heat \cite{Boyanovsky:2000zj}, which would have resulted in a $T^3\log T$ behavior of the
specific heat at leading order.

\section{Landau Fermi liquid theory \label{sfermi}}

In this section we will give a brief account of Landau's theory of Fermi liquids \cite{Landau:1956,Landau:1957},
with particular emphasis on the specific heat. Our presentation 
follows closely
Refs. \cite{Nozieres:1964,Negele:1988}, to which we refer for further details.

To begin with, let us consider a uniform, non-relativistic\footnote{The generalization of Landau's theory to relativistic
systems is discussed in \cite{Baym:1975va}.} gas of free \mbox{spin-${1\over2}$} fermions. The ground state consists of a Fermi sea,
corresponding to the occupation number
\begin{equation}
  n(k)=2\Theta(k_F-k),
\end{equation} 
where $k_F$ is the Fermi momentum. The total energy is given by
\begin{equation}
  E=\sum_{\bf k}{k^2\over2m}n(k).
\end{equation}
Now let us turn on interactions between the particles adiabatically. We \emph{assume} that the quantum mechanical states of the 
free system are gradually transformed into states of the interacting system. If this is the case, the interacting system is
called a (Landau) Fermi liquid.

 Next let us add an additional particle with momentum $k$ ($k>k_F$) to the free system, and then
switch on the interactions. The eigenstate of the interacting system, which is obtained in this way, is called a quasiparticle.
Denoting the energy of a quasiparticle with $\epsilon_{\bf k}$ and the chemical potential 
with $\mu$, one finds that
the lifetime of quasiparticles is proportional to $(\epsilon_{\bf k}-\mu)^{-2}$ \cite{Negele:1988,Quinn:1958} [see also 
Eq.~(\ref{imsl}) above]. Therefore the concept of a quasiparticle is well-defined  in the vicinity of the Fermi surface.
Since quasiparticles are adiabatically evolved from fermions, their distribution function at finite temperature is given
by the usual Fermi-Dirac distribution,
\begin{equation}
  n({\bf k})={1\over e^{(\epsilon_{\bf k}-\mu)/T}+1}. \label{f3}
\end{equation}

A weak perturbation of the interacting system will induce a change $\delta n({\bf k})$ in the occupation number. Landau postulated 
that the change in the total energy of the system is then given by\footnote{For simplicity we suppress spin labels.}
\begin{equation}
  \delta E=\sum_{\bf k}\epsilon^0_{\bf k}\delta n({\bf k})+{1\over2}\sum_{{\bf k},{\bf k}^\prime}f({\bf k},{\bf k}^\prime)
  \delta n({\bf k})\delta n({\bf k}^\prime),
\end{equation}
where $\epsilon^0_{\bf k}$ is the energy of a quasiparticle, and $f({\bf k},{\bf k}^\prime)$ is the quasiparticle interaction.
For isotropic systems 
the effective mass $m^*$ is defined as
\begin{equation}
 {k_F\over m^*}={d \epsilon_{\bf k}^0\over dk}\bigg|_{k=k_F}.
\end{equation}
The specific heat at constant volume and per unit volume is given by
\begin{equation}
  \cC_v={1\over V}\left({\partial E\over\partial T}\right)_V. \label{f6}
\end{equation}
Using Eqs. (\ref{f3})-(\ref{f6}) one finds that the specific heat of a Fermi liquid
is linear in the temperature at low temperature (as for a free Fermi gas),
$\cC_v=\gamma T$, where the coefficient $\gamma$ is given by \cite{Negele:1988}
\begin{equation}
  \gamma={1\over3}m^*k_F.
\end{equation}

Landau's theory is corroborated by results of the microscopic theory, i.e. quantum field theory.
For instance one finds also in quantum field theory that the specific heat of a cold fermionic system is
linear in the temperature at leading order in many cases.
The coefficient $\gamma$ can be computed e.g. from the 2PI effective action. For a theory with instantaneous four fermion 
interaction one finds \cite{Luttinger119} 
\begin{equation}
  \gamma={2\pi\over3}\imag\sum_{\bf k}{\partial\over\partial\omega}\log\left(\epsilon_{\bf k}-\omega+\real\Sigma_{\bf k}(\omega)+
  i\eta\right)\Big|_{\omega=\mu}, \label{f8}
\end{equation}
where $\Sigma_{\bf k}(\omega)$ is the fermion self energy.
Evaluating the derivative with respect to $\omega$, one finds that the coefficient is proportional to the inverse 
group velocity, 
\begin{equation}
  v_g^{-1}=1-{\partial\real\Sigma_{\bf k}\over\partial\omega}\bigg|_{
  \omega=\mu,\: k=\mathrm{on\:shell}}.
\end{equation}
It is clear, however, that Eq. (\ref{f8}) becomes meaningless if 
the group velocity vanishes at  $\omega=\mu$. In the previous chapter we did indeed encounter a
situation where this happens, namely for a system with long-range interactions. In that case Landau's
theory is not applicable, and the leading term of the specific heat will not be linear in the temperature.

\section{Specific heat from the entropy}

The specific heat $\cC_v$ at constant volume and per unit volume can be defined as the logarithmic derivative of the entropy density
with respect to temperature at constant 
number density,
\begin{equation}
  \cC_v=T\left(\partial \cS\over\partial T\right)_\cN,
\end{equation}
This can be rewritten in terms of derivatives at constant $T$ or $\mu$ in the following way \cite{LL:1958},
\begin{equation}
  {\cal C}_v=T\left[\left(\partial \cS\over\partial T\right)_\mu-\left(\partial \cN\over\partial T\right)_\mu^2
  \left(\partial \cN\over\partial \mu\right)_T^{-1}\right], \label{cvll}
\end{equation}
At low temperatures only the term with the entropy contributes \cite{Ipp:2003cj},
\begin{equation}
  \cC_v=T\left(\partial \cS\over\partial T\right)_\mu+\mathcal{O}(T^3). \label{x1}
\end{equation}

The thermodynamic potential for QCD is given by the following functional of the full 
propagators and self energies (assuming a ghost-free gauge) 
\cite{Luttinger:1960ua},
\begin{equation}
  \beta\Omega[D,S]={1\over2}\Tr\log D^{-1}-{1\over2}\Tr\Pi D-\Tr\log S^{-1}
  +\Tr\Sigma S+\Phi[D,S],
\end{equation}
where $S$ is the quark propagator, $\Sigma$ is the quark self energy, $D$ is the gluon propagator,
$\Pi$ is the gluon self energy, and $\Phi$ is a series of 2-particle-irreducible (skeleton) diagrams.

Using the fact that $\Omega[D,S]$ is stationary with respect to variations of $D$ and $S$, one
can derive an expression for the entropy which to two-loop order in the skeleton expansion is
entirely given by propagators and self energies \cite{Blaizot:2000fc,Vanderheyden:1998ph}.
We may neglect the contribution from longitudinal gluons, since they are subject to Debye screening
and give therefore only a contribution to the normal Fermi liquid part of the entropy.
Furthermore we may neglect antiparticle contributions in the fermionic sector since they would only lead to
exponentially suppressed terms, $\propto\exp(-\mu/T)$. 
Thus we find for the entropy density (with $P=-\Omega/V$)
\begin{eqnarray} 
  &&\!\!\!\!\!\!\!\!\!\cS=\left(\partial P\over\partial T\right)_\mu\simeq-2N_g\int {d^4Q\over(2\pi)^4}
  {\partial n_b(q_0)\over\partial T}
  \left(\imag\log D_T^{-1}-\imag\Pi_T\,\real D_T\right)\nonumber\\
  &&\quad-4NN_f\int {d^4K\over(2\pi)^4}{\partial n_f( k_0-\mu)\over\partial T}
  \left(\imag\log S_+^{-1}-\imag\Sigma_+\,\real S_+\right)+\cS^\prime, \qquad\label{x2}
\end{eqnarray}
where $D_T^{-1}=-q_0^2+q^2+\Pi_T$ and $S_+^{-1}=- k_0+k+\Sigma_+$ as in chapter \ref{csigma}.
For the evaluation of the imaginary parts we 
always assume that $q_0$ and $k_0$ are replaced by $q_0+i\eta$ and $k_0+i\eta$, respectively 
[see Eqs. (\ref{b8}) and (\ref{b11})].

In the original derivation of the anomalous specific heat in QED by Holstein \emph{et al.} \cite{Norton1}, only the term involving
$\imag\log S_+^{-1}$ had been taken into account in the formula for the entropy density. 
While such a formula for the entropy density is known to be correct for standard Fermi liquid systems \cite{Luttinger119}, 
it is not clear \emph{a priori} how to justify it for more complicated systems, which possibly exhibit
deviations from the Fermi liquid behavior.
 Therefore we will use the full
expression (\ref{x2}) to calculate the entropy density at the order $g^2T\log T$ in the following.
We will only neglect $\cS^\prime$, since it vanishes at two-loop order \cite{Vanderheyden:1998ph,Blaizot:2000fc},
and could therefore only give contributions which are suppressed by an additional factor of $g^2$.

\subsection{Quark part}
\begin{figure}
 \begin{center}
  \includegraphics{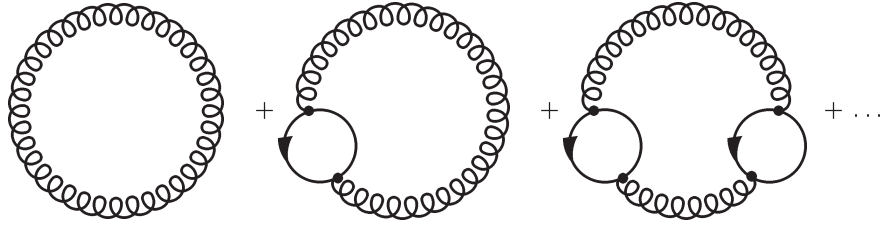}
 \end{center}
 \vspace{-.5cm}
 {\it\caption{Sum of the gluon ring diagrams. 
 \label{figring}}}
\end{figure}
In Eq. (\ref{x2}) we have the following contribution from the quarks,
\begin{eqnarray}
  &&\cS_{(\mathrm{q})}=-4NN_f\int{d^4K\over(2\pi)^4}
  {\partial n_f( k_0-\mu)\over\partial T}\nonumber\\
  &&\qquad\qquad\times\bigg[\imag\log\left(- k_0+k+\Sigma_+\right)
  -\imag\Sigma_+\,\real{1\over- k_0+k+\Sigma_+}\bigg]\nonumber\\
  &&\,\qquad\simeq-{1\over\pi^3}NN_f\int_0^{\infty}dk\,k^2\int_{-\infty}^\infty d k_0
  {\partial n_f( k_0-\mu)\over\partial T}\nonumber\\
  &&\qquad\qquad\times\bigg[\imag\log\left(- k_0+k\right)
  +\real\Sigma_+\,\imag{1\over- k_0+k}\bigg],\quad \label{x9}
\end{eqnarray}
where we have performed an expansion with respect to $\Sigma_+$, keeping only the free term
and the term corresponding to a single quark self energy insertion. Diagrammatically, the latter part corresponds
to the sum of the gluon ring diagrams, see Fig. \ref{figring}.
The free term gives the particle 
contribution to the free fermionic entropy density,
\begin{equation}
  \cS_{(\mathrm{q})}^{\mathrm{free}}\simeq NN_f{\mu^2 T\over3}.
\end{equation}
In the last term in Eq. (\ref{x9})
the factor $\imag\,1/(- k_0+k)$ forces the self energy to be on the light cone.
From Eq. (\ref{resit}) we obtain the following approximation of 
$\real\Sigma_+$  at leading order
\begin{equation}
  \real\Sigma_+\simeq -{g^2C_f\over12\pi^2}(k_0-\mu)\left[\log\left({M\over |k_0-\mu|}\right)
  +h\left({k_0-\mu\over T}\right)\right], \label{x21}
\end{equation}
with $M\propto \gf\mu$, and $h$ is some function. From the Fermi-Dirac distribution in Eq. (\ref{x9})
we see that $k_0-\mu$ is of the order $T$.
At leading logarithmic order we may therefore drop $h$ in Eq. (\ref{x21}), since it gives only a
$\cO(1)$ contribution to the result. This contribution would fix the scale under the logarithm,
but this scale will be determined anyway in Sec.~\ref{sechigher} using a slightly different approach.
Thus we get
\begin{equation}
  \cS_{(\mathrm{q})}^{NLO}={N_g\over\pi^2}\int_0^{\infty}dk\,k^2
  {\partial n_f(k-\mu)\over\partial T}
  {\gf^2\over12\pi^2}(k-\mu)\log\left({M\over |k-\mu|}\right).
\end{equation} 
We make the substitution $k=Tz+\mu$. 
The integral is dominated by small values of $z$, therefore we may send the lower integration limit
to $-\infty$. Then we obtain at order $T\log T$
\begin{equation}
  \cS_{(\mathrm{q})}^{NLO}
  ={\gf^2N_g\mu^2 T\over36\pi^2}\log\left({M\over T}\right). \label{x23}
\end{equation}
This result agrees with the one of Holstein \emph{et al.} \cite{Norton1} after correcting a factor of $4$ therein, as done
previously in \cite{Norton2}.
 
\subsection{Gluon part}

The gluon part $\cS_{({\rm g})}$ is given by the first line of Eq. (\ref{x2}).
It can again be interpreted in terms of ring diagrams, Fig. \ref{figring}. Using the
relation \cite{Blaizot:2000fc}
\begin{equation}
  \imag\log D^{-1}=\arctan\left({\imag\Pi\over\real D^{-1}}\right)-\pi\epsilon( q_0)
  \theta\left(-\real D^{-1}\right),
\end{equation}
we write $\cS_{({\rm g})}=\cS_{(\mathrm{cut})}+\cS_{(\Pi)}+\cS_{(\mathrm{pole})}$,
with
\begin{eqnarray}
  &&\cS_{(\mathrm{cut})}=2N_g\int {d^4Q\over(2\pi)^4}{\partial n_b(q_0)\over\partial T}
  \arctan\left({\imag\Pi_T\over q_0^2-q^2-\real\Pi_T}\right),\\
  &&\cS_{(\Pi)}=-2N_g\int {d^4Q\over(2\pi)^4}{\partial n_b(q_0)\over\partial T}
  \imag\Pi_T\,\real{1\over q_0^2-q^2-\Pi_T},\\
  &&\cS_{(\mathrm{pole})}=2N_g\int {d^4Q\over(2\pi)^4}{\partial n_b(q_0)\over\partial T}
  \pi\epsilon( q_0)\theta\left( q_0^2-q^2-\real\Pi_T\right).
\end{eqnarray}
For the cut term we use the approximation $ q_0\ll q$, because it can be checked that
including terms of higher order in $ q_0$ would only produce terms of higher order than
$T\log T$ (see Sec. \ref{sechigher}). In this region we have from Eqs. (\ref{repith}) and (\ref{impith})
\begin{equation}
  \Pi_T\simeq-i{\gf^2\mu^2 q_0\over 4\pi q}.
\end{equation}
Introducing an UV-cutoff $q_c$ for the moment, we obtain
\begin{equation}
  \cS_{(\mathrm{cut})}\simeq{N_g\over 2\pi^3}\int_{-\infty}^\infty d q_0 \int_{q_0}^{q_c}dq\,q^2\,
  {\partial n_b(q_0)\over\partial T}
  \arctan\left({\gf^2\mu^2 q_0\over4\pi q^3}\right).
\end{equation}
In order to evaluate this integral we make the substitution $y= q_0/T$, $x=4\pi q^3/(\gf^2\mu^2T)$.
Keeping only the term of order $T\log T$, we get the (finite) result
\begin{equation}
  \cS_{(\mathrm{cut})}\simeq{\gf^2N_g\mu^2T\over 36\pi^2}\log\left(M^\prime\over T\right). \label{x7}
\end{equation}
The determination of the constant $M^\prime$ would require a more accurate
calculation including the longitudinal mode (see Sec. \ref{sechigher}), which would
show that $M^\prime\propto \gf\mu$.

Next  let us evaluate $\cS_{(\Pi)}$. Following similar steps as in the computation of 
$\cS_{(\mathrm{cut})}$, we find 
\begin{eqnarray}
  &&\cS_{(\Pi)}\simeq2N_g\int_{-\infty}^\infty{d q_0\over2\pi}\int_{q_0}^{q_c}{dq\,q^2\over2\pi^2}
  {\partial n_b(q_0)\over\partial T}
  \left(-{\gf^2\mu^2 q_0\over4\pi q}\right){q^2\over q^4+
  \left({\gf^2\mu^2 q_0\over4\pi q}\right)^2}\nonumber\\
  &&\qquad\simeq-{\gf^2N_g\mu^2T\over 36\pi^2}\log\left(M^\prime\over T\right).
\end{eqnarray}
We observe that at order $T\log T$ this expression just cancels the contribution from Eq. (\ref{x7}).

\begin{figure}
 \begin{center}
  \includegraphics{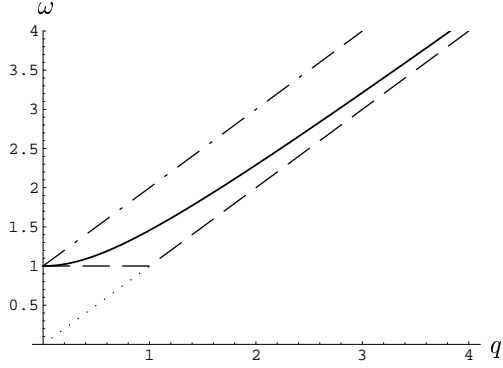}
 \end{center}
 {\it\caption{Illustration of the inequality (\ref{ineq}), with the plasma frequency $\omega_p$ normalized to $1$. 
 Continuous line: $\omega_T(q)$, dashed line: $\Theta(\omega_p-q)+q\Theta(q-\omega_p)$, dash-dotted line: $\omega_p+q$.
 \label{figineq}}}
\end{figure}

Finally let us consider the pole part. In the HDL approximation we have at low temperature
$\mu{\partial\tilde\Pi_T\over\partial\mu}\simeq2\tilde\Pi_T$, and therefore
\begin{eqnarray}
  &&\!\!\!\!\!\!\!
  \mu{\partial\cS_{(\mathrm{pole})}\over\partial\mu}=-2N_g\int {d^4Q\over(2\pi)^4}
  {\partial n_b(q_0)\over\partial T}
  \pi\epsilon( q_0)\,\delta\left( q_0^2-q^2-\real\tilde\Pi_T\right)2\,\real\tilde\Pi_T\nonumber\\
  &&\,\quad\qquad=-4\pi N_g\int {d^4Q\over(2\pi)^4}
  {\partial n_b(q_0)\over\partial T}
  ( q_0^2-q^2)\epsilon( q_0)\,\delta\left( q_0^2-q^2-\real\tilde\Pi_T\right),\qquad
\end{eqnarray}
where we have discarded contributions $\sim T^3$ which are negligible in the low-temperature limit.
Using \cite{LeB:TFT}
\begin{equation}
  \epsilon( q_0)\,\delta\left(\real \tilde D_T^{-1}\right)
  =Z_T(q)\left(\delta( q_0- \omega_T(q))-\delta( q_0+ \omega_T(q))\right)
\end{equation}
we find
\begin{equation}
  \mu{\partial\cS_{(\mathrm{pole})}\over\partial\mu}=-{2N_g\over\pi^2}\int_0^\infty dq\,q^2
  {\partial n_b( \omega_T(q))\over\partial T}( \omega_T(q)^2-q^2)Z_T(q).
\end{equation}
We would like to give an estimate for this integral.
With the plasma frequency $\omega_p=\gf\mu/(\pi\sqrt{3})$ we have the inequalities
\begin{equation}
   \Theta(\omega_p-q)+q\Theta(q-\omega_p) < \omega_T(q) < \omega_p+q \label{ineq}
\end{equation}
(see Fig. \ref{figineq}), and
\begin{equation}
  Z_T(q)<{1\over2q}
\end{equation}
(see Fig. \ref{figH2}). Since we assume $T\ll \omega_p\propto \gf\mu$, we can therefore estimate
\begin{eqnarray}
  &&\int_0^\infty dq\,q^4{\partial n_b(\omega_T(q))\over\partial T}Z_T(q)<
  \int_0^\infty dq\,q^4{\partial n_b(\omega_T(q))\over\partial T}{1\over2q}\nonumber\\
  &&\qquad<{1\over2}\int_0^{ \omega_p} dq\,q^3{\partial n_b(\omega_p)\over\partial T}+
  {1\over2}\int_{ \omega_p}^\infty dq\,q^3{\partial n_b(q)\over\partial T}
  \simeq { \omega_p^5\over8T^2}e^{- \omega_p/T},\qquad\qquad \label{i17}
\end{eqnarray}
and
\begin{eqnarray}
  &&\int_0^\infty dq\,q^2{\partial n_b(\omega_T(q))\over\partial T} \omega_T(q)^2Z_T(q)<
  \int_0^\infty dq\,q^2{\partial n_b(\omega_T(q))\over\partial T}( \omega_p+q)^2{1\over2q}\nonumber\\
  &&\qquad<{1\over2}\int_0^{ \omega_p} dq\,q( \omega_p+q)^2{\partial n_b(\omega_p)\over\partial T}+
  {1\over2}\int_{ \omega_p}^\infty dq\,q( \omega_p+q)^2{\partial n_b(q)\over\partial T}\nonumber\\
  &&\qquad\simeq {17 \omega_p^5\over24T^2}e^{- \omega_p/T}. \label{i18}
\end{eqnarray}
Apart from terms $\sim T^3$ which are dropped in the derivative with respect to $\mu$, this crude estimate
(which we shall refine in Sec. \ref{spole}) shows 
that the pole contribution is exponentially suppressed (essentially because
of $ \omega_T\ge \omega_p$).

\subsection{Result}

From the previous subsections we find the following result for the entropy density at low temperature,
\begin{equation}
  \cS=\cS_{(\mathrm{g})}+\cS_{(\mathrm{q})}\simeq NN_f{\mu^2T\over3}+{\gf^2N_g\mu^2 T\over36\pi^2}
  \log\left({M\over T}\right). \label{x13}
\end{equation}
From Eq. (\ref{x7}) we see that the $T\log T$ term can also be obtained by starting only with
\begin{equation}
  \cS\simeq NN_f{\mu^2T\over3}-2N_g\int {d^4Q\over(2\pi)^4}
  {\partial n_b(q_0)\over\partial T}
  \imag\log D_T^{-1}.
\end{equation}
This formula corresponds to integrating out the fermions, as has indeed been done in the approach of
\cite{Ipp:2003cj,Gan}. On the other hand, we see from Eqs. (\ref{x9}) and (\ref{x23}) that one
also gets the correct result by using the purely fermionic expression
\begin{equation}
  \cS\simeq-4NN_f\int {d^4K\over(2\pi)^4}{\partial n_f( k_0-\mu)\over\partial T}
  \imag\log (- k_0+k+\real\Sigma_+), \label{sf}
\end{equation}
which justifies the starting point of \cite{Norton1}.

\subsection{A note on the Sommerfeld expansion}
Since the inverse group velocity diverges on the Fermi surface, one might wonder
whether the coefficient of the linear power of temperature in the specific heat should also diverge.
Of course we have seen from explicit calculations that this is not the case. In order to see in which way
a straightforward application of Fermi liquid techniques goes wrong in our case, let us consider again
the entropy density. From Eq. (\ref{x9}) we have
\begin{equation}
  \cS^{NLO}=-{1\over\pi^2}NN_f\int_0^{\infty}d k_0\, k_0^2
  {\partial n_f( k_0-\mu)\over\partial T}\real\Sigma_+|_{T=0}.
\end{equation}
A naive Sommerfeld expansion, assuming (erroneously) that $\real\Sigma_+|_{T=0}$ is analytic at $ k_0=\mu$, 
would lead to
\begin{equation}
  \cS^{NLO}\simeq -NN_f{T\over3}{\partial\over\partial k_0}
  \left( k_0^2\real\Sigma_+\right)\Big|_{T=0,\, k_0=\mu}={\rm"}g^2\mu^2T\times\infty{\rm"}.
\end{equation}
Then we would have for the part of the entropy density which is linear in temperature
\begin{equation}
  \cS_{lin.}\sim NN_f{\mu^2T\over3}(1+{\rm"}g^2\infty{\rm"}), \label{x68}
\end{equation}
Of course this equation is \emph{wrong}, since the assumption that $\real\Sigma_+|_{T=0}$ 
is analytic at $ k_0=\mu$ is incorrect. The explicit calculation (see above) shows that instead of
a linear term with a singular coefficient, one obtains a $T\log T$ term with a finite coefficient.

\section{Specific heat from the energy density \label{secento}}
The energy density can be obtained from the expectation value of the energy momentum tensor,
\begin{equation}
  \mathcal{U}={1\over V}\int d^3x\,\langle T^{00}(x)\rangle,
\end{equation}
and the specific heat is then given by
\begin{equation}
  \cC_v=\left({d\mathcal{U}\over dT}\right)_\mathcal{N}. \label{cvu}
\end{equation}
Here the temperature derivative has to be taken at constant particle number density, 
in contrast to the calculation of the low temperature specific heat in the previous section, 
where all temperature derivatives had to be taken at constant chemical potential, 
see Eqs. (\ref{x1}), (\ref{x2}). Of course, both methods will ultimately yield the same result.

In \cite{Boyanovsky:2000zj} the specific heat is computed using the following formula for the
total energy density,
\begin{equation}
  \mathcal{U}=2\int d k_0\int {d^3k\over(2\pi)^3}n_f(k_0-\mu)\, k_0\,\rho_+( k_0,k), 
\end{equation}
where $\rho_+$ is the spectral density of the positive energy component of the quark propagator
(see below). It should be noted that this formula is incorrect even for a theory with only
instantaneous interactions of the type
\begin{equation}
  H_{int}={1\over2}\int d^3x\,d^3x^\prime\,\psi_\alpha^\dag({\bf x}, t)\psi_\beta^\dag({\bf x^\prime}, t)
  V_{\alpha\alpha^\prime,\beta\beta^\prime}({\bf x}-{\bf x^\prime})\psi_{\beta^\prime}({\bf x^\prime}, t)
  \psi_{\alpha^\prime}({\bf x}, t),
\end{equation}
in which case the correct formula reads \cite{Fetter:1971}
\begin{equation}
  \mathcal{U}=2\int d k_0\int {d^3k\over(2\pi)^3}n_f(k_0-\mu)\,{1\over2}( k_0+k)\rho_+( k_0,k).
  \label{u1}
\end{equation}
The anomalous behavior of the specific heat comes from dynamically screened interactions, 
whose non-instantaneous character cannot be neglected.
It would be
rather difficult to generalize Eq. (\ref{u1}) directly for non-instantaneous interactions, because
one would have to use an effective Hamiltonian which is nonlocal in time. 
If one naively tries to compute the anomalous specific heat from Eq. (\ref{u1}), the coefficient
of the logarithmic term turns out to be wrong by a factor ${1\over2}$.
Therefore, we
will follow a slightly different approach, using the energy momentum tensor of QCD without
integrating out the gluons.

The energy momentum tensor can be written as a sum of three pieces,
\begin{equation}
  T^{\mu\nu}=T^{\mu\nu}_{(\mathrm{q})}+T^{\mu\nu}_{(\rm{g})}+T^{\mu\nu}_{(\mathrm{int.})},
\end{equation}
corresponding to the quark part, the gluon part, 
and the interaction part. The contributions of these
parts will be evaluated in the following subsections. We will neglect
gluon self interactions and ghost contributions, since they give only higher order corrections
at low temperature.

\subsection{Quark part}
The quark part is given by
\begin{equation}
  T^{00}_{(\mathrm{q})}=i\sum_{f}\bar\psi_f\gamma^0\partial^0\psi_f, \label{f1}
\end{equation}
where we have written explicitly the sum over flavor space.
This is the (only) contribution which is taken into account by Boyanovsky and
de Vega \cite{Boyanovsky:2000zj}.
Let us repeat their calculation, but for simplicity without the renormalization group improvement
of the quark propagator proposed in \cite{Boyanovsky:2000bc}.
As for the entropy density is is sufficient to take into account only the positive energy component of the quark propagator.
Thus we find
\begin{equation}
  \mathcal{U}_{(\mathrm{q})}=
  2NN_f\int d k_0\int {d^3k\over(2\pi)^3}n_f(k_0-\mu)\, k_0\,\rho_+( k_0,k), 
  \label{q7}
\end{equation}
where the spectral density is defined as $\rho_+\equiv{1\over\pi}\imag S_+$.

In order to obtain the specific heat from Eq. (\ref{cvu}), we have to determine first the temperature
dependence of the chemical potential from the condition
\begin{equation}
  {d\mathcal{N}\over dT}\equiv{\partial\mathcal{N}\over\partial T}
  +{d\mu\over dT}{\partial\mathcal{N}\over\partial\mu}=0, \label{e10}
\end{equation}
where the particle number density $\mathcal N$ is given by
\begin{equation}
  \mathcal{N}=2NN_f\int d k_0\int {d^3k\over(2\pi)^3}n_f(k_0-\mu)\,\rho_+( k_0,k) \label{e11}
\end{equation}
(up to anti-particle contributions). We expand $\cN$ with respect to $g$,
\begin{equation}
  \cN=\cN_0+\gf^2\cN_2+\ldots
\end{equation} 
The free contribution $\cN_0$ is given by
\begin{equation}
  \cN_0=NN_f\left({\mu^3\over3\pi^2}+{\mu T^2\over3}\right). 
\end{equation}
In $\cN_2$ we are only interested in contributions which contain $\log(M/T)$. 
Such terms arise from infrared singularities from the gluon propagator, 
which are dynamically screened. This
corresponds to scattering processes of quarks which are close to the Fermi surface. Therefore the anomalous
terms come from the region $k\sim k_0\sim\mu$, where $\Sigma_+$ is given by (\ref{x21}) at leading order. At leading
logarithmic order we may again neglect the function $h$ in Eq. (\ref{x21}).
Subtracting the temperature independent part, we find then
\begin{equation}
  \gf^2\cN_2=2NN_f\int d k_0\int {d^3k\over(2\pi)^3}
  \left(n_f(k_0-\mu)-\Theta(\mu- k_0)\right)\delta( k_0-k-\real\Sigma_+)\bigg|_{\mathcal{O}(\gf^2)},
  \label{x40}
\end{equation}
where we have approximated the spectral density with a delta function, since the imaginary part of
$\Sigma_+$ is negligible compared to its real part. The integration can be performed easily, with the result 
\begin{equation}
  \gf^2\cN_2\simeq{\gf^2N_g\mu T^2\over 36\pi^2}\log\left(M\over T\right).
\end{equation}
We notice that this result is consistent with the result for the entropy density, Eq. (\ref{x13}).
Now we can solve Eq. (\ref{e10}) at low temperature, 
\begin{equation}
  {d\mu\over dT}=-{2\pi^2T\over3\mu}-{g^2C_fT\over18\mu}\log\left(M\over T\right).
  \label{x33}
\end{equation}
The approximate solution to this differential equation is given by
\begin{equation}
  \mu(T)\simeq\mu(0)\left(1-{\pi^2T^2\over3\mu(0)^2}-{g^2C_fT^2\over36\mu(0)^2}
  \log\left({M\over T}\right)\right). \label{x34}
\end{equation}
Eqs. (\ref{x33}) and (\ref{x34}) correctly reproduce the beginning 
of the perturbative expansions of 
Eqs. (2.37) and (2.38) in Ref. \cite{Boyanovsky:2000zj}.

For the specific heat we obtain from Eq. (\ref{q7}), following the same steps as in the calculation of
$d\mathcal{N}/dT$,
\begin{equation}
  \cC_{v(\mathrm{q})}\simeq NN_f\mu^2T+NN_f{\mu^3\over\pi^2}{d\mu\over dT}+{\gf^2N_g\mu^2T\over18\pi^2}
  \log\left({M\over T}\right). \label{x36}
\end{equation}
Using Eqs. (\ref{x33}) and (\ref{x34}) we find that the $T\log T$-terms cancel,
\begin{equation}
   \cC_{v(\mathrm{q})}\simeq NN_f{\mu^2T\over3},
\end{equation}
as stated in \cite{Boyanovsky:2000zj}. We should emphasize that this cancellation has nothing to do with the
nonperturbative renormalization group method which is employed in \cite{Boyanovsky:2000zj}. 
It should also be noted that this renormalization group method 
has been criticized recently \cite{Schafer:2004zf}, because the neglect of the $\beta$-function
in the renormalization group equation used in \cite{Boyanovsky:2000zj,Boyanovsky:2000bc} does not seem to be justified.

\subsection{Gluon part}
We now turn to the gluon part of the energy density, which has been explicitly neglected in Ref. \cite{Boyanovsky:2000zj}. 
This is given by
\begin{equation}
  T^{00}_{({\rm g})}={1\over2}\left({\bf E}^a\cdot{\bf E}^a+{\bf B}^a\cdot{\bf B}^a\right).
\end{equation}
Neglecting gluon self interactions, and keeping only the transverse part of the gluon propagator, we obtain
\begin{equation}
  \mathcal{U}_{({\rm g})}\simeq2N_g\int{d^4Q\over(2\pi)^4}n_b(q_0)\,\imag\left(\left( q_0^2+q^2\right)D_T\right).
  \label{e51}
\end{equation}
The pole contribution to this integral is again exponentially
suppressed, therefore we only have to consider the cut contribution. 
At low temperature we may neglect the temperature dependence of the gluon self energy.
Subtracting a less infrared sensitive contribution not involving the Bose distribution (corresponding
to the non-$n_b$-part in Sec. \ref{sechigher}), we find
\begin{equation}
  \mathcal{U}_{({\rm g})}\simeq
  2N_g\int{d^4Q\over(2\pi)^4}\left(n_b(q_0)+\Theta(- q_0)\right)( q_0^2+q^2)
  {{\gf^2\mu^2 q_0\over 4\pi q}\over\left( q_0^2-q^2\right)^2
  +\left({\gf^2\mu^2 q_0\over 4\pi q}\right)^2}. \label{u52}
\end{equation}
After the substitution $q_0=Ty$, $q^3=\gf^2\mu^2Tx/(4\pi)$ the integral can be readily done, with the result
\begin{equation}
  \mathcal{U}_{({\rm g})}\simeq{\gf^2N_g\mu^2T^2\over72\pi^2}\log(M/T),
\end{equation}
which gives the following contribution to the specific heat at order $T\log T$,
\begin{equation}
  \cC_{v({\rm g})}\simeq{\gf^2N_g\mu^2T\over36\pi^2}\log(M/T). \label{e52}
\end{equation}
Again the determination of the constant $M$ would require a more accurate calculation, similar
to the one in Sec. \ref{sechigher}.

\subsection{Interaction part}
The interaction part is given by
\begin{equation}
  T^{00}_{(\mathrm{int})}=g\sum_f\bar\psi_f\gamma^0A^0_aT_a\psi_f. 
\end{equation}
The expectation value of this term is essentially given by the
resummed gluon ring diagram as in  Fig. \ref{figring}. However, here only the longitudinal component of 
the gluon propagator appears in the loop
. This mode is subject to Debye screening, so
it can contribute only to the normal Fermi liquid part of the specific heat.

\subsection{Result}
From the previous subsections we see that the only contribution to the specific heat at
order $g^2T\log T$ when 
calculated along the lines of Ref. \cite{Boyanovsky:2000zj} comes from the gluon part, Eq. (\ref{e52}). 
While this confirms the observation of Ref. \cite{Boyanovsky:2000zj} that the quark contribution of 
order $g^2 T\log T$ cancels against a similar term from the temperature dependence of the 
chemical potential at fixed number density, it shows that the neglect of the gluon contribution to the
energy density in  \cite{Boyanovsky:2000zj} is not justified.

Our result for the specific heat from the energy density is consistent with
the result for the entropy density in Sec. \ref{secento}. For the entropy density  we found  at leading logarithmic order that
it was sufficient to take into account only the contribution of 
fermionic quasiparticles. For the energy density, on the other hand, the purely fermionic formula (\ref{q7})
is not sufficient to give the correct result. 

\section{Higher orders in the specific heat\label{sechigher}}

In this section we will evaluate higher terms in the low temperature specific heat which go
beyond the leading logarithmic approximation. A convenient starting point is the following expression for
the pressure, which becomes exact in the limit of large flavor number $N_f$ \cite{Moore:2002md,Ipp:2003zr,Ipp:2003jy}, 
\begin{eqnarray}
  &&\!\!\!\!\!\!\!\!\!\!\!\!
  P=NN_f\left({\mu^4\over12\pi^2}+{\mu^2T^2\over6}+{7\pi^2T^4\over180}\right)\nonumber\\
  &&-N_g\int {d^3q\over(2\pi)^3}\int_0^\infty {dq_0\over\pi}\bigg[2\left([n_b(q_0)+{1\over2}]\imag\log D_T^{-1}
  -{1\over2}\imag\log D_{\rm vac}^{-1}\right)\nonumber\\
  &&\quad+\left([n_b(q_0)+{1\over2}]\imag\log{D_L^{-1}\over q^2-q_0^2}
  -{1\over2}\imag\log{D_{\rm vac}^{-1}\over q^2-q_0^2}\right)\bigg]+{\cal O}(g^4\mu^4), \label{pres}
\end{eqnarray}
where the inverse gauge boson propagators are given by 
$D_T^{-1}=q^2-q_0^2+\Pi_T+\Pi_{\rm vac}$,  $D_L^{-1}=q^2-q_0^2+\Pi_L+\Pi_{\rm vac}$, and
$D_{\rm vac}^{-1}=q^2-q_0^2+\Pi_{\rm vac}$. The first line of Eq. (\ref{pres}) is
simply the pressure of a free fermionic quantum gas, and the remaining part 
is essentially given by the
sum of gluon ring diagrams in Fig. \ref{figring}. At finite $N_f$, Eq. (\ref{pres}) with $\Pi$ including also the leading contributions
from gluon loops still collects all infrared-sensitive contributions up to and including three-loop order \cite{Vuorinen:2003fs,Gerhold:2004tb}. 
We shall however find that all contributions from gluon loops to $\Pi$ are suppressed with $(T/\mu)^2$ in the specific heat, and
will thus be negligible for $T\ll\mu$ compared to the terms we shall keep. 

In the following we will always drop temperature independent terms in the pressure, since they do not
contribute to the specific heat at low temperature. We will refer to 
terms involving the bosonic distribution function $n_b$ as ``$n_b$-parts'', and to the 
remaining ones as ``non-$n_b$-parts''.

From the explicit form of the self energy (see Sec. \ref{secgluon} and \cite{Ipp:2003qt}) we see that
the $q$-integration naturally
splits into three regions:
$q<q_0$, $q_0<q<2\mu-q_0$, and $2\mu-q_0<q<2\mu+q_0$. We will consider these regions separately in the following.

\subsection{Transverse contribution, region II\label{secp2}}
The $n_b$-part of the contribution of the transverse gluons to the pressure is given by
\begin{equation}
  {P_T\over N_g}=-2\int {d^3q\over(2\pi)^3}\int_0^\infty{dq_0\over\pi}\,n_b(q_0)
  \imag\log(q^2-q_0^2+\Pi_T+\Pi_{\rm vac}). \label{cvh1}
\end{equation}
The cut contribution from region II is given by
\begin{equation}
  {P_T^\mathrm{II}\over N_g}=-2\int_0^\infty{dq_0\over\pi}\int_{q_0}^{2\mu-q_0} {dq\,q^2\over 2\pi^2}\,n_b(q_0)
  \arctan{\imag\Pi_T\over q^2-q_0^2+\real\Pi_T+\real\Pi_{\rm vac}}. \label{p2}
\end{equation}
As long as $T\ll\mu$ it is sufficient to take the self energy at zero temperature. 

The Bose-Einstein factor and the leading term in the gluon self energy
set the characteristic scales,
\begin{equation}
  q_0\sim T,\qquad q\sim(\gf^2\mu^2 T)^{1/3}.
\end{equation}
We want to perform an expansion with respect to a parameter $b$ defined as
\begin{equation}
  b:=\left({T\over \gf\mu}\right)^{1/3}.
\end{equation}
It turns out that the following approximation of the gluon self energy is sufficient through order
$T^3\log T$ in the entropy density,
\begin{eqnarray}
  &&\real\Pi_T(q_0,q)\simeq{\gf^2\over\pi^2}\left({\mu^2q_0^2\over q^2}-{\mu^2q_0^4\over3q^4}\right),\label{h55} \\
  &&\imag\Pi_T(q_0,q)\simeq{\gf^2\over4\pi}\left(-{\mu^2q_0\over q}+{\mu^2q_0^3\over q^3}-{q q_0\over4}\right). 
  \label{h56}
\end{eqnarray}
The first two terms in both lines are the leading terms of an expansion of the HDL self energy in
powers of $q_0$. Naively counting powers of $b$ in the integrand one would conclude that only these terms are 
responsible for the coefficients of the expansion in $b$, while the other terms in the self energy should give only
terms which are suppressed with additional powers of $g$ in the pressure. In principle this is correct, but it turns out that 
it is necessary to keep also
the last term in Eq. (\ref{h56}), which is beyond the HDL approximation. As we well see in Sec. \ref{upper} this
term gives a contribution in the hard region, where the approximation $q\ll\mu$ is not justified.

Introducing dimensionless integration variables 
$x$ and $y$ via 
\begin{equation}
  q_0=b^3\gf\mu y, \quad q=b \gf\mu (x/(4\pi))^{1/3}, \label{substt}
\end{equation}
we find after expanding the integrand with respect to $b$,
\begin{eqnarray}
  &&{P_T^\mathrm{II}\over N_g}\simeq {\gf^4\mu^4\over 12\pi^4}\int_0^\infty 
  dy{1\over e^y-1}\int_{x_{min}}^{x_{max}}dx
  \bigg[b^6\arctan\left({y\over x}\right)\nonumber\\
  &&+{y(\gf^2x^2-64y^2)\over8 (2\pi^2x)^{1/3}(x^2+y^2)}b^8+{32(2x)^{1/3}y^5\over\pi^{4/3}(x^2+y^2)^2}b^{10}
  \nonumber\\
  &&-{32y^5(24x^2y^2-8y^4+\pi^2(x^2+y^2)^2)\over3\pi^2x(x^2+y^2)^3}b^{12}+{\cal O}(b^{14})\bigg],
  \label{cvh4}
\end{eqnarray}
with $x_{min}=4\pi b^6y^3$ and $x_{max}=4\pi(2-b^3\gf y)^3/(b\gf)^3$.
In the coefficients of this expansion we have written down only those terms which do not
ultimately lead to terms that are suppressed by explicit positive powers of $\gf$.
The integrations are now straightforward, and we find\footnote{For the $y$-integration one needs the
following formulae valid for $\alpha>1$: 
\begin{eqnarray}
  &&\int_0^\infty dy{y^{\alpha-1}\over e^y-1}=\Gamma(\alpha)\zeta(\alpha),
  \quad\int_0^\infty dy{y^{\alpha-1}\log y\over e^y-1}=\Gamma'(\alpha)\zeta(\alpha)\nonumber
  +\Gamma(\alpha)\zeta'(\alpha).
\end{eqnarray}}
\begin{eqnarray}
  &&\!\!\!\!\!\!\!\!\!{P_T^\mathrm{II}\over N_g}\simeq \gf^4\mu^4
  \bigg[{b^6\over72\pi^2}\left(\log\left({32\pi\over(b\gf)^3}\right)+\gamma_E
  -{6\over\pi^2}\zeta^\prime(2)+{3\over2}\right)
  -{2^{2/3}\Gamma\left({8\over3}\right)\zeta\left({8\over3}\right)\over3\sqrt{3}\pi^{11/3}}b^8
  \nonumber\\
  &&\quad+{8\,2^{1/3}\Gamma\left({10\over3}\right)\zeta\left({10\over3}\right)
  \over9\sqrt{3}\pi^{13/3}}b^{10}
  +{16(\pi^2-8)\over45\pi^2}b^{12}\log b+\tilde c_T b^{12}+\mathcal{O}(b^{14})\bigg]. \label{cvh5}
\end{eqnarray}
The evaluation of the constant $\tilde c_T$ is a bit more involved because one has to sum up an infinite series
of contributions from the infrared region. This will be done in Sec. \ref{sect4}.

\subsection{Longitudinal contribution, region II}

The $n_b$-part of the contribution of the longitudinal gluons to the pressure is given by
\begin{equation}
  {P_L\over N_g}=-\int {d^3q\over(2\pi)^3}\int_0^\infty{dq_0\over\pi}\,n_b(q_0)
  \imag\log\left({q^2-q_0^2+\Pi_L+\Pi_{\rm vac}\over q^2-q_0^2}\right). \label{cvh14}
\end{equation}
As in the previous section the $q$-integration
decomposes into three parts. 
In the second region ($q_0<q<2\mu-q_0$) the characteristic scales are now
\begin{equation}
  q_0\sim T,\qquad q\sim \gf\mu.
\end{equation}
In a similar way as in the previous section, the gluon self energy can be approximated as
\begin{eqnarray}
  &&\real\Pi_L(q_0,q)\simeq{\gf^2\over\pi^2}\left(\mu^2-{2\mu^2q_0^2\over q^2}\right),\label{h57}\\
  &&\imag\Pi_L(q_0,q)\simeq{\gf^2\over2\pi}\left({\mu^2q_0\over q}-{\mu^2q_0^3\over q^3}-{q q_0\over4}\label{h58}\right). 
\end{eqnarray}
We introduce dimensionless
integration variables $y$ and $z$ via 
\begin{equation}
  q_0=b^3\gf\mu y, \quad q=\gf\mu z/\pi. \label{substl}
\end{equation}
Then we find after expanding
the integrand with respect to $b$,
\begin{eqnarray}
  &&\!\!\!\!\!\!\!\!\!\!\!\!\!\!\!\!\!{P_L^\mathrm{II}\over N_g}\simeq {\gf^4\mu^4\over 16\pi^2}\int_0^\infty 
  dy{1\over e^y-1}\int_{z_{min}}^{z_{max}}dz \nonumber\\
  &&\!\!\!\!\!\!\!\!\!\!\!\!\!\!\!\!\!\times\bigg[b^6{yz(-4\pi^2(1+z^2)+\gf^2z^4)\over\pi^4(1+z^2)^2}
  +b^{12}{y^3(\pi^2-12(z^2+1))\over3z(1+z^2)^3}+{\cal O}(b^{18})\bigg],\quad \label{cvh9}
\end{eqnarray}
with $z_{min}=b^3y\pi$ and $z_{max}=(2/\gf-b^3y)\pi$.
In the coefficients of this expansion we have written down only those terms which do not
ultimately lead to terms that are suppressed by explicit positive powers of $\gf$.
The integrations are now straightforward, and we find
\begin{eqnarray}
  &&{P_L^\mathrm{II}\over N_g}\simeq \gf^4\mu^4\Bigg[{b^6\over48\pi^2}\left(1+\log\left({\gf^2\over4\pi^2}\right)\right)
  \nonumber\\
  &&\qquad\quad+{\pi^2(12-\pi^2)\over240}b^{12}\log b+\tilde c_L b^{12}+\mathcal{O}(b^{18}\log b)\Bigg]. \label{cvh10}
\end{eqnarray}
As in the transverse sector, the determination of the constant $\tilde c_L$ requires
summation of contributions from the infrared region, which will be done in Sec. \ref{sect4}.

\subsection{Contributions from the upper integration limit \label{upper}}

In this subsection we will show that the HDL approximation for the gluon self energy
is sufficient to produce the series in $b$, apart from the order $b^6$,
where precisely two additional terms in the gluon the self energy 
are necessary, one in the transverse part and one in the longitudinal part.

First let us focus on the transverse part.
It is convenient to multiply the denominator and the numerator of the argument of
the arctangent in Eq. (\ref{p2}) with a factor $x^{1/3}/(\gf^2\mu^2b^2)$. Dropping numerical
factors, and also powers of $y$, we have then the following effective power counting
rules:
\begin{eqnarray}
  \!\!\!q^2 \!\!\!&\to&\!\!\! x, \label{uu1}\\
  \!\!\!q_0^2 \!\!\!&\to&\!\!\! x^{1/3}b^4,\\
  \!\!\!\gf^2\mu^2\left({q_0\over q}\right)^n \!\!\!&\to&\!\!\! 
  \left({b^2\over x^{1/3}}\right)^{n-1},\\
  \!\!\!\gf^2\mu^2\left({q_0\over q}\right)^n\left({q\over\mu}\right)^r
  \left({q_0\over\mu}\right)^s \left({T\over\mu}\right)^{2t} \!\!\!&\to&\!\!\!
  \left({b^2\over x^{1/3}}\right)^{n-1+{r+3s+6t\over2}}\left(\gf^2x\right)^{r+s+2t\over2},\qquad \label{u4}
\end{eqnarray}
where $n,r,s,t \in\mathbbm{N}_0$.
Here the third line corresponds to a HDL contribution, and the last line
corresponds to terms beyond the HDL approximation. At first sight, we would expect
that contributions from the last line (when $r+s+t>0$) could only produce terms which are 
suppressed with powers of $\gf$ in the pressure. A more careful analysis is required, however,
since the upper integration limit is proportional to $1/\gf^3$, which might invalidate
the naive power counting in the integrand.

After the Taylor expansion with respect to $b$, the term of 
order $b^{6+2M}$ in the integrand will
have the following general structure for $M\ge1$,
\begin{equation}
  \gf^4\mu^4b^6{\mathsf{P}_{2M}\over(x^2+y^2)^M}, \label{u6}
\end{equation}
where $\mathsf{P}_{2M}$ is of the order $b^{2M}$. We note that 
$\mathsf{P}_{2M}$ is a polynomial of degree $2M$ in the expressions on the right hand sides
of Eqs. (\ref{uu1})-(\ref{u4}). This can be seen from expanding first
$\real\Pi_T$ and $\imag\Pi_T$, and then the arctangent, with respect to $b$.
Each term in this polynomial contains at least one factor which comes 
from $\imag\Pi_T$, and at least one
factor which comes from $q^2-q_0^2+\real\Pi_T$. Explicitly, we find
contributions to the pressure of the following type from the region of large $x$ 
in (\ref{u6}),\footnote{The upper integration limit in Eq. (\ref{u7}) corresponds to
$q_{max}=2\mu$. Actually there exists no sharp upper cutoff if one 
takes the gluon self energy at finite temperature. There is only a 
``fuzzy'' cutoff which is  smeared
over a region of size $\Delta q\sim T$. However, this effect gives only corrections
of order $T/\mu\sim\gf b^3$.}
\begin{equation}
  U_M:=\gf^4\mu^4\int^{1/(b \gf)^3}dx {b^{6+2M}\over x^{2M}}x^{\alpha+{\beta\over3}}
  \prod_{i=1}^\gamma x^{-{c_i\over3}}(\gf^2x)^{d_i}
  \prod_{j=1}^\delta x^{-{e_j\over3}}(\gf^2x)^{f_j}, \label{u7}
\end{equation}
where $x^{\alpha+{\beta\over3}}$ comes from the free gluon propagator,
the first product comes from $\real\Pi_T$, and the second one from $\imag\Pi_T$.
The parameters $\alpha$, $\beta$, $\gamma$, $\delta$, $c_i$, $d_i$, $e_j$, and $f_j$
are natural numbers (including zero). 
The properties of $\mathsf{P}_{2M}$ listed above lead to the following set of
restrictions for these parameters,
\begin{eqnarray}
  4\beta+2\sum_{i=1}^\gamma c_i+2\sum_{j=1}^\delta e_j&=&2M,\label{u8}\\
  \alpha+\beta+\gamma+\delta&=&2M,\label{u9}\\
  \alpha+\beta+\gamma&\ge&1,\label{u10} \\
  \delta&\ge&1. \label{u11}
\end{eqnarray}
We can eliminate $\alpha$ and $\beta$ in Eq. (\ref{u7}) 
with the help of Eqs. (\ref{u8}) 
and (\ref{u9}). After performing the integration we obtain
\begin{equation}
  U_M\sim b^v\gf^w, \label{u13}
\end{equation}
where the exponents are given by
\begin{eqnarray}
  &&v=3\Big(1+M+\gamma+\delta-\sum_{i=1}^\gamma d_i-\sum_{j=1}^\delta f_j\Big),\\
  &&w=-3+M+3\gamma+3\delta-\sum_{i=1}^\gamma d_i-\sum_{j=1}^\delta f_j. \label{u14}
\end{eqnarray}
We are interested in contributions with $w\le0$. Therefore we obtain from (\ref{u11})
and (\ref{u14})
\begin{equation}
  {1\over3}\Big(3-M-3\gamma+\sum_{i=1}^\gamma d_i+\sum_{j=1}^\delta f_j\Big)\ge\delta\ge1. 
  \label{u15}
\end{equation}
Furthermore we get from $\beta\ge0$ using Eq. (\ref{u8})
\begin{equation}
  M-\sum_{i=1}^\gamma c_i-\sum_{j=1}^\delta e_j\ge0. \label{u16}
\end{equation}
Combining the inequalities (\ref{u15}) and (\ref{u16}) we get
\begin{equation}
  3\gamma+\sum_{i=1}^\gamma (c_i-d_i)+\sum_{j=1}^\delta (e_j-f_j)\le0. \label{u17}
\end{equation}
Comparing (\ref{u4}) and (\ref{u7}) we find $c_i-d_i\ge-1$, and therefore
\begin{equation}
  \sum_{i=1}^\gamma (c_i-d_i)\ge-\gamma. \label{u18}
\end{equation}
In order to proceed we have to take a closer look at the structure of $\imag\Pi_T$.
In the explicit zero temperature expression
\begin{equation}
  \imag\Pi_T={\gf^2\over48\pi}\left(-{12\mu^2q_0\over q}-3qq_0+{12\mu^2q_0^3\over q}
  +{2q_0^3\over q}+{q_0^5\over q^3}\right) \label{u19}
\end{equation}
we have $e_j=f_j=0$ for the first term, $e_j=f_j=1$ for the second one, 
and $e_j-f_j>0$ for the three remaining ones.
Finite temperature corrections to $\imag\Pi_T$ also have $e_j-f_j>0$.
Because of (\ref{u18}) and $\gamma\ge0$ 
we conclude that the inequality (\ref{u17}) can only be satisfied for $\gamma=0$ and 
for contributions which come from the first two terms of $\imag\Pi_T$ in Eq. (\ref{u19}). We write
\begin{equation}
  \delta=\delta_1+\delta_2,
\end{equation}
where $\delta_1$ and $\delta_2$ denote the numbers of factors in Eq. (\ref{u7}) which 
come from the first and the second term of $\imag\Pi_T$, respectively. We note that
\begin{equation}
  \sum_{j=1}^\delta e_j= \sum_{j=1}^\delta f_j=\delta_2. \label{u20}
\end{equation}
Therefore the inequalities (\ref{u15}) and (\ref{u16}) are reduced to
\begin{equation}
  {1\over3}(3-M+\delta_2)\ge\delta_1+\delta_2\ge1, \quad M-\delta_2\ge0. \label{u21}
\end{equation} 
The condition $\alpha\ge0$ gives with the help of Eqs. (\ref{u8}), (\ref{u9}) and (\ref{u20})
\begin{equation}
  3M-2\delta_1-\delta_2\ge0.
\end{equation}
The last inequality to be satisfied is (\ref{u10}), which gives (using Eq. (\ref{u9}))
\begin{equation}
  2M-\delta_1-\delta_2\ge1. \label{u23}
\end{equation}
The only solution of the system (\ref{u21})-(\ref{u23}) with non-negative $\delta_1$ and
$\delta_2$ is
\begin{equation}
  M=\delta_2=1, \quad \delta_1=0.
\end{equation}
We conclude that the only contribution from the upper integration limit that is not
suppressed with powers of $\gf$ arises from the following set of parameters in Eq. (\ref{u7}),
\begin{equation}
  (M,\alpha,\beta,\gamma,\delta)=(1,1,0,0,1), \quad (e_1,f_1)=(1,1),
\end{equation}
which gives $v=6$ and $w=0$ for the exponents in Eq. (\ref{u13}).
Explicitly, we have the following contribution to the pressure in this case,
\begin{equation}
  \gf^4\mu^4\int^{1/(b \gf)^3}dx\,b^8\gf^2x^{-1/3}\sim\gf^4\mu^4 b^6. \label{u26}
\end{equation}
We have thus shown that this is the only contribution in the whole series in $b$ (for $M\ge1$) 
which comes from the upper 
integration limit and which is not suppressed with powers of $\gf$. We conclude that
the naive power counting, as inferred from (\ref{uu1})-(\ref{u4}), is almost correct, 
which means that HDL self energy is sufficient to produce the series in $b$, 
with the sole exception of the term of order $b^6$. At this order
we get a contribution from a term in $\imag\Pi_T$ which is beyond the
HDL approximation [the second term in Eq. (\ref{u19})], as shown in Eq. (\ref{u26}).

The conclusions of this section can also be understood in a more direct way. 
With the definition $\Pi_T^{(2)}:=\Pi_T-\tilde\Pi_T$ we have
\begin{eqnarray}
  &&\imag\Pi_T^{(2)}\simeq -{\gf^2qq_0\over16\pi}, \qquad q\gg q_0,\\
  &&\imag\Pi_T^{(2)}\to0, \quad\qquad\qquad q\to q_0.
\end{eqnarray}
The expression in the first line is precisely the second term in Eq. (\ref{u19}). It gives the
following hard (two-loop) contribution to the pressure,
\begin{eqnarray}
  &&\!\!\!\!\!\!\!\!\!\!\!\!\!\!
  {1\over N_g}\biggr[P^{\mathrm{II,non-HDL}}_T-P^{\mathrm{II,non-HDL}}_T\Big|_{T=0}\biggr]\nonumber\\&&=
  -{1\over\pi^3}\int_0^\infty dq_0\, n_b(q_0)
  \int_{q_0}^{2\mu} dq\,q^2\,{\imag\Pi_T^{(2)}\over q^2-q_0^2} \simeq {\gf^2\mu^2 T^2
  \over48\pi^2}. \label{ht}
\end{eqnarray}
This is precisely the contribution which comes from the term proportional to $\gf^2$
in Eq. (\ref{cvh4}).
Since this two-loop expression is sensitive to hard scales, it is not 
surprising that the HDL approximation is not sufficient in this case. 
On the other hand, we see from the power counting rules (\ref{uu1})-(\ref{u4})
 that the HDL approximation is certainly good enough 
to produce all the terms in the series in $b$ that come from soft scales.
Of course one has to resum diagrams at soft scales, as we have done it in Sec. \ref{secp2}.

The arguments of the preceding paragraph can also be applied to the longitudinal case.
As in the transverse case the only contribution that is sensitive to hard momenta appears
at the two-loop level. Again we define $\Pi_L^{(2)}:=\Pi_L-\tilde\Pi_L$, which gives
\begin{eqnarray}
  &&\imag\Pi_L^{(2)}\simeq -{\gf^2qq_0\over8\pi}, \qquad q\gg q_0,\\
  &&\imag\Pi_L^{(2)}\to0, \quad\qquad\qquad q\to q_0.
\end{eqnarray}
We have the following two-loop contribution from hard momenta,
\begin{eqnarray}
  &&\!\!\!\!\!\!\!\!\!\!\!\!\!\!
  {1\over N_g}\biggl[P^{\mathrm{ II,non-HDL}}_L-P^{\mathrm {II,non-HDL}}_L\Big|_{T=0}\biggr]
  \nonumber\\&&=
  -{1\over2\pi^3}\int_0^\infty dq_0\, n_b(q_0)
  \int_{q_0}^{2\mu} dq\,q^2\,{\imag\Pi_L^{(2)}\over q^2-q_0^2} \simeq {\gf^2\mu^2 T^2
  \over48\pi^2}. \label{hl}
\end{eqnarray}
All the other terms in the series in $b$ are generated by the HDL approximation.

\subsection{Cut contributions from regions I and  III}
In the first region the imaginary parts of the complete self energies ($\Pi_T+\Pi_{\rm vac}$ and $\Pi_L+\Pi_{\rm vac}$)
vanish, therefore there is no cut contribution to the specific heat.

In the third region we have $q_0\sim T$, $q\sim2\mu$ and $dq\sim q_0\sim T$. 
If we take the gluon self energy at zero
temperature we can therefore make the following estimates for the transverse part, 
\begin{eqnarray}
  &&{P_T^{\rm III}\over N_g}=-{1\over\pi^3}\int_0^\infty dq_0\int_{2\mu-q_0}^{2\mu+q_0} dq\,q^2n_b(q_0)
  \arctan {\imag\Pi_T(q_0,q)\over q^2-q_0^2+\real\Pi_T(q_0,q)}\nonumber\\
  &&\qquad\sim \mu^2T^2\arctan{\gf^2\mu T\over \mu^2}\sim\gf^4\mu^4\times\gf b^9.
\end{eqnarray}
and for the longitudinal part,
\begin{equation}
 {P_L^{\rm III}\over N_g}\sim\mu^2T^2\arctan{\gf^2T^2\over \mu^2}\sim\gf^4\mu^4\times\gf^2 b^{12}.
\end{equation}
These are  contributions which are  suppressed by explicit powers of $\gf$. If one takes the gluon self energy at
finite temperature, the sharp boundaries of region III will be smeared by an amount of order
$\Delta q\sim T$, but this does not modify the above estimates.

\subsection{Pole part \label{spole}}

The pole contribution comes from region I, i.e. $q\le q_0$. In this region the HDL propagator has single poles for
$q_0\ge \omega_p$ at real $q_0=\omega_{T,L}(q)$, which allows us to carry out the $q_0$-integration. For the
$n_b$-part we find
\begin{eqnarray}
  &&\!\!\!\!\!\!\!\!\!\!\!\!\!\!\!
  {1\over N_g}\left[P^\mathrm{I}-P^\mathrm{I}\Big|_{T=0}\right]\nonumber\\
  &&=-{T\over2\pi^2}\int_0^\infty dq q^2
  \left[2\log\left(1-e^{-\omega_T(q)/T}\right)+\log\left({1-e^{-\omega_T(q)/T}\over1-e^{-q/T}}\right)\right].\qquad \label{pole1}
\end{eqnarray}
For $q\ll\omega_p$ we have
\begin{equation}
  \omega_T(q)=\omega_p\left(1+{3q^2\over5\omega_p^2}-{9q^4\over35\omega_p^4}
  +{\cal O}\left({q^6\over\omega_p^6}\right)\right).
\end{equation}
Inserting this expression into the Bose factor, we find the 
characteristic scale $q\sim\sqrt{\omega_p T}$, which is indeed consistent with
$q\ll\omega_p$ for small temperatures. The logarithm in Eq. (\ref{pole1}) can be expanded as
\begin{equation}
  \log\left(1-e^{-\omega_T(q)/T}\right)\simeq -e^{-\omega_T(q)/T}-{1\over2}e^{-2\omega_T(q)/T}+\ldots
\end{equation}
After the substitution $q=s\sqrt{\omega_p T}$ we obtain for the leading contribution of the transverse part
\begin{eqnarray}
  &&{1\over N_g}\left[P^\mathrm{I}_T-P^\mathrm{I}_T\Big|_{T=0}\right]\simeq {1\over\pi^2}\sqrt{\omega_p^3 T^5}
  e^{-\omega_p/T}\int_0^\infty ds\,s^2e^{-3s^2/5}\nonumber\\
  &&\qquad\qquad\qquad={5\over12}\sqrt{5\omega_p^3T^5\over3\pi^3} e^{-\omega_p/T}.
\end{eqnarray}
Using
\begin{equation}
  \omega_L(q)=\omega_p\left(1+{3q^2\over10\omega_p^2}-{3q^4\over280\omega_p^4}
  +{\cal O}\left({q^6\over\omega_p^6}\right)\right)
\end{equation}
an analogous calculation for the longitudinal part gives
\begin{equation}
  {1\over N_g}\left[P^\mathrm{I}_L-P^\mathrm{I}_L\Big|_{T=0}\right]\simeq 
  -{\pi^2\over90}T^4+{5\over6}\sqrt{5\omega_p^3T^5\over6\pi^3} e^{-\omega_p/T}. \label{polel}
\end{equation} 
We observe that all pole contributions are exponentially suppressed for $T\ll\omega_p$, with
the only exception of a $T^4$ contribution from the term $\imag\log(q^2-q_0^2)$ in the longitudinal sector, which contributes
the equivalent of an ideal-gas pressure of one bosonic degree of freedom, but with negative sign.

\subsection{Non-$n_b$ contribution}

Integrals without $n_b$ are less IR singular than the 
corresponding integrals with $n_b$, since the Bose distribution  behaves
like $T/q_0$ for small $q_0$. This means that within our perturbative accuracy no
resummation is necessary for the non-$n_b$ terms, 
and it is sufficient to consider the strictly perturbative two-loop
expression for $\cS_{\rm non-n_b}$. We can determine this contribution by the observation that
at two-loop order also the $n_b$ part is IR safe and given by
\begin{eqnarray}
  &&\!\!\!\!\!\!\!\!\!\!\!\!\!\!\!\!\!\!\!\!\!\!\!\!
  {P_{n_b}^{\rm 2-loop}\over N_g}=-\int {d^3q\over(2\pi)^3}\int_0^\infty {dq_0\over\pi}n_b(q_0)\imag 
  \left({2\Pi_T+\Pi_L\over q^2-q_0^2}\right)=\nonumber\\
  &&\!\!\!\!\!\!\!\!\!\!\!=-\int {d^3q\over(2\pi)^3}\int_0^\infty {dq_0\over\pi}n_b(q_0)
  \Big[{\pi\over 2 q}\delta(q-q_0)\real\Pi_G+{\cal P}{1\over q^2-q_0^2}
  \imag\Pi_G\Big], \label{pn2}
\end{eqnarray}
where $\Pi_G=2\Pi_T+\Pi_L$.
In order to evaluate Eq. (\ref{pn2})
we need $\real\Pi_G$ on the light cone. From Eq. (\ref{repig})
we find
\begin{equation}
  \real\Pi_G|_{q_0=q}={4\gf^2\over\pi^2}\int_0^\infty dk\,k\,n(k)=
  \gf^2\left({\mu^2\over\pi^2}+{T^2\over3}\right).
\end{equation}
Therefore we get the following contribution from the real part of the gluon self energy,
\begin{equation}
  {P_{n_b}^{\rm 2-loop,\,\real}\over N_g}=-{\gf^2\mu^2 T^2\over24\pi^2}+{\cal O}(T^4). \label{pre}
\end{equation}
Next we want to evaluate the contribution with $\imag\Pi_G$ in Eq. (\ref{pn2}).
First we approximate the gluon self energy with its zero zero temperature version.
(We will check the validity of this approximation below.)
The main contribution comes from region II, where
\begin{equation}
  {1\over q^2-q_0^2}\imag\Pi_G=-{\gf^2 q_0\over 4\pi q}.
\end{equation}
Inserting this expression into Eq. (\ref{pn2}) we find
\begin{equation}
{P_{n_b}^{\rm 2-loop,\,\imag}\over N_g}={\gf^2\mu^2 T^2\over24\pi^2}.  \label{pim}
\end{equation}
At this point we are able to check explicitly that at this order including finite
temperature corrections into the gluon self energy does not change the result.
From the integral representation of $\imag\Pi_G$ [Eqs. (\ref{imp1})-(\ref{imp4})] we find
\begin{eqnarray}
  &&\!\!\!\!\!\!\!\!\!\!\!\!\!\!\!\!\!
  {P_{n_b}^{\rm 2-loop,\,\imag}\over N_g}=-{\gf^2\over4\pi^4}\int_0^\infty dq_0\int_0^\infty dq\,q
  \,n_b(q_0)\nonumber\\
  &&\quad\times\left[\Theta(q-q_0)\int_{q+q_0\over2}^{q-q_0\over2}dk\,n(k)
  -\Theta(q_0-q)\int_{q_0-q\over2}^{q_0+q\over2}dk\,n(k)\right].
\end{eqnarray}
(Here we have dropped the vacuum contribution in $\imag\Pi_G$, since it would give no contribution in the following.)
After doing the $k$-integration we find for the second order susceptibility
\begin{eqnarray}
  &&\!\!\!\!\!\!\!\!\!\!\!\!\!\!\!\!\!\!\!\!\!\!\!\!\!\!\!\!\!\!\!\!\!\!\!\!\!\!\!
  {1\over N_g}{\partial^2P_{n_b}^{\rm 2-loop,\,\imag}\over\partial\mu^2}\bigg|_{\mu=0}=
  {\gf^2\over4\pi^4T}\int_0^\infty dq_0\int_0^\infty dq\,q\,n_b(q_0)\nonumber\\
  &&\qquad\qquad\times{\sinh\left(q\over2T\right)\sinh\left(q_0\over2T\right)
  \over\left[\cosh\left(q\over2T\right)+\cosh\left(q_0\over2T\right)\right]^2}.
\end{eqnarray}
We substitute $q=2T\log\alpha$ and $q_0=2T\log\beta$, which gives
\begin{equation}
  {1\over N_g}{\partial^2P_{n_b}^{\rm 2-loop,\,\imag}\over\partial\mu^2}\bigg|_{\mu=0}=
  {2\gf^2T^2\over\pi^4}\int_1^\infty d\alpha\int_1^\infty d\beta {(\beta^2-1)\log\beta
  \over(\alpha+\beta)^2(1+\alpha\beta)^2}={\gf^2T^2\over12\pi^2}. \label{p45}
\end{equation}
Comparing Eqs. (\ref{pim}) and (\ref{p45}) we see that at this order the gluon self energy
at zero temperature is sufficient to give the correct result. This could have been expected,
of course, since finite temperature corrections to the gluon self energy are suppressed
with $(T/\mu)^2$.

Adding up the two contributions (\ref{pre}) and (\ref{pim}), we find that 
they cancel precisely at the order $\gf^2\mu^2T^2$, which means that
\begin{equation}
  P_{n_b}^{\rm 2-loop}\simeq0.
\end{equation}
Therefore the two-loop non-$n_b$ contribution is equal to the standard perturbative result at
order $\gf^2\mu^2T^2$ \cite{Kapusta:1979fh,Kapusta:TFT},
\begin{equation}
  {P_{{\rm non-}n_b}\over N_g}=-{\gf^2\mu^2T^2\over16\pi^2}+{\cal O}\left(\gf^2\mu^4,\gf^2T^4\right). \label{pnon}
\end{equation}

\subsection{Evaluation of the coefficient of $T^4$ \label{sect4}} 
In this section we want to compute the complete coefficient of $T^4$ in the pressure. To this end
we will perform a systematic summation of IR enhanced contributions, in a similar way as in Sec. \ref{st3}.

First let  us consider the transverse part.
From the terms which are explicitly shown in Eq. (\ref{cvh4}) we find the following contribution to the
coefficient of $b^{12}$ in the pressure, 
\begin{eqnarray}
  &&\!\!\!\!\!\!\!\!\!\!\!\!\tilde c_T^{(1)}={1\over810\pi^2}\Big[-36\pi^2+\nonumber\\
  &&\qquad(\pi^2-8)\Big(248-96\gamma_E-9\pi^2+48\log(4\pi)
  +8640{\zeta^\prime(4)\over\pi^4}\Big)\Big]. \label{t42}
\end{eqnarray}
Evaluating explicitly the next few terms in the expansion of the integrand, 
one finds additional contributions of order $b^{12}$.
They arise from the fact that the $x$-integrations would be IR divergent, were it not for the cutoff
$x_{min}\propto b^6$. Therefore also terms which are formally of higher order in the integrand 
contribute to the order $b^{12}$ in the pressure.

Since $x_{min}$ depends on $\gf$ only through $b$, and since we are not interested
in terms in $P_T$ which contain $\gf$ explicitly, it is sufficient to take the HDL self energy in the following.
With the substitution (\ref{substt})  the transverse gluon self energy can be written as
\begin{equation}
  \tilde\Pi_T=\gf^2\mu^2H_T\left({b^2 y\over x^{1/3}}\right),
\end{equation}
with some function $H_T$. In this section we may
neglect the term $q^2-q_0^2$ from the free propagator in Eq. (\ref{p2}), because this term does not become singular for 
small $x$. 
After expansion of the integrand with respect to $b$ we
get then integrals of the type
\begin{equation}
  b^6\int_{x_{min}}dx \left({b^2y\over x^{1/3}}\right)^n \sim {b^{12}y^3\over{n-3}}. \label{t44}
\end{equation}
Now we see clearly that from arbitrary powers of $b$ in the integrand we get contributions to the order
$b^{12}$ in $P_T$. The case $n=3$ corresponds to the term of order $b^{12}\log b$, which we have 
evaluated already above.
As we are interested only in contributions from the IR region, we may take $\infty$ as upper integration limit in 
Eq. (\ref{t44}), since for $n>3$ we get then no 
contribution from the upper integration limit. [The cases $n<3$ have been evaluated explicitly in Eq. (\ref{t42}).]
Furthermore we see from Eq. (\ref{t44}) that from the $y$-integration we always get a factor
\begin{equation}
  \int_0^\infty {dy\,y^3\over{e^y-1}}={\pi^4\over15}.
\end{equation}
The complete coefficient can thus be written as
\begin{equation}
  \tilde c_T=\tilde c_T^{(1)}
  -{\pi^4\over15}{1\over12\pi^4 b^{12}}\int_{4\pi b^6}^\infty dx \sum_{n=14}^\infty {b^n\over n!}\left(\left[
  {\partial^n\over\partial b^n}b^6\arctan\left({\imag\tilde\Pi_T\over\real\tilde\Pi_T}\right)\right]\bigg|_{b=0,y=1}\right).
  \label{a5}
\end{equation}
This expression is in fact independent of $b$ (see Eq. (\ref{t44})). Therefore we may set simply $b=1$.
Summing up the (Taylor) series, we find after the substitution $x=4\pi z^3$
\begin{eqnarray}
  &&\tilde c_T=\tilde c_T^{(1)}-{\pi\over15}\int_1^\infty dz\Bigg({128+3\pi^4z^3-8\pi^2(2+3z^2)\over6\pi^3z}\nonumber\\
  &&\qquad+z^2
  \arctan\bigg[{\pi(1-z^2)\over 2z+(z^2-1)\log\left({z+1\over z-1}\right)}\bigg]\Bigg). \label{ct}
\end{eqnarray}
From this expression we see that the complete HDL self energy is necessary for this coefficient (and not
only the expansion for small $q_0$, which is sufficient for the fractional powers and the logarithmic terms).
The remaining integral over the parameter $z$ can probably not be done analytically. Numerically
one finds
\begin{equation}
  \tilde c_T=-0.00178674305\ldots
\end{equation}

The computation of the longitudinal part proceeds in a similar way. 
With the substitution (\ref{substl}) the longitudinal gluon self energy can be
written as
\begin{equation}
  \tilde\Pi_L=\gf^2\mu^2H_L\left({b^3 y\over x}\right),
\end{equation}
with some function $H_L$. In a similar way as in Eq. (\ref{t44}) we can make the estimate
\begin{equation}
   b^3\int_{x_{min,L}}dx\,x^2\left({b^3y\over x}\right)^n \sim {b^{12}y^3\over{n-3}}.
\end{equation}
In a similar way as above we obtain
\begin{equation}
  \tilde c_L=\tilde c_L^{(1)}
  -{\pi^4\over15}{1\over2\pi^6 b^{12}}\int_{\pi b^3}^\infty dx 
  \sum_{n=14}^\infty {b^n\over n!}\left(\left[
  {\partial^n\over\partial b^n}b^3x^2\arctan\left({\imag\tilde\Pi_L\over\real\tilde\Pi_L}\right)\right]\bigg|_{b=0,y=1}\right) \label{ww3}
\end{equation}
with
\begin{equation}
  \tilde c_L^{(1)}={1\over8640}\left(3\pi^4-2\pi^2(12-\pi^2)\left(-17+6\gamma_E-6\log(\pi)-540{\zeta^\prime(4)\over\pi^4}
  \right)\right).
\end{equation}
We have to be careful when summing up the Taylor series in Eq. (\ref{ww3}) for the following reason.
The function $\real\tilde\Pi_L$ has a zero at $b^3\simeq0.2653x$, and therefore
$\sigma(b):=\arctan(\imag\tilde\Pi_L/\real\tilde\Pi_L)$ is not an analytic function of $b$. We have to replace
$\sigma(b)$ with $\tilde\sigma(b):=-\arctan(\real\tilde\Pi_L/\imag\tilde\Pi_L)+{\pi\over2}$. $\sigma(b)$ and
$\tilde\sigma(b)$ have the same Taylor expansion around $b=0$, but only $\tilde\sigma$ is 
an analytic function of $b$. In this way we arrive at
\begin{eqnarray}
  &&\tilde c_L= 
  \tilde c_L^{(1)}-{\pi\over30}\int_1^\infty dz\Bigg({\pi(\pi^2+12(-1-z^2+z^3))\over24z}\nonumber\\
  &&\qquad+z^2\arctan\bigg[-{2z\over\pi}+{1\over\pi}\log\left({z+1\over z-1}\right)\bigg]\Bigg)
  \simeq 0.11902569216\ldots \label{cl}\qquad
\end{eqnarray}     

For the complete $T^4$ coefficient we also have to take into account the $\mathcal{O}(T^4)$ contribution from
Eq. (\ref{polel}). Thus we get finally
\begin{eqnarray}
  &&\!\!\!\!\!\!\!\!\!\!
  \tilde c=\tilde c_T+\tilde c_L-{\pi^2\over90}\nonumber\\
  &&\!\!\!\!\!={2048-256\pi^2-36\pi^4+3\pi^6\over2160\pi^2}\left(\gamma_E-90{\zeta^\prime(4)\over\pi^4}\right)\nonumber\\
  &&\!\!\!\!-{63488-9088\pi^2-648\pi^4+93\pi^6\over25920\pi^2}+{3\pi^4(12-\pi^2)\log\pi+128(\pi^2-8)\log(4\pi)\over2160\pi^2}
  \nonumber\\
  &&\!\!\!\!-{\pi\over30}\int_1^\infty dz\Bigg\{{1024+\pi^6-64\pi^2(2+3z^2)-12\pi^4(1+z^2-3z^3)\over24\pi^3z}\nonumber\\
  &&\!\!\!\!+2z^2\arctan\left[{\pi\left(1-z^2\right)\over2z+\left(z^2-1\right)\log\left({z+1\over z-1}\right)}\right]
  +z^2\arctan\left[-{2z\over\pi}+{1\over\pi}\log\left({z+1\over z-1}\right)\right]\Bigg\}\nonumber\\
  &&\!\!\!\!\!\!\simeq 0.0075766779858\ldots \label{ctilde}
\end{eqnarray}

\subsection{How to compute higher order coefficients}
Using the methods described above one can compute the coefficients of 
higher terms in the expansion of ${\cal C}_v$ with respect to $b$. This
is straightforward for the coefficients of the fractional powers and the 
logarithmic terms, where one only has to include higher orders
in the expansion of the HDL gluon self energy with respect to $q_0$, 
see Eqs. (\ref{h55}), (\ref{h56}), (\ref{h57}) and (\ref{h58}).
In this way one finds for the coefficients of $b^{14}$, $b^{16}$ 
and $b^{18}\log b$ in the pressure
\begin{eqnarray}
  c_{14}\!&=&\!{128\ 2^{2/3}(-70+9\pi^2)\Gamma\left({14\over3}\right)\zeta\left({14\over3}\right)
  \over243 \sqrt{3}\pi^{17/3}}\simeq 0.21345\ldots,\\
  c_{16}\!&=&\!{8\ 2^{1/3}(197120-28800\pi^2+729\pi^4)\Gamma\left({16\over3}\right)
  \zeta\left({16\over3}\right)\over3645\sqrt{3}\pi^{19/3}}\nonumber\\
  &\simeq& -0.75413\ldots,\\
  c_{18,\log}\!&=&\!{-94371840 + 15728640\pi^2 -606208\pi^4 - 7920\pi^8 + 2160\pi^{10} - 
  135\pi^{12}\over22680\pi^2}\nonumber\\
  &\simeq&18.61898\ldots.
\end{eqnarray}
We note in passing that the origin of the $b^6\log b$ term is quite different from 
the origin of the $b^{6n}\log b$ terms with $n>1$: the former arises from dynamical screening,
while all the remaining logarithmic terms come from integrals of the type $\int_{x_{min}} {dx\over x}$.

For the terms of the order $b^{12+6r}$, with $r\in\mathbbm{N}_0$ one has to sum up
again IR enhanced contributions, in a manner similar to the evaluation of the $b^{12}$-coefficient
described above. In order to see how these IR enhanced terms come about, let us make again
an estimate of the type (\ref{t44}), but now taking into account also the explicit $q^2$ and $q_0^2$ from the
inverse free propagator in
Eq. (\ref{p2}). Using the fact that $q/(\gf\mu)\propto bx^{1/3}$ and $q_0/(\gf\mu)\propto b^3 y$, we find
in the transverse sector contributions of the type
\begin{equation}
  b^6\int_{x_{min,T}}dx\left({b^2y\over x^{1/3}}\right)^n\left(b\,x^{1/3}\right)^{2p}
  \left(b^3y\right)^{2s}\sim {b^{12+6(p+s)}y^{3+2(p+s)}\over n-2p-3}. \label{hh4}
\end{equation}
When the denominator on the right hand side vanishes, the integrand on the left hand side
is equal to $b^{12+2(p+s)}y^{3+2(p+s)}/x$, which gives logarithmic terms upon integration
with respect to $x$ (which are not IR enhanced). 
The important point is, however, that arbitrarily high powers of $b$
in the integrand produce terms of the order $b^{12+6r}$, where $r$ is the total number of
$q^2$ and $q_0^2$ insertions in the integrand.

For the longitudinal part, the situation is completely analogous, as can be seen from
the estimate
\begin{equation}
  b^3\int_{x_{min,L}}dx\,x^2\left({b^3 y\over x}\right)^nx^{2p}(b^3y)^{2s}
  \sim {b^{12+6(p+s)}y^{3+2(p+s)}\over n-2p-3}.
\end{equation}

As an illustration let us compute the coefficient of $b^{18}$. From Eq. (\ref{hh4}) we
see that we get contributions from $(p,s)=(1,0)$ and $(p,s)=(0,1)$. Instead of simply setting
the explicit $q^2$ and $q_0^2$ to zero as in Eq. (\ref{a5}), we have to consider now the object
\begin{eqnarray}
  &&\left(q^2{\partial\over\partial\bar q^2}+q_0^2{\partial\over\partial\bar q_0^2}\right)
  \arctan\left({\imag\tilde\Pi_T(q_0,q)\over\bar q^2+\bar q_0^2+\real\tilde\Pi_T(q_0,q)}\right)
  \bigg|_{\bar q=\bar q_0=0}\nonumber\\
  &&\qquad=-{(q^2-q_0^2)\imag\tilde\Pi_T(q_0,q)\over\imag\tilde\Pi_T(q_0,q)^2+\real\tilde\Pi_T(q_0,q)^2}.
\end{eqnarray}
Therefore we obtain in place of Eq. (\ref{a5})
\begin{eqnarray}
  &&\!\!\!\!\!\!\!\!\!\!\!\!\!\!
  c_{18,T}=c_{18,T}^{(1)}+{8\pi^6\over63}
  {1\over12\pi^4 b^{18}}\int_{4\pi b^6}^\infty dx\nonumber\\
  &&\times\sum_{n=20}^\infty 
  {b^n\over n!}\left(\left[{\partial^n\over\partial b^n}
  {b^6\gf^2\mu^2\left(b^2(x/(4\pi))^{2/3}-b^6\right)\imag\tilde\Pi_T\over\imag\tilde\Pi_T^2+\real\tilde\Pi_T^2}
  \right]\Bigg|_{b=0,y=1}\right),\qquad \label{hh6}
\end{eqnarray}
where
\begin{eqnarray}
  &&c_{18,T}^{(1)}={16\over127575\pi^2}\Big[960(5760-960\pi^2+37\pi^4)\Big(2\gamma_E-1890
  {\zeta'(6)\over\pi^6}-\log(4\pi)\Big)\nonumber\\
  &&\quad\qquad-8(4158720-715680\pi^2+29921\pi^4)+675\pi^6\Big].
\end{eqnarray}
In a similar way as in Eq. (\ref{ct}), we can write Eq. (\ref{hh6}) explicitly as
\begin{eqnarray}
  &&c_{18,T}=c_{18,T}^{(1)}+{32\pi^4\over63}\int_1^\infty dz\Bigg[{1\over45\pi^6z}
  \big(-184320+30720\pi^2-1184\pi^4\nonumber\\
  &&\qquad\qquad+11520\pi^2z^2-960\pi^4z^2-720\pi^4z^4+45\pi^6z^6\big)\nonumber\\
  &&\qquad\qquad-{\pi^2z^5(z^2-1)^2\over\pi^2(z^2-1)^2
  +\left(2z+(z^2-1)\log\left({z+1\over z-1}\right)\right)^2}\Bigg].
\end{eqnarray}
Numerically, one finds readily
\begin{equation}
  c_{18,T}=7.97457343372231\ldots.
\end{equation}
For the longitudinal part, we find in a similar way
\begin{eqnarray}
  &&\!\!\!\!\!c_{18,L}=c_{18,L}^{(1)}+{8\pi^6\over63}
  {1\over2\pi^6 b^{18}}\int_{\pi b^3}^\infty dx\nonumber\\
  &&\qquad\times\sum_{n=20}^\infty 
  {b^n\over n!}\left(\left[{\partial^n\over\partial b^n}
  {b^3x^2\gf^2\mu^2\left((x/\pi)^2-b^6\right)\imag\tilde\Pi_L\over\imag\tilde\Pi_L^2+\real\tilde\Pi_L^2}
  \right]\Bigg|_{b=0,y=1}\right),\quad \label{hh11}
\end{eqnarray}
where
\begin{eqnarray}
  &&c_{18,L}^{(1)}={\pi^6\over181440}\Big[120(176-48\pi^2+3\pi^4)\Big(\gamma_E-945
  {\zeta'(6)\over\pi^6}-\log(\pi)\Big)\nonumber\\
  &&\quad\qquad-63584+16992\pi^2-1053\pi^4\Big].
\end{eqnarray}
Thus we obtain
\begin{eqnarray}
  &&c_{18,L}=c_{18,L}^{(1)}+{8\pi^4\over63}\int_1^\infty dz\Bigg[-{\pi^2\over192z}
  \big(176-48\pi^2+3\pi^4+96z^2\nonumber\\
  &&\qquad\quad-12\pi^2z^2+48z^4\big)+{\pi^2z^5\over\pi^2+
  \left(2z-\log\left({z+1\over z-1}\right)\right)^2}\Bigg]\nonumber\\
  &&\qquad\simeq8.59505395976\ldots.
\end{eqnarray}

To summarize, the temperature dependent part of the pressure at small temperature can be
expanded in a series with coefficients of the orders $g^4\mu^4 b^{6+n}$ and $g^4\mu^4 b^{6(1+n)}\log b$, 
with $n\in\mathbbm{N}_0$. In this  and the previous subsections we have described in detail how to
compute the coefficients of this series.
On the other hand, the computation of terms which are suppressed by explicit
powers of $\gf$ would be much more involved. In particular one would have to
include also diagrams beyond the large-$N_f$ formula (\ref{pres}).

\subsection{Specific heat}
Neglecting terms which are explicitly suppressed with powers of $\gf$, the pressure 
can be written as 
\begin{equation}
  P= P_{free}+\gf^4\mu^4f\left({T\over\gf\mu}\right)+{\cal O}(g^2\mu^4) \label{pc}
\end{equation}
with some function $f$. The specific heat (\ref{cvll}) can be written as a sum of two terms,
$\cC_v=\cC_{v,1}+\cC_{v,2}$, where
\begin{equation}
  \cC_{v,1}=T{\partial^2P\over\partial T^2},
\end{equation}
and
\begin{equation}
  \cC_{v,2}=-T \left({\partial^2 P\over\partial\mu\partial T}\right)^2\Big/
  {\partial^2 P\over\partial\mu^2}.
\end{equation}
Using the expression (\ref{pc}) for the pressure, and setting $T=b^3\gf\mu$, we find 
\begin{equation}
  (\cC_{v,1}- \cC_{v,1,free})/(\gf^3\mu^3)=b^3f''(b^3)
\end{equation}
and
\begin{equation}
  (\cC_{v,2}- \cC_{v,2,free})/(\gf^3\mu^3)=O(\gf^2).
\end{equation}
This means that beyond the free contribution 
$\cC_{v,2}$ only contains contributions which are suppressed with explicit powers
of $\gf$. Within our accuracy we may therefore neglect $\cC_{v,2}$ in the interaction
part of the specific heat.

Adding the transverse cut contribution [Eq.(\ref{cvh5})], the longitudinal cut contribution [Eq. ({\ref{cvh10})], 
the part of the pole contribution [Eq. (\ref{polel})] which is not exponentially suppressed, and the non-$n_b$ 
contribution [Eq. (\ref{pnon})], we arrive at our final result the the specific heat at low temperature\footnote{The
terms in the first two lines of Eq. (\ref{cv1}) beyond the leading logarithm
have been calculated for the first time by A. Ipp \cite{Ipp:2003qt}.}, 
\begin{eqnarray}
  &&\!\!\!\!\!\!\!\!\!\!\!\!
  {\mathcal \cC_v-\mathcal \cC_v^0\over N_g}={\gf^2\mu^2 T\over36\pi^2}\left(\log\left({4\gf\mu\over\pi^2T}\right)+\gamma_E
  -{6\over\pi^2}\zeta^\prime(2)-3\right)\nonumber\\
  &&-40{2^{2/3}\Gamma\left({8\over3}\right)\zeta\left({8\over3}\right)\over27\sqrt{3}\pi^{11/3}}
  T^{5/3}(\gf\mu)^{4/3}
  +560{2^{1/3}\Gamma\left({10\over3}\right)\zeta\left({10\over3}\right)
  \over81\sqrt{3}\pi^{13/3}}T^{7/3}(\gf\mu)^{2/3}\nonumber\\
  &&+{2048-256\pi^2-36\pi^4+3\pi^6\over180\pi^2}T^3
  \left[\log\left({\gf\mu\over T}\right)+c\right]\nonumber\\
  &&+{O}(T^{11/3}/(\gf\mu)^{2/3})
  +{O}(\gf^4\mu^2 T \log T),  \label{cv1}
\end{eqnarray}
where the constant $c$ is given by
\begin{equation}
  c={2160\pi^2\tilde c\over2048-256\pi^2-36\pi^4+3\pi^6} -{7\over12} \simeq-0.7300145937831\ldots,
\end{equation}
with $\tilde c$ given in Eq. (\ref{ctilde}).
We note that for sufficiently small temperatures there is a significant excess of the specific heat over its ideal-gas value,
whereas ordinary perturbation theory \cite{Kapusta:1979fh,Kapusta:TFT}
would have resulted in a low-temperature limit of $\cC_v/\cC_v^0=1-2\alpha_s/\pi$.

\subsection{HDL resummation}

As we have seen in the previous subsections, the nonanalytic terms in the low-temperature expansion of the entropy density
are determined by HDL contributions to the gluon self energy. Terms beyond the HDL approximation
are relevant for contributions from hard momenta $q\sim\mu$, yielding a term of order $\gf^2\mu^2T^2$
in the temperature-dependent part of the pressure, as shown in Eqs. (\ref{ht}) and (\ref{hl}).
In order to retain all contributions that are nonanalytic in $T$ at $T=0$, the HDL self energies
need to be kept unexpanded in
\begin{eqnarray}
  &&\!\!\!\!\!\!\!\!\!\!\!\!\!\!\!\!\!\!\!\!\!\!\!\!\!\!
  {1\over N_g}\left[P^\mathrm{HDL}-P^\mathrm{HDL}\Big|_{T=0}\right]
  =-{1\over2\pi^3}\int_0^\infty dq_0\,n_b(q_0)\int_0^{2\mu}dq\,q^2\nonumber\\
  &&\times\left[2\imag\log\left(q^2-q_0^2+\tilde\Pi_T\right)
  +\imag\log\left({q^2-q_0^2+\tilde\Pi_L\over q^2-q_0^2}\right)\right]. \label{hdlr1}
\end{eqnarray}
The sum of the transverse and longitudinal contribution decays like $q^{-3}$ for large $q$, therefore
we may send the upper integration limit of the $q$-integration to infinity. This just amounts to dropping
terms that are suppressed by explicit powers of $\gf$.
\begin{figure}
  \includegraphics[width=10cm]{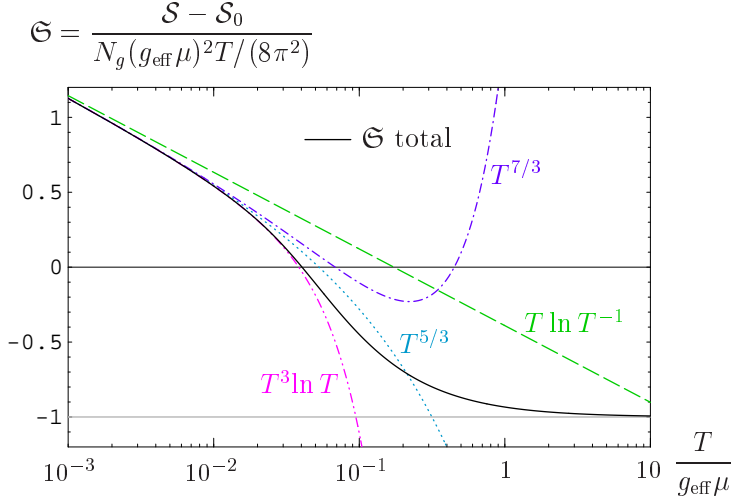}
  {\it\caption{The first few orders of the low-temperature series for the entropy density in comparison with the full
  HDL-resummed result (from \cite{Gerhold:2004tb}). \label{figent}}}
\end{figure}

In Eq. (\ref{hdlr1}) we have neglected terms like $\gf^2b^6\equiv(T/\mu)^2$ in $P/(\gf^4\mu^4)$, therefore we have to assume $T\ll\mu$.
This is however a weaker condition for the temperature than the condition $T\ll\gf\mu$ which we had in the previous subsections. 
The weaker condition is sufficient now, because we do not perform an expansion in $b$ in this subsection.
Adding the soft contribution from Eq. (\ref{hdlr1}), the hard contribution from Eqs. (\ref{ht}) and (\ref{hl}),
and the non-$n_b$ contribution from Eq. (\ref{pnon}), we arrive at the following result for the entropy density \cite{Gerhold:2004tb},
\begin{eqnarray}
  &&\!\!\!\!\!\!\!\!\!\!
  {1\over N_g}\left(\cS-\cS^0\right)=-{\gf^2\mu^2T\over24\pi^2}
  -{1\over2\pi^3}\int_0^\infty dq_0 {\partial n_b(q_0)\over\partial T}\int_0^{2\mu}dq\,q^2\nonumber\\
  &&\times\left[2\imag\log\left(q^2-q_0^2+\tilde\Pi_T\right)
  +\imag\log\left({q^2-q_0^2+\tilde\Pi_L\over q^2-q_0^2}\right)\right]
  +\mathcal{O}(\gf^4\mu^2T), \nonumber\\ \label{hdlr2}
\end{eqnarray}
where $\cS^0$ is the ideal-gas entropy density. The right hand side of Eq. (\ref{hdlr2}) is essentially given by one
universal function of the dimensionless variable $T/(\gf\mu)$, which we define through \cite{Gerhold:2004tb}
\begin{equation}
  {8\pi^2\over N_g \gf^2\mu^2 T}\left(\cS-\cS^0\right)=:\mathfrak{S}\left({T\over\gf\mu}\right)+\mathcal{O}(\gf^2).
\end{equation}
and which we have normalized such that the ordinary perturbative two-loop result \cite{Kapusta:1979fh,Kapusta:TFT} for the
low-temperature entropy density corresponds to $\mathfrak{S}=-1$. The function $\mathfrak{S}$ can be evaluated numerically. Fig \ref{figent}
shows a comparison of the first few orders of the low-temperature series with the full HDL result taken from 
\cite{Gerhold:2004tb}.
Further numerical results, in particular a comparison with non-perturbative large-$N_f$ results,
 can be found in \cite{Ipp:2003cj,Gerhold:2004tb}.

\section{Some remarks on neutrino emission from ungapped quark matter \label{sneutrino}}
A young neutron star loses energy mainly via neutrino emission from the bulk \cite{Yakovlev:2004iq,Schafer:2004jp}. 
 One generally distinguishes fast processes, which lead to neutrino emissivities proportional to $T^6$, and slow processes
which lead to neutrino emissivities proportional to $T^8$, see e.g. \cite{Yakovlev:2004iq} and references therein.
The dominant process 
for neutrino emission from ungapped quark matter is the quark analogon of the direct Urca process \cite{Schafer:2004jp,Iwamoto:1980eb,Iwamoto2},
\begin{eqnarray}
  d&\to&u+e^-+\bar\nu_e, \label{em1}\\
  u+e^-&\to& d+\nu_e. \label{em2}
\end{eqnarray} 
The cooling behavior is governed by the cooling equation
\begin{equation}
  \cC_v(T){\partial T\over\partial t}=-\epsilon(T),
\end{equation}
where $\epsilon(T)$ is the neutrino emissivity, which is given by \cite{Schafer:2004jp,Iwamoto:1980eb,Iwamoto2}
\begin{eqnarray}
  &&\epsilon(T)=3\sum_{\sigma_u,\sigma_d,\sigma_e}\int{d^3p_d\over(2\pi)^3}{1\over2E_d} \int{d^3p_u\over(2\pi)^3}{1\over2E_u}
  \int{d^3p_e\over(2\pi)^3}{1\over2E_e} \int{d^3p_\nu\over(2\pi)^3}{1\over2E_\nu} E_\nu\nonumber\\
  &&\qquad\times\bigg[ |M_\beta|^2(2\pi)^4\delta^4(P_d-P_u-P_e-P_\nu)\nonumber\\
  &&\qquad\qquad\times n_f(E_d-\mu_d)[1-n_f(E_u-\mu_u)][1-n_f(E_e-\mu_e)]\nonumber\\
  &&\qquad\quad+|M_{ec}|^2(2\pi)^4\delta^4(p_u+p_e-p_d-p_\nu)\nonumber\\
  &&\qquad\qquad\times n_f(E_u-\mu_u)n_f(E_e-\mu_e)[1-n_f(E_d-\mu_d)] \bigg], \label{emi}
\end{eqnarray}
where $\sigma_i$ are the fermion helicities, $E_i$ are the energies of the fermionic quasiparticles as determined from 
the poles of the propagators (neglecting the imaginary parts of the self energies), $M_\beta$ is the matrix
element for the reaction (\ref{em1}), and $M_{ec}$ is the matrix element for the reaction (\ref{em2}).

In \cite{Schafer:2004jp} it is shown that the emissivity through leading logarithmic order can be written as
\begin{eqnarray}
  &&\!\!\!\!\!\!\!\!\!\!\!\!\!\!\!\!\!\!\!\!\!\!\!\!\!\!\!
  \epsilon(T)={2g^2G_F^2\cos^2\theta_c\over\pi^7}\mu_e\mu_q^2T^6
  \int_{-\infty}^\infty dx_d\int_{-\infty}^\infty dx_u\int_{0}^\infty dx_\nu\,x_\nu^3\nonumber\\
  &&\!\!\!\!\!\!\!\!\!\!\!\!\!\!\!\!\!\!
  \times n_f(Tx_d)n_f(-Tx_u) n_f(T(x_u-x_d+x_\nu))\,v_g^{-1}(E_d)\,v_g^{-1}(E_u)\Big|_{E_i\to\mu_i+Tx_i}, \label{emi2}
\end{eqnarray} 
where $G_F$ is the Fermi coupling, $\theta_c$ is the Cabibbo angle, and $\mu_q\simeq\mu_u\simeq\mu_d$. The inverse
group velocity $v_g^{-1}$, as defined in Eq. (\ref{vg}), contains non-Fermi-liquid contributions, which we have
discussed in chapter \ref{csigma}.
As shown in \cite{Schafer:2004jp}
the non-Fermi-liquid corrections to the specific heat and the emissivity lead to a (modest) reduction of the
temperature at late times.

If one wants to go beyond the leading logarithmic accuracy of Ref. \cite{Schafer:2004jp}, at least the following ingredients will be
required. 
First, there will be  corrections to the emissivity from gluonic
corrections to the weak interaction vertex \cite{Schafer:2004jp}. It might also be necessary to generalize 
the formula (\ref{emi}) for the emissivity
in a way that takes into account the finite lifetime of the fermionic quasiparticles, since the
imaginary part of the quark self energy is of the same order as the term which fixes the scale under the logarithm in the real part.

As a first step towards a more complete computation of the emissivity, let us take the scale 
under the leading logarithm of the group velocity as given by Eqs. (\ref{resit}) and (\ref{fs1}),
and simply insert the resulting expression for the group velocity into Eq. (\ref{emi2}). Using the 
integral
\begin{eqnarray}
  &&\int_{-\infty}^\infty dx_1\int_{-\infty}^\infty dx_1\int_{0}^\infty dx_3\, n_f(Tx_1)n_f(-Tx_2) n_f(T(x_2-x_1+x_3))\nonumber\\
  &&\qquad\times{\partial\over\partial\alpha}\left(\Li_{\alpha}(-e^{-x_1})+\Li_{\alpha}(-e^{x_1})\right)\Big|_{\alpha=0}
  =-105.303\ldots
\end{eqnarray}
we would then find
\begin{equation}
  \epsilon(T)\simeq {457\over2520\pi}g^2G_F^2\cos^2\theta_c\,\mu_q^2\mu_eT^6\left(1+{2g^2\over9\pi^2}\log\left({4.295m\over T}\right)
  \right).
\end{equation}
While the constant under the logarithm will certainly be modified in a more complete computation as discussed above, 
it is already interesting to note that
this constant is about an order of magnitude larger than the corresponding constant in the specific heat, which we find to be
approximately equal to $0.282$ from Eq. (\ref{cv1}).

\chapter{Color superconductivity \label{ccsc}}
\section{Basics of superconductivity}
Since its experimental discovery in 1911, 
superconductivity has been extensively studied both in experimental and theoretical condensed matter physics. 
The first microscopic theory of superconductivity was developed
by Bardeen, Cooper and Schrieffer (BCS) in 1957 \cite{Bardeen:1957mv}, see e.g. 
\cite{Fetter:1971,Schrieffer:1964,Mahan:1981} for pedagogical reviews. The key
observation is that the electron-phonon interaction induces an effective electron-electron interaction which is
attractive\footnote{Using Wilsonian renormalization group methods \cite{Wilson:1973jj}, 
it can be shown that the four fermion operator corresponding to the scattering of two electrons with opposite momenta
is the only marginal operator in the vicinity of the Fermi surface, all other fermionic interactions
being irrelevant \cite{Polchinski:1992ed}.}.
This leads to the formation of Cooper pairs. The ground state, which is different from the ground state of a
normal Fermi liquid, is then given by
\begin{equation}
  |\Psi\rangle=\prod_{\bf k} \left(u_{\bf k}+v_{\bf k}c_{{\bf k}\uparrow}^\dag c_{-{\bf k}\downarrow}^\dag\right)
  |0\rangle, \label{csc1}
\end{equation}
where $c_{{\bf k}\uparrow}^\dag$ and $c_{{\bf k}\downarrow}^\dag$ are creation operators for electrons with 
spin up and spin down, respectively.
The coefficients $u_{\bf k}$ and $v_{\bf k}$ are determined by minimization of the free energy. 
The state (\ref{csc1}) can be characterized by a non-vanishing ``di-electron'' condensate,
\begin{equation}
  \langle\psi\psi\rangle\neq0, \label{csc3}
\end{equation}
which leads to a gap in the fermionic quasiparticle spectrum.
A condensate like (\ref{csc3}) breaks spontaneously the $U(1)$ invariance of electrodynamics.
Therefore the photon will acquire a mass via the Higgs mechanism. This leads to the
Meissner effect, namely the screening of magnetic fields inside a superconductor.


\section{Basics of color superconductivity}

Let us begin with an ensemble of free quarks. At finite density
the ground state of this system is simply given by a Fermi sphere. 
Now let us turn on interactions. At asymptotic densities at first sight a
purely perturbative treatment might seem appropriate because the interactions are ``weak'' due to
asymptotic freedom. However, it is important to note 
that single gluon exchange is attractive in the color antitriplet channel\footnote{This can be seen as follows.
The tree level scattering amplitude of two quarks is equal to the tree level scattering amplitude of two electrons
in QED, times the QCD factor
\begin{equation}
  T^a_{ik}T^a_{jl}=-{1\over3}(\delta_{ik}\delta_{jl}-\delta_{il}\delta_{jk})
  +{1\over6}(\delta_{ik}\delta_{jl}+\delta_{il}\delta_{jk}). \label{sg}
\end{equation}
Since the electron-electron interaction is repulsive, the negative sign of the first term in Eq. (\ref{sg})
implies that the quark-quark interaction is attractive in the antitriplet channel.}. This has dramatic
consequences, since an arbitrarily weak attractive interaction leads to the Cooper instability of
the Fermi surface, and the ground state will contain diquark condensates,
\begin{equation}
  \Phi^{ij,fg,\alpha\beta}\sim\langle\psi_c^{i,f,\alpha}({\bf p})\bar\psi^{i,g,\beta}(-{\bf p})\rangle\neq0,
  \label{psipsi}
\end{equation}
where $i$, $j$ are fundamental color indices,  $f$, $g$ are flavor indices, $\alpha$, $\beta$
are spinor indices, and
$\psi_c$ denotes the charge conjugated spinor, $\psi_c=C\bar\psi^T$.
By virtue of its similarity to superconductivity in ordinary condensed matter phy\-sics, 
this phenomenon was termed \emph{color superconductivity}.

In the usual notation for the irreducible SU(3) representations one has
\begin{equation}
  {\bf 3} \otimes {\bf 3} = {\bf 6} \oplus {\bf\bar 3}. 
\end{equation}
Therefore one cannot construct a color singlet from the expectation value (\ref{psipsi}).
This means that global color symmetry is spontaneously broken by the diquark condensate.

Quarks have color, flavor and spin as internal degrees of freedom. Therefore quite
many different color superconducting phases could be possible, depending on the color, flavor and spin structure of the
expectation value (\ref{psipsi}).
At very high densities the phase with the lowest free energy is the
so-called color flavor locked (CFL) phase \cite{Alford:1998mk,Schafer:1999fe,Shovkovy:1999mr,Evans:1999at}, 
where up, down and strange quarks 
contribute to the pairing on an equal footing,
\begin{equation}
  \Phi_\mathrm{CFL}\propto \varepsilon^{ijA}\varepsilon^{fgA}=\delta^{if}\delta^{jg}-\delta^{ig}\delta^{jf}.
\end{equation} 
The symmetry breaking pattern of the CFL phase is (neglecting electromagnetism and quark masses for the moment)
\begin{equation}
  SU(3)_c\times SU(3)_L\times SU(3)_R \times U(1)_B\to SU(3)_{c+L+R}\times {\mathbbm Z}_2. \label{sb}
\end{equation}
Here $SU(3)_c$ corresponds to global color symmetry, $SU(3)_{R,L}$ correspond to right and left handed
flavor symmetry, and $U(1)_B$ corresponds to baryon number conservation. These symmetries are broken down to
the diagonal subgroup $SU(3)_{c+L+R}$ times a ${\mathbbm Z}_2$ group that corresponds to $\psi\to-\psi$.
In the CFL phase all eight gluons acquire a mass through the Higgs mechanism \cite{Son:1999cm,Rischke:2000ra,Casalbuoni:2000na}.
The electromagnetic $U(1)$ symmetry is also broken, but there remains an unbroken
$\tilde U(1)$ \cite{Alford:1998mk,Alford:1999pb}, whose generator is given by $\tilde Q_\mathrm{CFL}=Q+{2\over\sqrt{3}}T_8$, with 
$Q=\mathrm{diag} (-{1\over3},-{1\over3},{2\over3})$ in flavor 
space, and $T_8={1\over2\sqrt{3}}\mathrm{diag}(1,1,-2)$ in color space. 

As in every field theoretical model with spontaneous symmetry breaking, Nambu-Goldstone bosons appear in the CFL phase
\cite{Son:1999cm,Casalbuoni:1999wu,Rho:1999xf,Manuel:2000wm,Zarembo:2000pj,Beane:2000ms}.
Because of the symmetry breaking pattern (\ref{sb}) there are 17 broken generators, which would lead to 17 Nambu-Goldstone
bosons. Eight of these, corresponding to the breaking of $SU(3)_c$, are ``eaten'' by the gluons. 
The ``surviving'' Nambu-Goldstone bosons are an octet corresponding to flavor symmetry breaking and a singlet corresponding
to baryon number breaking.
While this singlet boson is strictly massless, 
the octet bosons become massive if finite quark
masses are taken into account. Furthermore there is a light singlet from $U(1)_A$ breaking.

At lower densities, the strange quark will eventually decouple, which should lead to a two-flavor color
superconducting (2SC) phase \cite{Bailin:1983bm,Alford:1997zt,Rapp:1997zu}, where only up and down quarks participate in the pairing,
\begin{equation}
  \Phi_\mathrm{2SC}\propto \varepsilon^{ij3}\varepsilon^{fg3}.
\end{equation}
In the 2SC phase the global color symmetry group $SU(3)_c$ is broken down to $SU(2)_c$, 
but flavor symmetry remains unbroken. Therefore there are five would-be Nambu-Goldstone bosons 
\cite{Casalbuoni:2000cn,Miransky:2001sw,Rischke:2002rz}, which are ``eaten'' by the gluons.
The gluons with adjoint colors $1$-$3$ are massless, while the gluons with adjoint colors $4$-$8$
acquire a mass through the Higgs mechanism \cite{Rischke:2000qz,Casalbuoni:2001ha}.
As in the CFL phase the electromagnetic $U(1)$ symmetry is broken, but again an unbroken
$\tilde U(1)$ survives \cite{Casalbuoni:2000cn,Schafer:1999pb}, whose generator is now given by 
$\tilde Q_\mathrm{2SC}= {1\over3\sqrt{2}}({\bf 1}-2\sqrt{3}T_8)$. 

In Ref. \cite{Alford:2002kj} it was argued that the 2SC phase will be disfavored if one
takes into account the constraints from color and electric charge neutrality at finite strange
quark mass $m_s$. To date a completely reliable description of the phase structure of QCD at densities below the
CFL regime is still lacking. In the following we will list some of the phases that
have been proposed in the literature.

Computing the meson masses from the low energy effective Lagrangian in the CFL phase,
one finds that the masses of the $K^+$ and $K^0$ become imaginary if $m_s$ exceeds
a certain value. This indicates the formation of a kaon condensate 
\cite{Schafer:2003vz,Schafer:2000ew,Bedaque:2001je,Kaplan:2001qk}.

The charge neutrality condition and the finite strange quark mass
lead to a mismatch of the Fermi momenta of the different quark flavors. This might give rise to
so-called gapless superconducting phases (g2SC \cite{Huang:2004ik,Shovkovy:2004me,Shovkovy:2003uu,Huang:2003xd} and 
gCFL \cite{Alford:2003fq,Alford:2004hz,Fukushima:2004zq,Alford:2004zr}), where the fermionic quasiparticle dispersion laws
are modified, some of them corresponding to gapless excitations. In Refs. \cite{Huang:2004bg,Huang:2004am,Casalbuoni:2004tb} however, it was found
that the Meissner masses are imaginary in these phases. This would indicate that the gapless phases are unstable, if no mechanism
can be found which makes the Meissner masses real quantities.

So far we have assumed only spin zero diquark condensates. The gaps resulting from spin one condensates
are much smaller than in the spin zero case.
Nevertheless spin one condensates might play a certain role at intermediate densities, 
since they may be formed from quarks of the same flavor, see Refs.
\cite{Bailin:1983bm,Pisarski:1999bf,Pisarski:1999tv,Schafer:2000tw,Schmitt:2002sc,Schmitt:2004hg}.

The mismatch of the Fermi momenta 
might also induce a so-called LOFF (Larkin-Ovchin\-nikov-Fulde-Ferrell) pairing \cite{Alford:2000ze,Fulde:1964,Larkin:1964,
Bowers:2002xr,Casalbuoni:2003wh}. 
The LOFF state is characterized by Cooper pairs with non-zero total momentum, and
therefore translational and rotational symmetry are spontaneously broken. This leads to crystalline structures
that might have relevant consequences for astro\-physics \cite{Rajagopal:2000wf}.


\section{The color superconductivity gap equation}

The gap equation for a color superconducting phase can be derived from the Schwinger-Dyson equation
for the quark two-point function, allowing for 
a non-vanishing expectation value (\ref{psipsi}). At this point it
is convenient to introduce a Nambu-Gor'kov basis for the quark spinors, which
is given by $\Psi=(\psi, \psi_c)^T$, $\bar\Psi=(\bar\psi, \bar\psi_c)$. 
The inverse quark propagator in this basis is given by \cite{Bailin:1983bm,Wang:2001aq,Manuel:2000nh}
\begin{equation}
  S^{-1}=\left(
    \begin{array}{cc} \Qs+\mu\gamma_0
    +\Sigma & \Phi^- \\
    \Phi^+ & \Qs-\mu\gamma_0+\bar\Sigma 
    \end{array}
  \right), \label{sc1}
\end{equation}
where $\Phi^\pm$ are the gap functions, related by
$\Phi^-(Q)=\gamma_0 [\Phi^+(Q)]^\dag\gamma_0$, and $\Sigma(Q)$ is the quark self energy, with
$\bar\Sigma(Q)=C[\Sigma(-Q)]^T C^{-1}$. Flavor and fundamental color indices are suppressed in Eq. (\ref{sc1}).

In order to solve the Schwinger-Dyson equation a suitable approximation scheme has to be chosen.
To date only the exponent and the leading contribution to the prefactor of the gap [see Eq. (\ref{gap}) below] have been determined 
consistently, where the following set of approximations turned out to suffice:
\begin{itemize}
 \item{replace the full quark-gluon vertex with its tree level counterpart (at least in Coulomb gauge)
  \cite{Brown:2000eh}},
 \item{replace the full gluon propagator with its HDL approximation \cite{Rischke:2001py}},
 \item{replace the quark self energy
   with the leading logarithmic result of the normal phase (see Eq.~(\ref{resit})) \cite{Wang:2001aq}.}
\end{itemize}
A rather lengthy calculation \cite{ Rischke:2003mt,Pisarski:1999tv, Wang:2001aq}
then yields the following gap equation for the 2SC phase 
(neglecting the gauge dependent part)
\begin{equation}
  \phi^+(\epsilon_{\bf k},k)={g^2\over18\pi^2}\int_0^\delta d(q-\mu)Z(\epsilon_{\bf q})
  {\phi^+(\epsilon_{\bf q},q)\over\epsilon_{\bf q}}\tanh\left({\epsilon_{\bf q}\over 2T}\right)
  {1\over2}\log\left({b^2\mu^2\over|\epsilon_{\bf q}^2-\epsilon_{\bf k}^2|}\right),
\end{equation}
where $\epsilon_{\bf k}=\sqrt{(k-\mu)^2+|\phi^+|^2}$, $\delta\sim\mathcal{O}(g\mu)$ is a cutoff, $b=256\pi^4(2/( N_f g^2))^{5/2}$, and
\begin{equation}
  Z(E-\mu)=\left[1-{\partial\real\Sigma_+\over\partial (E-\mu)}\right]^{-1}
\end{equation}
is usually called ``wave function renormalization factor'' in this context (in the previous chapters we referred to this quantity
as group velocity). With the definition
\begin{equation}
  x={g\over3\sqrt{2}\pi}\log\left({2b\mu\over k-\mu+\epsilon_{\bf k}}\right)
\end{equation} 
the solution of the gap equation at zero temperature can be written as 
\cite{ Brown:1999aq,Pisarski:1999tv, Wang:2001aq,Son:1998uk,Schafer:1999jg, Hong:1999fh, Hsu:1999mp}
\begin{equation}
  \phi^+=2 b b_0^\prime\mu\exp\left(-{3\pi^2\over\sqrt{2}g}\right)F(x), \label{gap}
\end{equation}
where $b_0^\prime=\exp[-(\pi^2+4)/8]$. $F(x)$ is quite a complicated function, which is given
explicitly in \cite{Wang:2001aq}. At the Fermi surface $F(x)$ is equal to $1+\mathcal{O}(g^2)$, 
and away from the Fermi surface it decreases rapidly.

The parametric dependence of the gap on the coupling constant as shown in Eq. (\ref{gap})
was first derived by Son \cite{Son:1998uk} using renormalization group techniques.
It is different from the BCS result $\phi^+\sim\exp(-1/g^2)$, which is a consequence of
the fact that quasistatic chromomagnetic interactions are only dynamically screened.

The result for the gap can also be derived within the framework of high density effective theory (HDET),
in which the only relevant fermionic degrees of freedom are those in the 
vicinity of the Fermi surface \cite{Nardulli:2002ma,Hong:1998tn,Hong:1999ru,Schafer:2003jn,Reuter:2004kk}.

In the literature also the possible gauge dependence of the prefactor of the gap, which would indicate
an inconsistent approximation scheme, has been discussed. In \cite{Rajagopal:2000rs} the gap equation was
solved numerically in a general covariant gauge, with the result that large gauge dependences
occur for $g>g_c\sim0.8$. In Ref. \cite{Hou:2004bn}, however, it was argued that in a covariant gauge 
one has to include the quark gluon 
vertex correction in order to obtain a gauge independent result. In \cite{Pisarski:2001af}
it was shown that the prefactor of the gap is gauge independent in a general Coulomb gauge, if the gap is
evaluated on the quasiparticle mass shell. In the next section we will give a general proof that the
fermionic quasiparticle dispersion laws in a color superconductor are gauge independent \cite{Gerhold:2003js}.

\section{Gauge independence of fermionic dispersion laws}
The inverse quark propagator in the Nambu-Gor'kov basis, as shown in Eq. (\ref{sc1}),
is the momentum space version of the second derivative of
the effective action,
\begin{equation}
  {\delta^2\Gamma\over\delta\bar\Psi(x)\delta\Psi(y)}\Big|_{\psi=\bar\psi=A_i^a=0,
  A_0^a=\bar A_0^a},
  \label{sc2}
\end{equation}
where $\Psi=(\psi, \psi_c)^T$, $\bar\Psi=(\bar\psi, \bar\psi_c)$, and $\bar A_0^a$
is the expectation value of $A_0^a$, which is in general non-vanishing (see Sec. \ref{secg}).

It should be noted that the doubling of fermionic fields in
terms of $\Psi$ and $\bar\Psi$ is just a notational convenience here;
the effective action itself should be viewed as depending only
on either $(\psi,\bar \psi)$ or the set $\Psi=(\psi, \psi_c)^T$.

The gauge dependence identity for the effective action [Eq. (\ref{gauge})]
can be written as
\begin{eqnarray}
  &&\!\!\!\!\!\!\!\!\!\!\!\!\!\!\!\!\delta\Gamma =\int d^4x\left({\delta\Gamma\over\delta\psi(x)}
  \delta X_{(\psi)}(x)-\delta X_{(\bar\psi)}(x)
  {\delta\Gamma\over\delta\bar\psi(x)}+{\delta\Gamma\over\delta A^{a\mu}(x)}
  \delta X_{(A)}^{a\mu}(x)\right) \nonumber\\
  &&\!\!\!\!\!\!\!\equiv\int d^4x\left({\delta\Gamma\over\delta\psi(x)}
  \delta X_{(\psi)}(x)+{\delta\Gamma\over\delta\psi_c(x)}
  \delta X_{(\psi_c)}(x)+{\delta\Gamma\over\delta A^{a\mu}(x)}
  \delta X_{(A)}^{a\mu}(x)\right),\quad\label{sc3}
\end{eqnarray}
Eq. (\ref{sc3}) can be cast in a more compact form using the DeWitt
notation, but now only for the fermions,
\begin{equation}
  \delta\Gamma= \Gamma_{,i}\delta X^i+\int d^4x{\delta\Gamma\over\delta A^{a\mu}(x)}
  \delta X_{(A)}^{a\mu}(x), 
  \label{sc5}
\end{equation}
where $i=(\psi(x), \psi_c(x))^T$, $\bar i=(\bar\psi(x), \bar\psi_c(x))$,
and the comma denotes functional derivation. Taking the second derivative of
(\ref{sc5}), setting $\psi=\bar\psi=A^a_i=0$, $A^a_0=\bar A^a_0$,
and using the fact that
\begin{equation} 
  {\delta\Gamma\over\delta A^a_0}\Big|_{\psi=\bar\psi=A^a_i=0,A^a_0=\bar A^a_0}=0,
\end{equation}
we obtain a gauge dependence identity for the inverse propagator (\ref{sc2}),
\begin{equation} \label{gdiip1}
  \delta\Gamma_{i\bar j}=-\Gamma_{k\bar j}\delta X^k_{\,,i}+
  \Gamma_{ki}\delta X^{k}_{\,,\bar j}+\int d^4x{\delta\Gamma_{i\bar j}\over\delta A^{a0}(x)}
  \delta X_{(A)}^{a0}(x).
\end{equation}

Up to this point, our functional relations are completely general
and apply also to the case of inhomogeneous color superconducting phases. 
We do not attempt to cover
the complications this case may add to the question of gauge
independence, but continue by assuming translational invariance.
This allows us to introduce $\delta\bar A^{a0}:=-\delta X_{(A)}^{a0}(x=0)$
and to write (\ref{gdiip1}) as
\begin{equation}
  \delta\Gamma_{i\bar j}=-\Gamma_{k\bar j}\delta X^k_{\,,i}-
  \delta X^{\bar k}_{\,,\bar j}\Gamma_{i\bar k}-{\partial\Gamma_{i\bar j}\over\partial \bar A^{a0}}
  \delta\bar A^{a0}.
  \label{e6}
\end{equation}
Furthermore, we can transform Eq. (\ref{e6}) into momentum space,
\begin{equation} 
  \delta\Gamma_{i\bar j}(Q)+\delta\bar A^{a0}{\partial\Gamma_{i\bar j}(Q)\over\partial \bar A^{a0}}
  =-\Gamma_{k\bar j}(Q)\delta X^k_{\,,i}(Q)-
  \delta X^{\bar k}_{\,,\bar j}(Q)\Gamma_{i\bar k}(Q), \label{sc7}
\end{equation}
where the indices $i$ and $\bar i$ from now on comprise only color, flavor, 
Dirac and Nambu-Gor'kov indices. 
Using Eq. (\ref{c1})
we obtain 
\begin{equation}
  \delta\det(\Gamma_{i\bar j})+\delta\bar A^{a0}
  {\partial\over\partial \bar A^{a0}}\det(\Gamma_{i\bar j})
  \equiv \delta_{\rm tot}\det(\Gamma_{i\bar j})
  =-\det(\Gamma_{i\bar j})[\delta X^k_{\,,k}
  +\delta X^{\bar k}_{\,,\bar k}]. \label{sc8}
\end{equation}
The left hand side of this identity is the total variation 
\cite{Nielsen:1975fs,Aitchison:1983ns} of the determinant of the inverse
quark propagator, with the first term
corresponding to the explicit variation of the gauge fixing function, and the second term coming from the
gauge dependence of $\bar A_0^a$.

Since the determinant is equal to the product of the eigenvalues, Eq. (\ref{sc8}) implies 
that the location of the singularities of the quark propagator 
is gauge independent, provided that the singularities of $\delta X^k_{\,,k}$ do not coincide
with those of the quark propagator. 
The singularity structure of $\delta X$ may be discussed in a similar way as in the case of
the Higgs model, see Sec. \ref{sechig}. The quantity
$\delta X$ is
1PI up to a full ghost propagator, and up to gluon tadpole insertions
(see Ref. \cite{Kobes:1990dc} for the explicit diagrammatic structure)
\footnote{It should be noted that $\delta X^k_{\,,k}$ will not contain any 
gluon tadpoles if its perturbative expansion is constructed using the 
shifted field $A^\prime=A-\bar A$.}.
As in \cite{Kobes:1990dc,Kobes:1990xf} one may argue that
the singularities of the ghost propagator are not
correlated to the singularities of the quark propagator. 
Gauge independence of the zeros of the inverse fermion propagator
then follows provided that
also the 1PI parts of $\delta X$ have
no singularities coinciding with the singularities of the fermion propagator.
An important caveat in fact comes from massless poles in the
unphysical degrees of freedom of the gauge boson propagator,
which are typical in covariant gauges and which can give rise
to spurious mass shell singularities as encountered in the case
of hot QCD in \cite{Baier:1991dy}.
But, as was pointed out in \cite{Rebhan:1992ak},
these apparent gauge dependences are avoided if the quasiparticle
mass-shell is approached with a general infrared cut-off such
as finite volume, and this cut-off lifted only in the end, i.e.,
after the mass-shell limit has been taken.

The determinant which appears in (\ref{sc8}) 
is taken with respect to color, flavor,
Dirac and Nambu-Gor'kov indices. The determinant in Nambu-Gor'kov space may be evaluated
explicitly using Eq. (\ref{c2}).
From Eq. (\ref{sc1}) we obtain in this way
\begin{eqnarray}
  \det(\Gamma_{i\bar j})&=&\det\big[(\Qs-\mu\gamma_0+\bar\Sigma)
  (\Qs+\mu\gamma_0+\Sigma)\nonumber\\
  &&\qquad-(\Qs-\mu\gamma_0+\bar\Sigma)
  \Phi^-(\Qs-\mu\gamma_0+\bar\Sigma)^{-1}\Phi^+
  \big].\qquad \label{sc10}
\end{eqnarray}
The inverse of the matrix of which the determinant is taken here appears, of course, in the ordinary
quark propagator, which is obtained by inverting (\ref{sc1}),
\begin{eqnarray}
  &&\!\!\!\!\!\!\!\!\!\cG ^+(Q)=\big[(\Qs-\mu\gamma_0+\bar\Sigma)
  (\Qs+\mu\gamma_0+\Sigma)\nonumber\\
  &&\qquad\quad-(\Qs-\mu\gamma_0+\bar\Sigma)
  \Phi^-(\Qs-\mu\gamma_0+\bar\Sigma)^{-1}\Phi^+
  \big]^{-1}(\Qs-\mu\gamma_0+\bar\Sigma).\qquad
\end{eqnarray}

At leading order, when the quark self energy $\Sigma$ can be neglected, (\ref{sc10})
can be approximated by
\begin{equation}
  \det(\Gamma_{i\bar j})\simeq\det\left[(\Qs-\mu\gamma_0)
  (\Qs+\mu\gamma_0)-(\Qs-\mu\gamma_0)
  \Phi^-(\Qs-\mu\gamma_0)^{-1}\Phi^+\right].
\end{equation}
This expression can be rewritten as
\begin{eqnarray}
  \det(\Gamma_{i\bar j})&\simeq&\det\Big[\left(q_0^2-(q-\mu)^2-
  (\phi_{r+}^+)^\dag\phi_{r+}^+\right)\mathcal P_{r+}^+\nonumber\\
  &&\quad\ +\left(q_0^2-(q-\mu)^2-(\phi_{l-}^+)^\dag\phi_{l-}^+\right)
  \mathcal P_{l-}^+\nonumber\\
  &&\quad\ +\left(q_0^2-(q+\mu)^2-(\phi_{r-}^-)^\dag\phi_{r-}^-\right)
  \mathcal P_{r-}^-\nonumber\\
  &&\quad\ +\left(q_0^2-(q+\mu)^2-(\phi_{l+}^-)^\dag\phi_{l+}^-\right)
  \mathcal P_{l+}^-\Big], \label{det}
\end{eqnarray}
where the projection operators introduced in  \cite{Pisarski:1999av} are given by
\begin{equation}
  \mathcal{P}_{r,l\pm}^\pm=\mathcal{P}_{r,l}\mathcal{P}_\pm({\bf q})\Lambda^\pm({\bf q}), \label{proj}
\end{equation}
where $\mathcal{P}_{r,l}={1\over2}(1\pm\gamma_5)$ is the chirality projector, $\mathcal{P}_\pm({\bf q})={1\over2}
(1\pm\gamma_5\gamma_0\gamma^i\hat{ q}^i)$ is the helicity projector, and $\Lambda^\pm({\bf q})$ is the
energy projector given in Eq. (\ref{ep}). (Note that 
$\mathcal{P}^+_{r-}=\mathcal{P}^+_{l+}=\mathcal{P}^-_{r+}=\mathcal{P}^-_{l-}=0$
in the massless limit which we are considering here \cite{Pisarski:1999av}.)
If we assume for simplicity that $\phi^\dag\phi$ is diagonal with respect to color and 
flavor indices, the determinant in Eq. (\ref{det}) can be readily evaluated,
\begin{eqnarray}
  \det(\Gamma_{i\bar j})&\simeq&\prod_f\left(q_0^2-(q-\mu)^2-
  |\phi_{r+}^{+(f)}|^2\right)\left(q_0^2-(q-\mu)^2-
  |\phi_{l-}^{+(f)}|^2\right)\nonumber\\
  &&\!\!\times\left(q_0^2-(q+\mu)^2-
  |\phi_{r-}^{-(f)}|^2\right)\left(q_0^2-(q+\mu)^2-
  |\phi_{l+}^{-(f)}|^2\right),\qquad
\end{eqnarray}
giving the well known branches of the quasiparticle excitation spectrum 
for each value of the color-flavor index $f$ \cite{Pisarski:1999av}.

At leading order, when the quark self energy can be neglected, 
Eqs. (\ref{sc8}) and (\ref{sc10}) imply that the gap function is gauge independent
on the quasiparticle mass shell.
It should be noted, however, that at higher orders only dispersion
relations obtained from (\ref{sc10}), which also include the quark self energy $\Sigma$,
can be expected to be gauge independent.

We note furthermore that for $\Phi^\pm\to0$ the determinant (\ref{sc10}) becomes
\begin{equation}
  \det\left[(\Qs+\mu\gamma_0+\Sigma)(\Qs-\mu\gamma_0+\bar\Sigma)\right]
  =\det(\cG ^+_0(Q)^{-1})\times\det(\cG ^+_0(-Q)^{-1}),
  \label{sc15}
\end{equation}
where $\cG ^+_0(Q)=[\Qs+\mu\gamma_0+\Sigma]^{-1}$. 
Therefore, in this case the gauge independence proof, of course, boils down to the gauge 
independence proof for the singularities of the quark propagator
without diquark condensates.

\section{Computation of $\delta X$}

\begin{figure}
 \begin{center}
  \includegraphics{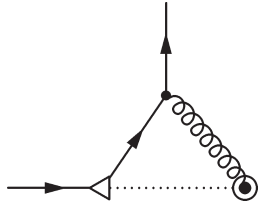}
 \end{center}
 \vspace{-5mm}
 {\it\caption{Feynman diagram for $\delta X_{(1)}$ (see Eq. (\ref{dx})). \label{dxfig}}}
\end{figure}

The aim of this subsection is the evaluation of $\delta X^k_{\,,k}$ for the 2SC phase in covariant gauge. 
This is interesting because the existing calculations in the literature give gauge dependent results
for the gap in this gauge\footnote{In \cite{Hong:2003ts} the gap equation was considered in a non-local gauge, with the result
that the ``best''  value of the gauge parameter in covariant gauge should be $\xi\simeq {1\over3}$, but then the
result for the gap does not agree with the well-established result in Coulomb gauge \cite{Pisarski:2001af}.}.

We will focus on possible
singularities of $\delta X^k_{\,,k}$, since they might invalidate the above proof of gauge independence.
At one-loop level $\delta X^k_{\,,k}$ contains terms like 
\begin{eqnarray}
  &&\delta X_{(1)}(K) := g^2\int {d^4 Q\over i(2\pi)^4} \mathrm{Tr}_{D,c,f}\bigg[T_a
  {1\over P^2} {1\over\sqrt{2\xi}}P^\mu D_{\mu\nu}(P)
  \cG ^+(Q)\gamma^\nu T_a\nonumber\\
  &&\qquad\qquad\qquad+T_a^T{1\over P^2} {1\over\sqrt{2\xi}}P^\mu D_{\mu\nu}(P) \cG ^-(Q)\gamma^\nu T_a^T\bigg], \label{dx}
\end{eqnarray}
where $P=K-Q$,  $D_{\mu\nu}$ is the gluon propagator, $\cG^\pm$ is the quark propagator (see below), and we have used
\begin{eqnarray}
  D^i_\alpha &\rightarrow& -igT_a\psi,\ igT_a^T\psi_c \\
  \mathcal G^\alpha_{\ \beta} &\rightarrow& {1\over P^2}\delta^{ab},\\
  \delta F^\alpha \propto F^\alpha &\rightarrow& {1\over\sqrt{2\xi}}P^\mu A_\mu^a.
\end{eqnarray}
Eq. (\ref{dx}) corresponds to the diagram in Fig. \ref{dxfig}, where the vertex with a circle denotes $\delta F^\alpha$
and the small triangle denotes $D_\alpha^i$.

In the 2SC phase the gap matrix is given by \cite{Alford:1997zt,Pisarski:1999tv} 
\begin{equation}
  [\Phi^+]_{ij}^{fg}=\varepsilon^{fg}\varepsilon_{ij3}(\phi^+(\mathcal{P}^+_{r+}-\mathcal{P}^+_{l-})
 +\phi^-(\mathcal{P}^-_{r-}-\mathcal{P}^-_{l+})),
\end{equation}
where $f$, $g$ are flavor indices and $i$, $j$ are fundamental color indices. The $\mathcal{P}$'s are the
projection operators defined in Eq. (\ref{proj}). We have assumed for simplicity that the right-handed and
the left-handed gap functions are equal up to a sign \cite{Pisarski:1999tv}. 
For the quark propagator $\cG^\pm(Q)$ one finds then \cite{Rischke:2000qz}
\begin{equation}
  [\cG^\pm]_{ij}^{fg}=\delta^{fg}\left[(\delta_{ij}-\delta_{i3}\delta_{j3})G^\pm+\delta_{i3}\delta_{j3}G^\pm_0\right], \label{g}
\end{equation} 
with \cite{Rischke:2000qz}
\begin{eqnarray}
  G^\pm(Q)&=&\sum_{e=\pm}{q_0\mp(\mu-eq)\over q_0^2-(\mu-eq)^2-|\phi^e|^2}\Lambda^{\pm e}
  ({\bf q})\gamma_0,\label{g1}\\
  G^\pm_0(Q)&=&\sum_{e=\pm}{q_0\mp(\mu-eq)\over q_0^2-(\mu-eq)^2}\Lambda^{\pm e}
  ({\bf q})\gamma_0, \label{g0}
\end{eqnarray}
where $\Lambda^\pm$ are the energy projectors given in Eq. (\ref{ep}). 
We shall consider only the part of $\delta X_{(1)}$ which contains the 
propagator $G^\pm$, because it is this part which depends on the condensate. 
We will call this part $\delta X_{(11)}$.
In the HDL approximation the gluon propagator obeys the following Ward identity,
\begin{equation}
  P^\mu D_{\mu\nu}(P)=\xi{P_\nu\over P^2}.
\end{equation}  
Thus we obtain after taking the color trace, neglecting the antiparticle propagator and the
imaginary part of the gap function \cite{Pisarski:1999tv},
and dropping an irrelevant prefactor
\begin{eqnarray}
  &&\!\!\!\!\!\!\!\!
  \delta X_{(11)}(K)=2g^2\sqrt{\xi}\int {d^4 Q\over i(2\pi)^4}{1\over(P^2)^2}{q_0\over q_0^2
  -(\mu-q)^2-(\phi^+)^2}\mathrm{Tr}_D[\Lambda^+_{\bf q}\gamma_0\gamma_\nu]P^\nu\nonumber\\
  &&\!\!\!\!=4g^2\sqrt{\xi}\int {d^4 Q\over i(2\pi)^4}{1\over(p_0^2-p^2)^2}{q_0(p_0-{\bf p}
  \cdot{\bf \hat  q})\over q_0^2-(\mu-q)^2-(\phi^+)^2}=\nonumber\\
  &&\!\!\!\!=4g^2\sqrt{\xi}\int {d^4 Q\over i(2\pi)^4}{-iq_4(i(k_4-q_4)-kt+q)
  \over[(k_4-q_4)^2+(k^2+q^2-2kqt)]^2(q_4^2+(\mu-q)^2+(\phi^+)^2)},\quad\nonumber\\
\end{eqnarray}  
where $t=\cos\theta$ and $q_4=-iq_0$. The integral is dominated by $\theta\simeq0$, therefore we may set 
$t=1$ in the numerator. We set $k=\mu$ and define $\zeta=q-\mu$.
After carrying out the $q_4$-integration (assuming that $\phi^+$ is almost independent of $q_4$)
we perform the analytical continuation $k_4\to ik_0$ and take the on-shell limit, $k_0\to\phi^+$.
Then the $\zeta$-integral (with UV cutoff $\pm\mu$) gives terms proportional to $\log(k_0^2-(\phi^+)^2)$.
 Thus there may be indeed a mass shell singularity in $\delta X$, at least within our approximations. It is however only a logarithmic
singularity, which is harmless because on the right hand side of the gauge dependence identity we
have expressions like
 \begin{equation}
  (k_0^2-(\phi^+)^2)\left(\log(k_0^2-(\phi^+)^2)+\mathrm{finite}\right),
\end{equation}
which vanish for $k_0\rightarrow\phi$.
We conclude that 
no IR cutoff should be necessary to
assure gauge independence of the dispersion laws (at zero temperature).

An explicit calculation shows indeed that a simple IR cutoff for the gauge dependent part of the
gluon propagator in the gap equation is not sufficient to remove the gauge dependence of the
gap in covariant gauge. In Ref. \cite{Hou:2004bn} it has been argued that this gauge dependence will
only disappear after the inclusion of vertex corrections in the gap equation.

\section{Gluon tadpoles \label{secg}}
In a color superconductor global color symmetry is dynamically broken by the
diquark condensate. Therefore there is no symmetry
which forbids the existence of gluon tadpoles, cf. the discussion in Sec. \ref{sglob}.  This section
is devoted to a detailed analysis of the tadpole diagrams.

\subsection{Leading order tadpole diagram (2SC)}

\begin{figure}
 \begin{center}
  \includegraphics{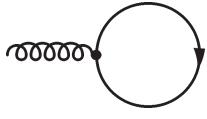}
  \vspace{-2mm}
  {\it\caption{Leading order tadpole diagram. \label{fig0}}}
 \end{center}
\end{figure}


Let us consider the one-loop tadpole diagram as shown in Fig. \ref{fig0},
\begin{equation}
  \mathcal{T}^a=-{g\over2}\int {d^4Q\over i(2\pi)^4}\mathrm{Tr}_{D,c,f,NG}[\hat\Gamma_0^a
   S(Q)]. \label{ac1}
\end{equation} 
Here $S(Q)$ is the quark propagator in the Nambu-Gor'kov basis, 
which is given by Eq. (\ref{sc1}). At leading order we may neglect the quark self energy $\Sigma$.
The quark-gluon vertex $\hat\Gamma_0^a$ is given by 
\begin{equation}
  \hat\Gamma_0^a=\left(\begin{array}{cc}\Gamma_0^a &0\\0&\bar\Gamma_0^a
  \end{array}\right),
\end{equation}
with $\Gamma_0^a=\gamma_0 T^a$ and $\bar\Gamma_0^a=-\gamma_0 (T^a)^T$.
The trace in (\ref{ac1}) has to be taken with respect to Dirac, color, flavor and 
Nambu-Gor'kov indices.
First we evaluate the trace over Nambu-Gor'kov space which gives
\begin{equation}  
  \mathcal{T}^a=-{g\over2}\int {d^4Q\over i(2\pi)^4}\mathrm{Tr}_{D,c,f}[\Gamma_0^a
  \cG ^+(Q)+\bar\Gamma_0^a\cG ^-(Q)],
\end{equation}
where $\cG^\pm$ is given in Eq. (\ref{g}) for the 2SC phase.
Evaluating the trace over flavor and color space 
we get
\begin{equation}  
  \mathcal{T}^a=g(T^a)_{33}\int {d^4Q\over i(2\pi)^4}\mathrm{Tr}_D
  [\gamma_0 (G^+-G^-- G_0^++G_0^-)].
\end{equation}
Assuming that $\phi^-\simeq0$ \cite{Pisarski:1999tv} and that $\phi^+$
has negligible four-momentum dependence in the vicinity of
the quasiparticle pole we obtain
\begin{eqnarray}
  \mathcal{T}^a&\!\simeq\!&-4g(T^a)_{33}\int {d^4Q\over i(2\pi)^4}\left({\mu-q\over q_0^2-(\mu-q)^2
  -|\phi^+|^2}-{\mu-q\over q_0^2-(\mu-q)^2}
  \right)\nonumber\\
  &\!\simeq\!&{g\over\pi^2}(T^a)_{33}\int_0^\infty dq\,q^2
  (\mu-q)\left({1\over\sqrt{(\mu-q)^2+|\phi^+|^2}}-{1\over|\mu-q|}\right).\qquad
\end{eqnarray}
In order to obtain an order of magnitude estimate, we make the approximation 
$\phi^+(q)\simeq\phi^+_0\Theta(2\mu-q)$ 
with $\phi^+_0=const.$ \cite{Rischke:2000qz}. 
Then the $q$-integration can be readily performed, with the result
\begin{equation}
  \mathcal{T}^a\simeq-{2g\over\pi^2}(T^a)_{33}\, \mu\,(\phi_0^+)^2
  \,\mathrm{log}\left({\phi^+_0\over2\mu}\right)
  +{\mathcal O}((g\mu\phi_0^+)^2),
\quad (T^a)_{33} = -\delta^{a8}/\sqrt3.
 \label{e27}
\end{equation}
This result is in fact of order $\mu(\phi_0^+)^2$, 
because $\mathrm{log}(\phi_0^+/(2\mu))$ is of order $1/g$ [see Eq. (\ref{gap})].

\subsection{Gluon tadpoles for CFL with $m_s\neq0$}
We would like to evaluate the gluon tadpole in the CFL phase with a non-vanishing strange
quark mass $m_s$.
The inverse Nambu-Gor'kov propagator takes the same form as in Eq. (\ref{sc1}) (with $\Sigma\to0$ at leading order),
where the free inverse propagators are now given by
\begin{eqnarray}
  &&([\cG^+_0]^{-1})^{rs}_{ij}=[(\Qs+\mu\gamma_0)(\delta^{rs}-\delta^{r3}\delta^{s3})+
  (\Qs+\mu\gamma_0-m_s)\delta^{r3}\delta^{s3}]\delta_{ij}, \\
  &&([\cG^-_0]^{-1})^{rs}_{ij}=[(\Qs-\mu\gamma_0)(\delta^{rs}-\delta^{r3}\delta^{s3})+
  (\Qs-\mu\gamma_0-m_s)\delta^{r3}\delta^{s3}]\delta_{ij},\qquad\qquad
\end{eqnarray}
Again the lower indices are color indices and the upper ones are flavor indices.
At finite $m_s$ the Dirac structure of the gap is more complicated than in the
massless case \cite{Pisarski:1999av}, but for $m_s\ll\mu$
we may assume that the additional terms are suppressed 
at least with $m_s/\mu$ \cite{Fugleberg:2002rk},
\begin{equation}
  [\Phi^+]^{rs}_{ij}\simeq\phi^+\Lambda^+_{\bf q}\gamma_5
  (\delta_i^r\delta_j^s-\delta_j^r\delta_i^s)+\mathcal{O}\left(\frac{m_s}{\mu}\right).
\end{equation} 
As usual we have $\Phi^-=\gamma_0[\Phi^+]^\dag\gamma_0$.
The inverse full quark propagator can be written as
\begin{equation}
  [\cG^+]^{-1}=[\cG^+_0]^{-1}-\Phi^-\cG^-_0\Phi^+
  =\left(\begin{array}{ccccccccc}
  b_1 & & & & b_2 & & & & b_3 \\
  & b_4 & & & & & & & \\
  & & b_5 & & & & & & \\
  & & & b_4 & & & & & \\
  b_2 & & & & b_1 & & & & b_3 \\
  & & & & & b_5 & & & \\
  & & & & & & b_6 & & \\
  & & & & & & & b_6 & \\
  b_3 & & & & b_3 & & & & b_7 \end{array} \right). \label{ms5}
\end{equation}
The nine rows and columns of this matrix correspond to color and flavor indices, namely
(color, flavor)=(1,1),(1,2),(1,3),(2,1),(2,2),(2,3),(3,1),(3,2), (3,3).
The $b_i$ are matrices in Dirac space, given explicitly by
\begin{eqnarray}
  b_1&=&\Qs+\mu\gamma_0+\gamma_0\phi^\dag\gamma_0[\Qs-\mu\gamma_0]^{-1}\phi
        +\gamma_0\phi^\dag\gamma_0[\Qs-\mu\gamma_0-m_s]^{-1}\phi,\nonumber\\
  b_2&=&\gamma_0\phi^\dag\gamma_0[\Qs-\mu\gamma_0-m_s]^{-1}\phi,\nonumber \\
  b_3&=&\gamma_0\phi^\dag\gamma_0[\Qs-\mu\gamma_0]^{-1}\phi,\nonumber \\
  b_4&=&\Qs+\mu\gamma_0+\gamma_0\phi^\dag\gamma_0[\Qs-\mu\gamma_0]^{-1}\phi,\nonumber \\
  b_5&=&\Qs+\mu\gamma_0-m_s+\gamma_0\phi^\dag\gamma_0[\Qs-\mu\gamma_0]^{-1}\phi,\nonumber \\
  b_6&=&\Qs+\mu\gamma_0+\gamma_0\phi^\dag\gamma_0[\Qs-\mu\gamma_0-m_s]^{-1}\phi,\nonumber\\
  b_7&=&\Qs+\mu\gamma_0-m_s+2\gamma_0\phi^\dag\gamma_0[\Qs-\mu\gamma_0]^{-1}\phi,
\end{eqnarray}
with $\phi=\phi^+\Lambda^+_{\bf q}\gamma_5$.
The full quark propagator is obtained by inverting the matrix in (\ref{ms5}),
\begin{equation}
  \!\!\!\!\!\!\!\cG^+=\left(\begin{array}{ccccccccc}
  a_1 & & & & a_2 & & & & a_3 \\
  & a_5 & & & & & & & \\
  & & a_6 & & & & & & \\
  & & & a_5 & & & & & \\
  a_2 & & & & a_1 & & & & a_3 \\
  & & & & & a_6 & & & \\
  & & & & & & a_7 & & \\
  & & & & & & & a_7 & \\
  a_4 & & & & a_4 & & & & a_8 \end{array} \right),\qquad
\end{equation}
with 
\begin{eqnarray}
  a_1&=&(b_1-b_2)^{-1}+a_2,\nonumber\\
  a_2&=&(-2b_3+b_7b_3^{-1}(b_1+b_2))^{-1}(b_3-b_7b_3^{-1}b_2)(b_1-b_2)^{-1},\nonumber\\
  a_3&=&(2b_3-b_7b_3^{-1}(b_1+b_2))^{-1},\nonumber\\
  a_4&=&-b_3^{-1}(b_2(b_1-b_2)^{-1}+(b_1+b_2)a_2),\nonumber\\
  a_5&=&b_4^{-1},\nonumber\\
  a_6&=&b_5^{-1},\nonumber\\
  a_7&=&b_6^{-1},\nonumber\\
  a_8&=&-b_3^{-1}(b_1+b_2)a_3.
\end{eqnarray}
Similar expressions can be obtained for $\cG^-:=[[\cG^-_0]^{-1}-\Phi^+\cG^+_0\Phi^-]^{-1}$.

The gluon tadpole is given by
\begin{equation}
  \mathcal{T}^a:=-{g\over2}\int {d^4Q\over i(2\pi)^4}
  \mathrm{Tr_D}\gamma_0\left([\cG^+]_{ij}^{rs}(T^a)_{ji}-[\cG^-]_{ij}^{rs}(T^a)_{ij}\right)
  \delta^{rs}.
\end{equation}
Taking the trace over color and flavor space we find that the only non-vanishing tadpole 
is $\mathcal{T}^8$,
\begin{equation}
 \mathcal{T}^8=-{g\over2\sqrt{3}}\int {d^4Q\over i(2\pi)^4}\mathrm{Tr_D}\gamma_0(a_1+a_5+a_6-2a_7-a_8)
  -(\textrm{similar terms from } G^-).
\end{equation}
We perform a Taylor expansion in $m_s$, keeping only the leading term which is of order $m_s^2$.
With the approximation $\phi^-\simeq0$ we get
\begin{eqnarray}
  &&\!\!\!\!\!\!\!\!\!\!\!\!\!\!\!\!\!\!\!\!\!\!
  \mathcal{T}^a\simeq\delta^{a8}
  \frac{gm_s^2}{2\sqrt{3}\pi^3}\int_{-\infty}^\infty dq_4\int_0^\infty dq\nonumber\\
  &&\times\frac{q\,(\phi^+)^2\left(q_4^2-3(q-\mu)^2-8(\phi^+)^2\right)}
  {\left(q_4^2+(q-\mu)^2+(\phi^+)^2\right)^2
  \left(q_4^2+(q-\mu)^2+4(\phi^+)^2\right)}.
\end{eqnarray}
This integral seems to be proportional to $(\phi^+)^2$. It turns out, however, that the factor $(\phi^+)^2$ 
cancels against a factor $1/(\phi^+)^2$ from the result of the integration. 
This is related to the fact that for $\phi^+\rightarrow0$ the integral
would contain an infrared singularity, which is regulated by the gap \cite{Kryjevski:2003cu}.

We make the approximation $\phi^+\simeq\phi^+_0\Theta(2\mu-q)$. 
After the integration we take the limit $\phi^+_0\rightarrow0$, 
since we assume $m_s\gg\phi^+_0$. This leads to \cite{Gerhold:2004ja}
\begin{equation}
  \mathcal{T}^8\simeq -\mu m_s^2{4g\over9\sqrt{3}\pi^2}(21-8\log2). \label{tc}
\end{equation}

\subsection{Color neutrality}

To address 
the question of color neutrality \cite{Alford:2002kj,Shovkovy:2003uu,Iida:2000ha,Amore:2001uf,Steiner:2002gx,Neumann:2002jm} we consider 
the partition function
\begin{equation}
  \exp(-\Omega/T)=\int\mathcal{D}\varphi\,\exp(-S[\varphi]),
\end{equation}
where $\varphi$ denotes the set of all fields, and $S[\varphi]$ is the QCD action (including gauge 
fixing terms and ghosts). Following the argument given 
in \cite{Khlebnikov:1996vj} it is easy to see that the system described by this 
partition function is color neutral, at least if one chooses a gauge fixing which
does not involve $A_0^a$, for instance Coulomb gauge:
The fields $A_0^a$ appear in the action as Lagrange multipliers for the Gauss law constraint
\cite{ItzZ:QFT}.
 Therefore in the path integral the integration over the zero-momentum modes $A_{0,{\bf p}=0}^a$  
produces delta functions, $\delta(N_a)$, where $N_a$ are the color charges. This means that
only color neutral field configurations contribute to the partition function, q.e.d.

The gluon tadpole in the 2SC phase [Eq. (\ref{e27})] induces a non-vanishing expectation value for the gluon field
$A_0^8$, 
\begin{equation}
  \bar A^0_{8}\sim m_D^{-2}\mu(\phi^+)^2\sim (\phi^+)^2/(g^2\mu), \qquad \mathrm{(2SC)}\label{A0}
\end{equation}
where $m_D$ is the corresponding leading-order Debye mass \cite{Rischke:2000qz}.
This expectation value acts as an effective chemical potential for the color number 8,
\begin{equation}
  \mu_8=g \bar A^0_8\sim (\phi^+)^2/(g\mu). \qquad \mathrm{(2SC)}
\end{equation}
It may be noted that the chemical potential $\mu_8$ which has been found by requiring
color neutrality in an NJL model is also proportional to $\phi^2$ \cite{Steiner:2002gx}. 
In a similar way one finds in the CFL phase from Eq. (\ref{tc})
\begin{equation}
  \mu_8\sim{m_s^2/\mu}, \qquad \mathrm{(CFL)}
\end{equation}
which agrees again qualitatively with the result of the NJL model calculation \cite{Steiner:2002gx}.
One should emphasize, however,
that whereas in NJL models color neutrality has to be imposed as an additional condition,
color neutrality is guaranteed automatically in QCD by the integration over the $A_0^a$ 
zero-momentum modes.      

The tadpole diagrams we computed above correspond to the first derivative of the effective action at \emph{vanishing}
mean field,
\begin{equation}
  \mathcal {T}^a\sim\frac{\delta\Gamma}{\delta A_0^a}\bigg|_{A=0}.
\end{equation}
In \cite{Dietrich:2003nu} the expectation value of $A_0^8$ is computed self-consistently from the Yang-Mills equation,
\begin{equation} 
  \frac{\delta\Gamma}{\delta A_0^a}\bigg|_{A=\bar A}=0.
\end{equation}
This approach is completely equivalent to our approach, which can be seen from the Taylor expansion
\begin{equation}
  \frac{\delta\Gamma}{\delta A_0^a}\bigg|_{A=0}=
  \frac{\delta\Gamma}{\delta A_0^a}\bigg|_{A=\bar A}-
  \frac{\delta^2\Gamma}{\delta A_0^a\delta A_0^b}\bigg|_{A
  =\bar A}\bar A_0^b
  +\ldots
\end{equation}
On the right hand side the first term vanishes, while the second term is essentially
the inverse gluon propagator at zero momentum times the expectation value of $A$.


The tadpole diagram in Fig. \ref{fig0} which we evaluated above corresponds to the fermionic part of
the color charge density,
\begin{equation}
  \rho^a_{(q)}=\sum_{f=1}^{N_f}\bar\psi_f T^a\gamma_0\psi_f.   \label{rf}
\end{equation}
One might ask whether the gluons could also contribute to the charge density. In 
Ref. \cite{Alford:2002kj} it was argued that the gluonic contribution should vanish, since the gluonic part of the
charge density
\begin{equation}
  \rho^a_{(gl)}=f^{abc}A^b_i F^c_{i0}
\end{equation}
contains the chromoelectric field strength, which vanishes in any (super-)con\-ducting system. The aim
of the next subsection  is to corroborate this argument by analyzing the tadpole
diagram with a gluon loop (Fig. \ref{fig1}) \cite{Gerhold:2004sk}.

\subsection{Gluon self energy and gluon loop tadpole (2SC)\label{sec1}}

\begin{figure}
 \begin{center}
  \includegraphics{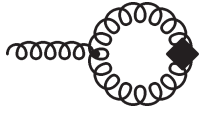}
  \vspace{-2mm}
  {\it\caption{Tadpole diagram with resummed gluon propagator. \label{fig1}}}
 \end{center}
\end{figure}

Let us consider the
tadpole diagram in Fig. \ref{fig1}, where the rhombus indicates a
resummed gluon propagator.
In \cite{Rischke:2000qz} integral representations for the 
various components of the gluon self energy in the 2SC phase have been derived. 
(The Feynman diagram for the gluon self energy at leading order looks like the second one of Fig. \ref{figgluon}, with the
normal fermion propagators now replaced by Nambu-Gor'kov propagators.)
It turns out the the gluon self energy in the 2SC phase has some
off-diagonal  components with respect to the color indices, therefore the tadpole diagram  in Fig. \ref{fig1} does not vanish 
\emph{a priori}.
The non-vanishing components of the gluon self energy in the 2SC phase are
\begin{equation}
  \Pi_{11}=\Pi_{22}=\Pi_{33}, \quad \Pi_{44}=\Pi_{55}=\Pi_{66}=\Pi_{77}, \quad
  \Pi_{45}=-\Pi_{54}=\Pi_{67}=-\Pi_{76}.
\end{equation}
In Ref. \cite{Rischke:2000qz} the gluon self energy is diagonalized by a unitary
transformation in color space. For our purposes it is more convenient not to perform this
transformation, since we want to keep the structure constants totally antisymmetric.
For the $44$ and $45$ components of the self energy analytical results can be obtained rather easily, as we
will demonstrate in the following. We shall always assume that the energy and the momentum of the gluon are
much smaller than $\mu$, because if there is any contribution to the tadpole diagrams, it will come from soft
gluons\footnote{For $p_0\gg\phi$ the gluon self energy is the same as in the normal phase. Since
it is then proportional to $\delta_{ab}$, it gives no contribution to the tadpole diagram. Furthermore the 
typical gluon momenta relevant for the tadpole diagram would be of the order of the Debye or Meissner masses
(or even smaller).}.
In the notation of Ref. \cite{Rischke:2000qz} we have
\begin{eqnarray}
  &&\!\!\!\!\!\!\!\!\!\!\!\!\!\!\!\!\!\!\!\!\!\!
  \Pi_{44}^{00}=-{1\over2}g^2\int {d^3k\over(2\pi)^3}\sum_{e_1,e_2=\pm}
  (1+e_1e_2\,{\bf\hat k}_1\cdot{\bf\hat k}_2)\nonumber\\
  &&\!\!\!\!\!\!\!\!\!\!\!\!
  \times\left(n_1^0(1-n_2)+(1-n_1^0)n_2\right) \left[{1\over p_0+\epsilon_1^0+\epsilon_2+i\eta}
  -{1\over p_0-\epsilon_1^0-\epsilon_2+i\eta}
  \right], \label{pi44}
\end{eqnarray}
and
\begin{eqnarray}
  &&\!\!\!\!\!\!\!\!\!\!\!\!\!\!\!\!\!\!\!\!\!\!
  -i\Pi_{45}^{00}\equiv\hat\Pi^{00}=-{1\over2}g^2\int {d^3k\over(2\pi)^3}\sum_{e_1,e_2=\pm}
  (1+e_1e_2\,{\bf\hat k}_1\cdot{\bf\hat k}_2)\nonumber\\
  &&\!\!\!\!\!\!\!\!\!\!\!\!
  \times \left(n_1^0(1-n_2)-(1-n_1^0)n_2\right)
  \left[{1\over p_0+\epsilon_1^0+\epsilon_2+i\eta}
  -{1\over p_0-\epsilon_1^0-\epsilon_2+i\eta}
  \right]. \label{pi45}
\end{eqnarray}
Here we have used the notations of Ref. \cite{Rischke:2000qz}, 
\begin{equation}
  \epsilon_i=\sqrt{(\mu-e_i k_i)^2+|\phi_i|^2},
\end{equation}
\begin{equation}
  n_i={\epsilon_i+\mu-e_ik_i\over 2\epsilon_i},
\end{equation}
where $\phi_i\equiv \phi^{e_i}(\epsilon_i, k_i)$ is the on-shell gap function. The superscript ``0'' means $\phi_i\to0$.
First let us evaluate $\imag \Pi_{44}^{00}$. Without loss of 
generality we assume $p_0>0$. 
For $p_0<\mu$ the only contribution arises from $e_1=e_2=+$. With
${\bf k}_1={\bf k}+{\bf p}$, ${\bf k}_2={\bf k}$, ${\bf\hat k}\cdot{\bf\hat p}=t$, $k_1=\sqrt{k^2+p^2+2pt}$,
$\xi=k-\mu$, $\phi:=\phi_2$ we find
\begin{eqnarray}
  &&\!\!\!\!\!\!\!\!\imag\Pi_{44}^{00}=-{g^2\over8\pi}\int_{-\mu}^\infty d\xi\,(\mu+\xi)^2
  \int_{-1}^1dt\,\Theta(p_0-\phi)\left(1+{\mu+\xi+pt\over k_1}\right)\nonumber\\
  &&\quad\times\Theta\!\left(p_0-\sqrt{\xi^2+\phi^2}\right)
  \bigg[{\sqrt{\xi^2+\phi^2}+\xi\over2\sqrt{\xi^2+\phi^2}}\,
  \delta\left(p_0-\mu+k_1-\sqrt{\xi^2+\phi^2}\right)\nonumber\\
  &&\qquad+{\sqrt{\xi^2+\phi^2}-\xi\over2\sqrt{\xi^2+\phi^2}}\,
  \delta\left(p_0+\mu-k_1-\sqrt{\xi^2+\phi^2}\right)\bigg].
\end{eqnarray}
Because of the step function in the second line 
we have $\xi\ll\mu$ for $p_0\ll\mu$. Therefore it is sufficient to
approximate $\phi$ with its value at the Fermi surface, which we denote with $\phi_0$. 
Rewriting the delta functions
as $\delta(t-\ldots)$, we find for $p\ll\mu$
\begin{eqnarray}
  &&\!\!\!\!\!\!\!\!\!\!\!\imag\Pi_{44}^{00}=-\Theta(p_0-\phi_0){g^2\over16\pi p}
  \int_{-\sqrt{p_0^2-\phi_0^2}}^{\sqrt{p_0^2-\phi_0^2}} d\xi\,(\mu+\xi)^2\int_{-1}^1dt\nonumber\\
  &&\!\!\!\!\!\!\!\!\!\!\times\bigg[{\left(\kappa+\xi\right)\left(2\mu-p_0+pt+\xi+\kappa\right)
  \over\kappa\left(\mu+\xi\right)}
  \Theta\!\left(p+p_0+\xi-\kappa\right)
  \Theta\!\left(p-p_0-\xi+\kappa\right)\delta(t-{\bar t}_1)\nonumber\\
  &&\!\!\!\!\!\!\!\!\!+{\left(\kappa-\xi\right)\left(2\mu+p_0+pt+\xi-\kappa\right)
  \over\kappa\left(\mu+\xi\right)}\Theta\!\left(p+p_0-\xi-\kappa\right)
  \Theta\!\left(p-p_0+\xi+\kappa\right)\delta(t-{\bar t}_2)\bigg]\nonumber\\
\end{eqnarray}
with
$
  \kappa=\sqrt{\xi^2+\phi_0^2}
$
and
\begin{eqnarray}
  &&{\bar t}_1={2\mu\left(-p_0-\xi+\kappa\right)
  +\phi_0^2-p^2+p_0^2-2p_0\kappa\over2p(\mu+\xi)},\\ 
  &&{\bar t}_2={2\mu\left(p_0-\xi-\kappa\right)
  +\phi_0^2-p^2+p_0^2-2p_0\kappa\over2p(\mu+\xi)}.
\end{eqnarray}
We make the substitution $\xi\to-\xi$ in the term with ${\bar t}_1$, which leads to
\begin{eqnarray}
  &&\!\!\!\!\!\!\!\!\imag\Pi_{44}^{00}=-\Theta(p_0-\phi_0){g^2\over16\pi p} 
  \int_{-\sqrt{p_0^2-\phi_0^2}}^{\sqrt{p_0^2-\phi_0^2}} d\xi\,\Theta\!\left(p+p_0-\xi-\kappa\right)
  \Theta\!\left(p-p_0+\xi+\kappa\right)\nonumber\\
  &&\times{1\over\kappa}\left[(4\mu^2-p^2+(p_0+2\xi)^2)(-\xi+\kappa)-\phi_0^2(2p_0+3\xi-\kappa)\right].
  \label{gls9}
\end{eqnarray} 
The $\xi$-integration is now straightforward, and we find
\begin{eqnarray}
  &&\!\!\!\!\!\!\!\!\!\!\!\!\!\!\!\!\!\!\!
  \imag\Pi_{44}^{00}=-\Theta(p_0-\phi_0)\Theta(-p_0+\sqrt{p^2+\phi_0^2})\nonumber\\
  &&\qquad\times{g^2\over24\pi p}\sqrt{p_0^2-\phi_0^2}\left(12\mu^2-3p^2+p_0^2-\phi_0^2\right) \nonumber\\
  &&\quad-\Theta(p_0-\sqrt{p^2+\phi_0^2}){g^2\mu^2\phi_0^2\over 2\pi(p_0^2-p^2)}\nonumber\\
  &&\qquad\times\left[1+{-6\phi_0^2p_0^2(p_0^2-p^2)
  +3(p_0^2-p^2)^3+\phi_0^4(p^2+3p_0^2)\over 12\mu^2(p_0^2-p^2)^2}\right].
\end{eqnarray}
This result is a generalization of the one-loop gluon self energy at zero temperature in the
normal phase given in \cite{Ipp:2003qt} (see also Sec. \ref{secgluon}). As in the normal phase, the leading part of the
gluon self energy is of the order $g^2\mu^2$ (for $p_0,p\ll\mu$).
In order to maintain consistency in the following, we shall keep only these leading terms,
since contributions from lower powers of $\mu$ might mix with contributions from diagrams
with a higher number of loops. Denoting this approximation with a tilde, we have
\begin{eqnarray}
  &&\imag\tilde\Pi_{44}^{00}=-\Theta(p_0-\phi_0){g^2\mu^2\over2\pi}\bigg[\Theta(-p_0+\sqrt{p^2+\phi_0^2})
  {\sqrt{p_0^2-\phi_0^2}\over p}\nonumber\\
  &&\quad\qquad\qquad+\Theta(p_0-\sqrt{p^2+\phi_0^2}){\phi_0^2\over p_0^2-p^2}\bigg].
\end{eqnarray}
In the following we will refer to this  analogon of the HDL approximation as the ``leading order'' approximation.
$\imag\Pi^{00}_{44}$ is an odd function of $p_0$, therefore the real part can be
calculated with the following dispersion relation\footnote{In principle it is not sufficient to take
only the part of Eq. (\ref{pi44}) where $e_1=e_2=+$ when computing the real part of the gluon self energy.  However,
as argued in \cite{Rischke:2001py},  for $p_0,p\ll\mu$ antiparticles (being always far from the ``Fermi surface'')
only give a constant term in the transverse gluon self energy, and this constant is the same in the
superconducting phase and in the normal phase (at leading order).} \cite{Rischke:2001py}
\begin{equation}
  \real\tilde\Pi_{44}^{00}(p_0,p)={1\over\pi}{\mathcal P}\int_0^\infty d\omega\,\imag\tilde\Pi_{44}^{00}(\omega,p)
  \left({1\over \omega+p_0}+{1\over\omega-p_0}\right).
\end{equation}
In contrast to the $11$, $22$, $33$ and $88$ components of the gluon self energy,
the principal value integral can be performed analytically in this case, 
\begin{eqnarray}
   &&\!\!\!\!\!\!\!\!\!\!\!\!\!\!\!\!\!\!\!\!
  \real\tilde\Pi_{44}^{00}= -{g^2\mu^2\over\pi^2}\Bigg[\Theta\left(\phi_0^2-p_0^2\right)
  \left(1-{\sqrt{\phi_0^2-p_0^2}\over p}
  \arctan{p\over\sqrt{\phi_0^2-p_0^2}}\right)\nonumber\\
  &&+\Theta\left(p_0^2-\phi_0^2\right)\left(1+{\sqrt{p_0^2-\phi_0^2}\over2 p}\log\bigg|{\sqrt{p_0^2-\phi_0^2}-p
  \over\sqrt{p_0^2-\phi_0^2}+p}\bigg|\right)\nonumber\\
  &&+{\phi_0^2\over 2(p^2-p_0^2)}\log\bigg|{\phi_0^2+p^2-p_0^2\over\phi_0^2}\bigg|\Bigg].
\end{eqnarray}
As a consistency check, we may extract the Debye mass in the normal and in the superconducting phase,
\begin{eqnarray}
  -\lim_{p\to0}\lim_{p_0\to0}\lim_{\phi_0\to0}\real\tilde\Pi_{44}^{00}(p_0,p)
  \!\!\!&=&\!\!\!{g^2\mu^2 \over\pi^2},\\
  -\lim_{p\to0}\lim_{p_0\to0}\real\tilde\Pi_{44}^{00}(p_0,p)
  \!\!\!&=&\!\!\!{g^2\mu^2 \over2\pi^2},
\end{eqnarray}
which are the standard results for $N_f=2$ \cite{Rischke:2000qz}. 
For the other Lorentz components of the self energy one 
finds following the same steps as above
\begin{eqnarray}
  &&\!\!\!\!\!\!\!\!
  \imag\tilde\Pi_{44}^{0i}=-\Theta(p_0-\phi_0){g^2\mu^2\over4\pi p^2}\hat{p}^i\bigg[\Theta(p-\upsilon)
  \left(2p_0\upsilon+\phi_0^2\log{p_0-\upsilon\over p_0+\upsilon}\right)\nonumber\\
  &&\ \qquad\qquad+\Theta(\upsilon-p)\,\phi_0^2\left({2pp_0\over p_0^2-p^2}+\log{p_0-p\over p_0+p}\right)
  \bigg],\\
  &&\!\!\!\!\!\!\!\!\imag\tilde\Pi_{44}^{ij}=\Theta(p_0-\phi_0){g^2\mu^2\over4\pi p^3}\nonumber\\
  &&\times\bigg[(\delta^{ij}-\hat{p}^i\hat{p}^j)
  \bigg(\Theta(p-\upsilon)\left(\upsilon(p_0^2-p^2+\phi_0^2)
  +\phi_0^2p_0\log{p_0-\upsilon\over p_0+\upsilon}\right)\nonumber\\
  &&\qquad+\Theta(\upsilon-p)\phi_0^2\left(2p+p_0\log{p_0-p\over p_0+p}\right)\bigg)\nonumber\\
  &&\quad-2\hat{p}^i\hat{p}^j \bigg(\Theta(p-\upsilon)\left(\upsilon(p_0^2+\phi_0^2)
  +\phi_0^2p_0\log{p_0-\upsilon\over p_0+\upsilon}\right)\nonumber\\
  &&\qquad+\Theta(\upsilon-p)\phi_0^2\left({(2p_0^2-p^2)p\over p_0^2-p^2}
  +p_0\log{p_0-p\over p_0+p}\right)\bigg)\bigg],
\end{eqnarray}
where  $\upsilon=\sqrt{p_0^2-\phi_0^2}$.
Again the real parts could be obtained analytically via dispersion relations, but the resulting
expressions are rather unwieldy and we shall refrain from writing them down here.

The off-diagonal components of the gluon self energy can be evaluated in an analogous way. In place of Eq. (\ref{gls9})
we find
\begin{eqnarray} 
  &&\!\!\!\!\!\!\!\!\imag\hat\Pi^{00}=-\Theta(p_0-\phi_0){g^2\over4\pi p} 
  \int_{-\sqrt{p_0^2-\phi_0^2}}^{\sqrt{p_0^2-\phi_0^2}} d\xi\,\Theta\!\left(p+p_0-\xi-\kappa\right)
  \Theta\!\left(p-p_0+\xi+\kappa\right)\nonumber\\
  &&\times
  {\mu\over\kappa}\left(\phi_0^2-(p_0+2\xi)(\kappa-\xi)\right),
\end{eqnarray}
which gives
\begin{equation}
  \imag\hat\Pi^{00}=-\Theta(p_0-\sqrt{p^2+\phi_0^2})g^2\mu {p_0\phi_0^2(\phi_0^2+p^2-p_0^2)
  \over 2\pi(p_0^2-p^2)^2}.
\end{equation}
We observe that in this one-loop result 
there is a term which is linear in $\mu$, but there is no term of order
$g^2\mu^2$. Therefore we have at our order of accuracy
\begin{equation}
  \imag\tilde{\hat\Pi}^{\raisebox{-5pt}{\scriptsize{00}}}=0.
\end{equation}
In fact this can be seen rather directly from the ($e_1=e_2=+$)-component of Eq. (\ref{pi45}): 
if we set $d^3k\to 2\pi\mu^2d\xi dt$ 
we find at the order $g^2\mu^2$ that the integrand is odd with respect to $\xi\to-\xi$, $t\to-t$, and therefore
the integral vanishes at this order. This argument also holds for the real part,  
$\real\tilde{\hat\Pi}^{\raisebox{-5pt}{\scriptsize{00}}}=0$.
In the same way it can be shown that $\tilde{\hat\Pi}^{\raisebox{-5pt}{\scriptsize{0i}}}=0$
and $\tilde{\hat\Pi}^{\raisebox{-5pt}{\scriptsize{ij}}}=0$.

This implies that the resummed gluon propagator is diagonal in the color indices at leading order. Since the
three-gluon vertex is antisymmetric in the color indices at tree level, 
we conclude that the tadpole diagram of Fig. (\ref{fig1}) vanishes at the order of our computation. This means that
at this order the expectation value of the gluon field, Eq. (\ref{A0}), is
not changed by the tadpole diagram in Fig. (\ref{fig1}).   
Therefore the effective chemical potential is determined completely by the
quark part of the color charge density at the order of our computation, whereas the gluonic part is negligible at this order.

We note that there is no contribution from the tadpole diagram with
a ghost loop. The reason is that there is no direct coupling between quarks and ghosts in the QCD Lagrangian, therefore the
ghost self energy vanishes at the order $g^2\mu^2$, and the ghost propagator is proportional to the unit matrix in color space at leading order.
Furthermore we remark that e.g. in covariant or Coulomb gauge in
particular the gauge dependent part of the gluon propagator is diagonal in the color indices at leading order.
Therefore the effective chemical potential is gauge independent at this order.

\subsection{Gluon vertex correction (2SC)\label{sec2}}

\begin{figure}
 \begin{center}
  \includegraphics{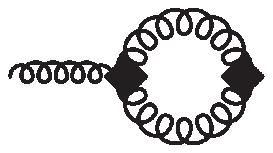} \qquad\quad
  \includegraphics{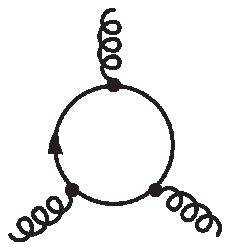}
  \vspace{-1mm}
  {\it\caption{Tadpole diagram with resummed gluon propagator and three-gluon vertex correction. \label{fig2}}}
 \end{center}
\end{figure}

We would like to check whether the above result is modified if one replaces the
tree level gluon vertex with the one-loop vertex correction, as shown in Fig. \ref{fig2}. 
We remark that the tadpole diagram in  Fig. \ref{fig2} corresponds to a higher order correction to
the \emph{fermionic} part of the charge density [Eq. (\ref{rf})].
Let us evaluate the one-loop three-gluon vertex of Fig. \ref{fig2} in the limit where one of the gluon 
momenta 
approaches zero. 
After taking the trace with respect to Nambu-Gor'kov, color and flavor indices we find
\begin{eqnarray}
  &&\!\!\!\!\!\!\!\!\!\!\!\!\!\!\!\!\!\!\!\!\!
  \Gamma_{485}^{\mu0\nu}=-{ig^3\over\sqrt{3}} \int {d^4K\over i(2\pi)^4}\Tr_D\bigg[
  -\gamma^\mu G^+(K_1) \gamma^0 G^+(K_1) \gamma^\nu G_0^+(K_2) \nonumber\\
  &&-2 \gamma^\mu G_0^+(K_1)\gamma^0 G_0^+(K_1) \gamma^\nu  G^+(K_2) 
  -\gamma^\mu G^-(K_1)\gamma^0 G^-(K_1) \gamma^\nu  G_0^-(K_2) \nonumber\\
  &&-2\gamma^\mu G_0^-(K_1)\gamma^0 G_0^-(K_1) \gamma^\nu  G^-(K_2) 
  +\gamma^\mu\Xi^+(K_1)\gamma^0 \Xi^-(K_1) \gamma^\nu  G_0^-(K_2)\nonumber\\ 
  &&+\gamma^\mu\Xi^-(K_1)\gamma^0 \Xi^+(K_1) \gamma^\nu  G_0^+(K_2) \bigg], \label{v1} \\
  &&\!\!\!\!\!\!\!\!\!\!\!\!\!\!\!\!\!\!\!\!\!
  \Gamma_{181}^{\mu0\nu}=-{2g^3\over\sqrt{3}} \int {d^4K\over i(2\pi)^4}\Tr_D\bigg[
  \gamma^\mu G^+(K_1)\gamma^0 G^+(K_1) \gamma^\nu  G^+(K_2)  \nonumber\\
  &&-\gamma^\mu G^-(K_1)\gamma^0 G^-(K_1) \gamma^\nu  G^-(K_2) 
  + \gamma^\mu\Xi^+(K_1)\gamma^0 \Xi^-(K_1) \gamma^\nu  G^-(K_2) \nonumber\\ 
  &&-\gamma^\mu \Xi^-(K_1)\gamma^0 \Xi^+(K_1) \gamma^\nu  G^+(K_2) 
  +\gamma^\mu \Xi^+(K_1)\gamma^0 G^+(K_1) \gamma^\nu  \Xi^-(K_2) \nonumber\\ 
  &&-\gamma^\mu \Xi^-(K_1)\gamma^0 G^-(K_1) \gamma^\nu  \Xi^+(K_2) 
  -\gamma^\mu G^-(K_1)\gamma^0 \Xi^+(K_1) \gamma^\nu  \Xi^-(K_2)  \nonumber\\
  &&+\gamma^\mu G^+(K_1)\gamma^0 \Xi^-(K_1) \gamma^\nu  \Xi^+(K_2) \bigg] ,\\
  &&\!\!\!\!\!\!\!\!\!\!\!\!\!\!\!\!\!\!\!\!\!
  \Gamma_{484}^{\mu0\nu}=-{g^3\over\sqrt{3}} \int {d^4K\over i(2\pi)^4}\Tr_D\bigg[
  \gamma^\mu G^+(K_1)\gamma^0 G^+(K_1) \gamma^\nu  G_0^+(K_2)  \nonumber\\
  && -2\gamma^\mu G_0^+(K_1)\gamma^0 G_0^+(K_1) \gamma^\nu  G^+(K_2) 
  -\gamma^\mu G^-(K_1)\gamma^0 G^-(K_1) \gamma^\nu  G_0^-(K_2)  \nonumber\\
  &&+2\gamma^\mu G_0^-(K_1)\gamma^0 G_0^-(K_1) \gamma^\nu  G^-(K_2) 
  +\gamma^\mu\Xi^+(K_1)\gamma^0 \Xi^-(K_1) \gamma^\nu  G_0^-(K_2)  \nonumber\\
  && -\gamma^\mu\Xi^-(K_1)\gamma^0 \Xi^+(K_1) \gamma^\nu  G_0^+(K_2)  \bigg],\\
  &&\!\!\!\!\!\!\!\!\!\!\!\!\!\!\!\!\!\!\!\!\!
   \Gamma_{888}^{\mu0\nu}=-{2g^3\over3\sqrt{3}} \int {d^4K\over i(2\pi)^4}\Tr_D\bigg[
  \gamma^\mu G^+(K_1)\gamma^0 G^+(K_1) \gamma^\nu  G^+(K_2)  \nonumber\\
  &&-4\gamma^\mu G_0^+(K_1)\gamma^0 G_0^+(K_1) \gamma^\nu  G_0^+(K_2) 
  -\gamma^\mu G^-(K_1)\gamma^0 G^-(K_1) \gamma^\nu  G^-(K_2)  \nonumber\\
  &&+4\gamma^\mu G_0^-(K_1)\gamma^0 G_0^-(K_1) \gamma^\nu  G_0^- (K_2) 
  +\gamma^\mu \Xi^+(K_1)\gamma^0 \Xi^-(K_1) \gamma^\nu  G^-(K_2)  \nonumber\\
  &&-\gamma^\mu \Xi^-(K_1)\gamma^0 \Xi^+(K_1) \gamma^\nu  G^+(K_2) 
  -\gamma^\mu \Xi^+(K_1)\gamma^0 G^+(K_1) \gamma^\nu  \Xi^-(K_2)  \nonumber\\
  &&+\gamma^\mu \Xi^-(K_1)\gamma^0 G^-(K_1) \gamma^\nu  \Xi^+(K_2) 
  +\gamma^\mu G^-(K_1)\gamma^0 \Xi^+(K_1) \gamma^\nu  \Xi^-(K_2)  \nonumber\\
  &&-\gamma^\mu G^+(K_1)\gamma^0 \Xi^- (K_1) \gamma^\nu \Xi^+(K_2)  \bigg], \label{v4}
\end{eqnarray}
where $K_1=K+P$ and $K_2=K$.
Furthermore one finds the relations
\begin{eqnarray}
  &&\!\!\!\!\!\!\!\!\!\!\!\!\!\!\!\!\!\!
  \Gamma_{485}=-\Gamma_{584}=\Gamma_{687}=-\Gamma_{786}, \quad\Gamma_{181}=\Gamma_{282}=\Gamma_{383},\nonumber\\
  &&\Gamma_{484}=\Gamma_{585}=\Gamma_{686}=\Gamma_{787}.
\end{eqnarray}
The propagators $G^\pm$ and $G^\pm_0$ in Eqs. (\ref{v1})-(\ref{v4}) are defined in
Eqs. (\ref{g1}) and (\ref{g0}), and $\Xi^\pm$ is given by \cite{Rischke:2000qz}
\begin{eqnarray}
  &&\Xi^+(K)=-\sum_{h=r,l}\sum_{e=\pm}{\phi^e_h(K)\over k_0^2-(\mu-ek)^2-|\phi_h^e|^2}\mathcal{P}_{-h}\Lambda^{-e}_{\bf k},\\
  &&\Xi^-(K)=-\sum_{h=r,l}\sum_{e=\pm}{[\phi^e_h(K)]^*\over k_0^2-(\mu-ek)^2-|\phi_h^e|^2}\mathcal{P}_{h}\Lambda^{e}_{\bf k}.
\end{eqnarray}
(In the following we will assume again $|\phi_r^e|=|\phi_l^e|=\phi^e$ \cite{Rischke:2000qz}.)
In order to evaluate  Eqs. (\ref{v1})-(\ref{v4}) we need the following Dirac traces,
\begin{eqnarray}
  T_{(1)}^{\mu\nu}&=&\Tr_D\left[\gamma^\mu\Lambda^{e_1}_{{\bf k}_1}\gamma^0\gamma^0\Lambda^{e_2}_{{\bf k}_1}
  \gamma^0\gamma^\nu\Lambda^{e_3}_{{\bf k}_2}\gamma^0\right],\\ 
  T_{(2)}^{\mu\nu}&=&\Tr_D\left[\gamma^\mu\mathcal{P}_{-h_1}\Lambda^{-e_1}_{{\bf k}_1}\gamma^0
  \mathcal{P}_{h_2}\Lambda^{e_2}_{{\bf k}_1}\gamma^\nu\Lambda^{-e_3}_{{\bf k}_2}\gamma^0\right],\\ 
  T_{(3)}^{\mu\nu}&=&\Tr_D\left[\gamma^\mu\mathcal{P}_{-h_1}\Lambda^{-e_1}_{{\bf k}_1}\gamma^0
  \Lambda^{e_2}_{{\bf k}_1}\gamma^0\gamma^\nu\mathcal{P}_{h_2}\Lambda^{e_3}_{{\bf k}_2}\right],\\ 
  T_{(4)}^{\mu\nu}&=&\Tr_D\left[\gamma^\mu\Lambda^{e_1}_{{\bf k}_1}\gamma^0\gamma^0
  \mathcal{P}_{h_1}\Lambda^{e_2}_{{\bf k}_1}\gamma^\nu\mathcal{P}_{-h_2}
  \Lambda^{-e_3}_{{\bf k}_2}\right]. 
\end{eqnarray}
One finds
\begin{eqnarray}
  &&T^{00}_{(1)}=\delta_{e_1e_2}\left(1+e_1e_3{\bf\hat k}_1\cdot{\bf\hat k}_2\right),\\
  &&T^{0i}_{(1)}=T^{io}_{(1)}=\delta_{e_1e_2}\left(e_1{\hat k}^i_1+e_3{\hat k}^i_2\right),\\
  &&T^{ij}_{(1)}=\delta_{e_1e_2}\left[\delta^{ij}\left(1-e_1e_3{\bf\hat k}_1\cdot{\bf\hat k}_2\right)+e_1e_3
  \left(\hat{k}^i_1\hat{k}^j_2+\hat{k}^i_2\hat{k}^j_1\right)\right].
\end{eqnarray}
$T_{(2)}$,$T_{(3)}$ and $T_{(4)}$ contain a common factor ${1\over2}\delta_{h_1h_2}$, which we do not write
in the following. Then we find
\begin{eqnarray}
  &&T^{00}_{(1)}= T^{00}_{(2)}= T^{00}_{(3)}= T^{00}_{(4)},\\
  &&T^{0i}_{(1)}= -T^{0i}_{(2)}= T^{0i}_{(3)}= -T^{0i}_{(4)},\\
  &&T^{i0}_{(1)}= T^{i0}_{(2)}= -T^{i0}_{(3)}= T^{i0}_{(4)},\\
  &&T^{ij}_{(1)}= T^{ij}_{(2)}=- T^{ij}_{(3)}=- T^{ij}_{(4)}.
\end{eqnarray}
After performing the energy integration
one finds for instance
\begin{eqnarray}
  &&\!\!\!\!\!\!\!\!\!
  \Gamma^{000}_{485}(P,0,-P)={ig^3\over2\sqrt{3}}\int {d^3k\over(2\pi)^3}\sum_{e_1,e_2=\pm}
  (1+e_1e_2\,{\bf\hat k}_1\cdot{\bf\hat k}_2)\nonumber\\ 
   &&\!\!\!\!\!\times\bigg[\bigg({1\over\left(p_0-\epsilon_1-\epsilon_2^0+i\eta\right)^2}
   +{1\over\left(p_0+\epsilon_1+\epsilon_2^0+i\eta\right)^2}\bigg)
  {2\phi_1^2+3\xi_1^2-3\xi_1\epsilon_1\mathrm{sgn}(\xi_2)
  \over\epsilon_1^2}\nonumber\\
  &&\ \ +\bigg({1\over p_0-\epsilon_1-\epsilon_2^0+i\eta}
   -{1\over p_0+\epsilon_1+\epsilon_2^0+i\eta}\bigg){\phi_1^2\over\epsilon_1^3}\bigg], \label{gls19}\\
  &&\!\!\!\!\!\!\!\!\!
  \Gamma^{000}_{484}(P,0,-P)= -{g^3\over2\sqrt{3}}\int {d^3k\over(2\pi)^3}\sum_{e_1,e_2=\pm}
  (1+e_1e_2\,{\bf\hat k}_1\cdot{\bf\hat k}_2)\nonumber\\
  &&\!\!\!\!\!\times\bigg[\bigg({1\over\left(p_0-\epsilon_1-\epsilon_2^0+i\eta\right)^2}
   +{1\over\left(p_0+\epsilon_1+\epsilon_2^0+i\eta\right)^2}\bigg)
  {-\mathrm{sgn}(\xi_2)(2\phi_1^2+3\xi_1^2)+3\xi_1\epsilon_1\over\epsilon_1^2}\nonumber\\
  &&\ \ -\bigg({1\over p_0-\epsilon_1-\epsilon_2^0+i\eta}
   -{1\over p_0+\epsilon_1+\epsilon_2^0+i\eta}\bigg){\mathrm{sgn}(\xi_2)\phi_1^2\over\epsilon_1^3}\bigg],
  \label{gls21}
\end{eqnarray}
where ${\bf k}_1={\bf k}+{\bf p}$, ${\bf k}_2={\bf k}$, and $\xi_i=e_ik_i-\mu$. 
The integrals can be evaluated in a similar way as in the previous section. 
It is easy to see that $\Gamma^{\mu0\nu}_{484}(P,0,-P)$ vanishes at the order $g^3\mu^2$.
In general (see also \cite{Casalbuoni:2001ha}) one finds that 
$\Gamma^{\mu0\nu}_{abc}(P,0,-P)$ is non-vanishing at the order $g^3\mu^2$ only for those combinations of color
indices where $f_{abc}$ is non-vanishing. In the previous section we have seen that the gluon 
propagator is diagonal in the color indices at leading order.
Therefore the tadpole diagram in Fig. \ref{fig2} also vanishes at this order.

\subsection{Mixing with Nambu-Goldstone bosons (2SC)\label{sec3}}

The symmetry breaking pattern of the 2SC phase is  $SU(3)_c\times U(1)_B\to SU(2)\times\tilde U(1)_B$ 
\cite{Alford:1997zt}.
Therefore five massless (would-be) Nambu-Goldstone (NG) bosons appear in this phase, which correspond to fluctuations of the
diquark condensate. 
As shown in \cite{Casalbuoni:2000cn,Miransky:2001sw,Rischke:2002rz}, one finds the following 
effective action for gluons, ghosts and NG bosons after integrating out the quarks, 
\begin{eqnarray}
  &&\!\!\!\!\!\!\!\!
  \Gamma=\int d^4x\,\mathcal{L}_\mathrm{(g)}+{1\over2}\Tr\log\left({\mathcal S}^{-1}+\gamma_\mu\Omega^\mu\right)
  \nonumber\\
  &&=\int d^4x\,\mathcal{L}_\mathrm{(g)}
  +{1\over2}\Tr\log{\mathcal S}^{-1}+{1\over2}\Tr\left({\mathcal S}\gamma_\mu\Omega^\mu\right)
  -{1\over4}\Tr\left({\mathcal S}\gamma_\mu\Omega^\mu{\mathcal S}\gamma_\nu\Omega^\nu\right)\nonumber\\
  &&\quad+{1\over6}\Tr\left({\mathcal S}\gamma_\mu\Omega^\mu{\mathcal S}\gamma_\nu\Omega^\nu
  {\mathcal S}\gamma_\rho\Omega^\rho\right)+\ldots, {\label{seff}}
\end{eqnarray}
where the gluonic part of the Lagrangian is e.g. in covariant gauge given by
\begin{equation}
  \mathcal{L}_\mathrm{(g)}=-{1\over4}F_{\mu\nu}^a F^{\mu\nu a}+{1\over2\lambda}(\partial^\mu A_\mu^a)^2
  -\bar c^a\partial^\mu D_\mu^{ab}c^b,
\end{equation}
where $F^{\mu\nu}_a$ is the usual field strength of the gluons, $c$ and $\bar c$ are the (anti-)ghost fields,
and $D_\mu^{ab}$ is the covariant derivative in the adjoint representation (see e.g. \cite{ItzZ:QFT}).
The quark propagator $S$ reads, in the notation of Ref.
\cite{Rischke:2002rz},
\begin{equation}
    S=\left(
    \begin{array}{cc} [G_0^+]^{-1}
    & \Phi^- \\
    \Phi^+ & [G_0^-]^{-1}
    \end{array}
  \right)^{-1}, 
\end{equation}
and $\Omega_\mu$ is given by
\begin{equation}
   \Omega^\mu(x,y)=-i\left(
    \begin{array}{cc} \omega^\mu(x)
    & 0 \\
    0 & -(\omega^\mu)^T(x)
    \end{array}
  \right)\delta^4(x-y).
\end{equation}
Here $\omega^\mu$ is the (Lie algebra valued) Maurer-Cartan one-form introduced in \cite{Casalbuoni:2000cn},
\begin{equation}
  \omega^\mu={\mathcal V}^\dag\left(i\partial^\mu+g A_a^\mu T_a\right){\mathcal V},
\end{equation}
where
\begin{equation}
  {\mathcal V}=\exp\left[i\left(\sum_{a=4}^7\varphi_aT_a+{1\over\sqrt{3}}\varphi_8 B\right)\right]
\end{equation}
parametrizes the coset space $SU(3)_c\times U(1)_B/SU(2)\times\tilde U(1)_B$ \cite{Miransky:2001sw}, 
$B=({\bf 1}+\sqrt{3}T_8)/3$ is a generator orthogonal to the one of $\tilde U(1)_B$, and
$\varphi_a$ are the NG bosons.
We may expand $\omega^\mu$ in powers of the fields,
\begin{eqnarray}
  &&\!\!\!\!\!\!\!\!\!\!
  \omega^\mu=-{1\over\sqrt{3}}\partial^\mu\varphi_8B+\bigg[g A^\mu_a-\partial^\mu\tilde\varphi_a
  -gf_{abc}A^\mu_b\tilde\varphi_c-{g\over3}f_{ab8}A^\mu_b\varphi_8
  -{1\over2}f_{abc}\tilde\varphi_b\partial^\mu\tilde\varphi_c\nonumber\\
  &&\qquad-{1\over6}f_{a8c}\varphi_8\partial^\mu\tilde\varphi_c
  -{1\over6}f_{ab8}\tilde\varphi_b\partial^\mu\varphi_8+\ldots\bigg]T_a
\end{eqnarray}
where $\tilde\varphi_a\equiv\varphi_a$ for $a=4,5,6,7$ and $\tilde\varphi_a\equiv0$ otherwise.

Let us examine the various terms in the effective action (\ref{seff}). First consider the term 
linear in $\Omega$, which contains the  one-loop gluon tadpole\footnote{The $\varphi$-tadpole vanishes because 
the external external momentum of the tadpole diagram is 
zero.} (Fig. \ref{fig0}).
In principle this term also gives contributions to higher $n$-point functions, since $\Omega$ contains arbitrarily high powers
of $\varphi$. In particular, there is a contribution to the 
bosonic self energy, but this effect is negligible at leading order since the one-loop tadpole [Eq. (\ref{e27})]
is only linear in $\mu$.

The term quadratic in $\Omega$  contains the bosonic self energy. As in section \ref{sec2} one finds that at leading order
the self energy is diagonal in the color indices. There are mixed terms between gluons and NG bosons, which
could be eliminated by choosing a suitable t'Hooft gauge \cite{Rischke:2002rz}. This is not necessary for our purposes,
but we note that choosing this t'Hooft gauge would not alter our conclusion, since also in the t'Hooft gauge of Ref. \cite{Rischke:2002rz}
the gluon, NG and ghost propagators are diagonal in the color indices at leading order. The gluon and NG propagators in the 
t'Hooft gauge are given explicitly in Eqs. (54) and (55) of Ref. \cite{Rischke:2002rz}. At leading order the propagators
(and in particular the gauge dependent parts of the propagators)
are diagonal in the color indices also without the unitary transformation in color space that is employed in Ref. \cite{Rischke:2002rz}.

The term quadratic in $\Omega$ also contains  
new types of three-boson vertices  proportional to $f_{abc}$, which involve at least one NG boson.

The term cubic in $\Omega$ is only non-vanishing at the order $g^3\mu^2$ 
for those color indices where $f_{abc}\neq0$. In this thesis
we do not consider vertices with more than three legs, since these would appear only in two-loop 
tadpole diagrams.

To summarize, we find at leading order as in the previous subsections that the bosonic propagators are
diagonal in the color indices, and that three-boson vertices of the type  $V^{\mu0\nu}_{abc}(P,0,-P)$ are non-zero 
only for those combinations of 
color indices where $f_{abc}\neq0$. We conclude that the tadpole diagram in Fig.~(\ref{fig2}) vanishes
at the order of our computation, even if we take into account NG bosons.

\subsection{Tadpoles with bosonic loops in the CFL phase \label{sec4}}
In the case $m_s=0$ the situation
for the diagrams in Figs. (\ref{fig1}), (\ref{fig2}) is similar to the 2SC phase.
In the CFL phase the one-loop gluon self energy is proportional to the unit matrix in color space \cite{Rischke:2000ra}
(as in the normal phase). Therefore the
diagram in Fig. \ref{fig1} vanishes identically. 
For the gluon vertex correction one finds after taking the trace with respect to Nambu-Gor'kov, color and
flavor indices 
\begin{eqnarray}
  &&\!\!\!\!\!\!\!\!\!\!\!\!\!\!\!
  \Gamma^{\mu0\nu}_{abc}(P,0,-P)=-{g^3\over24}\int {d^4K\over i(2\pi)^4}\Tr_D \bigg\{if_{abc}\bigg[\gamma^\mu G_1^+(K_1)\gamma^0
  G_2^+(K_1)\gamma^\nu G_2^+(K_2)\nonumber\\
  &&+\gamma^\mu G_1^-(K_1)\gamma^0G_2^-(K_1)\gamma^\nu G_2^-(K_2) 
  +\gamma^\mu G_2^+(K_1)\gamma^0 G_1^+(K_1)\gamma^\nu G_2^+(K_2)\nonumber\\
  &&+\gamma^\mu G_2^-(K_1)\gamma^0G_1^-(K_1)\gamma^\nu G_2^-(K_2) 
  +\gamma^\mu G_2^+(K_1)\gamma^0G_2^+(K_1)\gamma^\nu G_1^+(K_2)\nonumber\\ 
  &&+\gamma^\mu G_2^-(K_1)\gamma^0G_2^-(K_1)\gamma^\nu G_1^-(K_2) 
  +6\gamma^\mu G_2^+(K_1)\gamma^0G_2^+(K_1)\gamma^\nu G_2^+(K_2)\nonumber\\ 
  &&+6\gamma^\mu G_2^-(K_1)\gamma^0G_2^-(K_1)\gamma^\nu G_2^-(K_2)
  +\gamma^\mu \Xi_2^+(K_1)\gamma^0\Xi_2^-(K_1)\gamma^\nu G_1^-(K_2)\nonumber\\
  &&+\gamma^\mu \Xi_2^-(K_1)\gamma^0\Xi_2^+(K_1)\gamma^\nu G_1^+(K_2)
  -\gamma^\mu \Xi_2^+(K_1)\gamma^0\Xi_2^-(K_1)\gamma^\nu G_2^-(K_2)\nonumber\\
  &&-\gamma^\mu \Xi_2^-(K_1)\gamma^0\Xi_2^+(K_1)\gamma^\nu G_2^+(K_2)
  +\gamma^\mu \Xi_1^-(K_1)\gamma^0\Xi_2^+(K_1)\gamma^\nu G_2^+(K_2)\nonumber\\
  &&+\gamma^\mu \Xi_1^+(K_1)\gamma^0\Xi_2^-(K_1)\gamma^\nu G_2^-(K_2)
  +\gamma^\mu \Xi_2^-(K_1)\gamma^0\Xi_1^+(K_1)\gamma^\nu G_2^+(K_2)\nonumber\\
  &&+\gamma^\mu \Xi_2^+(K_1)\gamma^0\Xi_1^-(K_1)\gamma^\nu G_2^-(K_2)
  +2\gamma^\mu \Xi_2^-(K_1)\gamma^0\Xi_2^+(K_1)\gamma^\nu G_2^+(K_2)\nonumber\\
  &&+2\gamma^\mu \Xi_2^+(K_1)\gamma^0\Xi_2^-(K_1)\gamma^\nu G_2^-(K_2) 
  +\gamma^\mu \Xi_1^+(K_1)\gamma^0G_2^+(K_1)\gamma^\nu \Xi_2^-(K_2)\nonumber\\
  &&+\gamma^\mu \Xi_1^-(K_1)\gamma^0G_2^-(K_1)\gamma^\nu \Xi_2^+(K_2)
  +\gamma^\mu \Xi_2^+(K_1)\gamma^0G_2^+(K_1)\gamma^\nu \Xi_1^-(K_2)\nonumber\\
  &&+\gamma^\mu \Xi_2^-(K_1)\gamma^0G_2^-(K_1)\gamma^\nu \Xi_1^+(K_2)
  +2\gamma^\mu \Xi_2^+(K_1)\gamma^0G_2^+(K_1)\gamma^\nu \Xi_2^-(K_2)\nonumber\\
  &&+2\gamma^\mu \Xi_2^-(K_1)\gamma^0G_2^-(K_1)\gamma^\nu \Xi_2^+(K_2) 
  +\gamma^\mu \Xi_2^-(K_1)\gamma^0G_1^-(K_1)\gamma^\nu \Xi_2^+(K_2)\nonumber\\
  &&+\gamma^\mu \Xi_2^+(K_1)\gamma^0G_1^+(K_1)\gamma^\nu \Xi_2^-(K_2)
  -\gamma^\mu \Xi_2^-(K_1)\gamma^0G_2^-(K_1)\gamma^\nu \Xi_2^+(K_2)\nonumber\\
  &&-\gamma^\mu \Xi_2^+(K_1)\gamma^0G_2^+(K_1)\gamma^\nu \Xi_2^-(K_2)
  +\gamma^\mu G_1^-(K_1)\gamma^0\Xi_2^+(K_1)\gamma^\nu \Xi_2^-(K_2)\nonumber\\
  &&+\gamma^\mu G_1^+(K_1)\gamma^0\Xi_2^-(K_1)\gamma^\nu \Xi_2^+(K_2)
  -\gamma^\mu G_2^-(K_1)\gamma^0\Xi_2^+(K_1)\gamma^\nu \Xi_2^-(K_2)\nonumber\\
  &&-\gamma^\mu G_2^+(K_1)\gamma^0\Xi_2^-(K_1)\gamma^\nu \Xi_2^+(K_2) 
  +\gamma^\mu G_2^+(K_1)\gamma^0\Xi_1^-(K_1)\gamma^\nu \Xi_2^+(K_2)\nonumber\\
  &&+\gamma^\mu G_2^-(K_1)\gamma^0\Xi_1^+(K_1)\gamma^\nu \Xi_2^-(K_2)
  +\gamma^\mu G_2^+(K_1)\gamma^0\Xi_2^-(K_1)\gamma^\nu \Xi_1^+(K_2)\nonumber\\
  &&+\gamma^\mu G_2^-(K_1)\gamma^0\Xi_2^+(K_1)\gamma^\nu \Xi_1^-(K_2) 
  +2\gamma^\mu G_2^+(K_1)\gamma^0\Xi_2^-(K_1)\gamma^\nu \Xi_2^+(K_2)\nonumber\\
  &&+2\gamma^\mu G_2^-(K_1)\gamma^0\Xi_2^+(K_1)\gamma^\nu \Xi_2^-(K_2) \bigg]\nonumber\\
  &&\!\!\!\!\!\!\!\!\!\!\!\!\!+d_{abc}\bigg[\gamma^\mu G_1^+(K_1)\gamma^0G_2^+(K_1)\gamma^\nu G_2^+(K_2)\nonumber\\
  &&-\gamma^\mu G_1^-(K_1)\gamma^0G_2^-(K_1)\gamma^\nu G_2^-(K_2) 
  +\gamma^\mu G_2^+(K_1)\gamma^0 G_1^+(K_1)\gamma^\nu G_2^+(K_2)\nonumber\\
  &&-\gamma^\mu G_2^-(K_1)\gamma^0G_1^-(K_1)\gamma^\nu G_2^-(K_2) 
  +\gamma^\mu G_2^+(K_1)\gamma^0G_2^+(K_1)\gamma^\nu G_1^+(K_2)\nonumber\\ 
  &&-\gamma^\mu G_2^-(K_1)\gamma^0G_2^-(K_1)\gamma^\nu G_1^-(K_2) 
  +6\gamma^\mu G_2^+(K_1)\gamma^0G_2^+(K_1)\gamma^\nu G_2^+(K_2)\nonumber\\ 
  &&-6\gamma^\mu G_2^-(K_1)\gamma^0G_2^-(K_1)\gamma^\nu G_2^-(K_2)
  +\gamma^\mu \Xi_2^+(K_1)\gamma^0\Xi_2^-(K_1)\gamma^\nu G_1^-(K_2)\nonumber\\
  &&-\gamma^\mu \Xi_2^-(K_1)\gamma^0\Xi_2^+(K_1)\gamma^\nu G_1^+(K_2)
  -\gamma^\mu \Xi_2^+(K_1)\gamma^0\Xi_2^-(K_1)\gamma^\nu G_2^-(K_2)\nonumber\\
  &&+\gamma^\mu \Xi_2^-(K_1)\gamma^0\Xi_2^+(K_1)\gamma^\nu G_2^+(K_2)
  +\gamma^\mu \Xi_1^-(K_1)\gamma^0\Xi_2^+(K_1)\gamma^\nu G_2^+(K_2)\nonumber\\
  &&-\gamma^\mu \Xi_1^+(K_1)\gamma^0\Xi_2^-(K_1)\gamma^\nu G_2^-(K_2)
  +\gamma^\mu \Xi_2^-(K_1)\gamma^0\Xi_1^+(K_1)\gamma^\nu G_2^+(K_2)\nonumber\\
  &&-\gamma^\mu \Xi_2^+(K_1)\gamma^0\Xi_1^-(K_1)\gamma^\nu G_2^-(K_2)
  +2\gamma^\mu \Xi_2^-(K_1)\gamma^0\Xi_2^+(K_1)\gamma^\nu G_2^+(K_2)\nonumber\\
  &&-2\gamma^\mu \Xi_2^+(K_1)\gamma^0\Xi_2^-(K_1)\gamma^\nu G_2^-(K_2) 
  +\gamma^\mu \Xi_1^+(K_1)\gamma^0G_2^+(K_1)\gamma^\nu \Xi_2^-(K_2)\nonumber\\
  &&-\gamma^\mu \Xi_1^-(K_1)\gamma^0G_2^-(K_1)\gamma^\nu \Xi_2^+(K_2)
  +\gamma^\mu \Xi_2^+(K_1)\gamma^0G_2^+(K_1)\gamma^\nu \Xi_1^-(K_2)\nonumber\\
  &&-\gamma^\mu \Xi_2^-(K_1)\gamma^0G_2^-(K_1)\gamma^\nu \Xi_1^+(K_2)
  +2\gamma^\mu \Xi_2^+(K_1)\gamma^0G_2^+(K_1)\gamma^\nu \Xi_2^-(K_2)\nonumber\\
  &&-2\gamma^\mu \Xi_2^-(K_1)\gamma^0G_2^-(K_1)\gamma^\nu \Xi_2^+(K_2) 
  +\gamma^\mu \Xi_2^-(K_1)\gamma^0G_1^-(K_1)\gamma^\nu \Xi_2^+(K_2)\nonumber\\
  &&-\gamma^\mu \Xi_2^+(K_1)\gamma^0G_1^+(K_1)\gamma^\nu \Xi_2^-(K_2)
  -\gamma^\mu \Xi_2^-(K_1)\gamma^0G_2^-(K_1)\gamma^\nu \Xi_2^+(K_2)\nonumber\\
  &&+\gamma^\mu \Xi_2^+(K_1)\gamma^0G_2^+(K_1)\gamma^\nu \Xi_2^-(K_2)
  +\gamma^\mu G_1^-(K_1)\gamma^0\Xi_2^+(K_1)\gamma^\nu \Xi_2^-(K_2)\nonumber\\
  &&-\gamma^\mu G_1^+(K_1)\gamma^0\Xi_2^-(K_1)\gamma^\nu \Xi_2^+(K_2)
  -\gamma^\mu G_2^-(K_1)\gamma^0\Xi_2^+(K_1)\gamma^\nu \Xi_2^-(K_2)\nonumber\\
  &&+\gamma^\mu G_2^+(K_1)\gamma^0\Xi_2^-(K_1)\gamma^\nu \Xi_2^+(K_2) 
  +\gamma^\mu G_2^+(K_1)\gamma^0\Xi_1^-(K_1)\gamma^\nu \Xi_2^+(K_2)\nonumber\\
  &&-\gamma^\mu G_2^-(K_1)\gamma^0\Xi_1^+(K_1)\gamma^\nu \Xi_2^-(K_2)
  +\gamma^\mu G_2^+(K_1)\gamma^0\Xi_2^-(K_1)\gamma^\nu \Xi_1^+(K_2)\nonumber\\
  &&-\gamma^\mu G_2^-(K_1)\gamma^0\Xi_2^+(K_1)\gamma^\nu \Xi_1^-(K_2) 
  +2\gamma^\mu G_2^+(K_1)\gamma^0\Xi_2^-(K_1)\gamma^\nu \Xi_2^+(K_2)\nonumber\\
  &&-2\gamma^\mu G_2^-(K_1)\gamma^0\Xi_2^+(K_1)\gamma^\nu \Xi_2^-(K_2) \bigg]\bigg\},
\end{eqnarray}
where $K_1=K+P$, $K_2=K$, and \cite{Rischke:2000ra}
\begin{eqnarray} 
  G^\pm_n(K)&=&\sum_{e=\pm}{k_0\mp(\mu-ek)\over k_0^2-(\mu-ek)^2-\phi^e_{(n)}(K)^2}\Lambda^{\pm e}
  ({\bf k})\gamma_0,\\
  \Xi^\pm_n(K)&=&\mp\sum_{e=\pm}{\phi^e_{(n)}(K)\over k_0^2-(\mu-ek)^2-\phi^e_{(n)}(K)^2}\Lambda^{\mp e}({\bf k})\gamma_5.
\end{eqnarray}
(As in the 2SC case we have assumed that right-handed and left-handed gap functions are real and equal up to
a minus sign \cite{Rischke:2000ra}.)
Here $\phi_{(1)}$ is the singlet gap
and $\phi_{(2)}(={1\over2}\phi_{(1)})$ is the octet gap \cite{Rischke:2000ra,Zarembo:2000pj}.
It is straightforward to show that in
the three-gluon vertex correction $\Gamma_{abc}^{\mu0\nu}(P,0,-P)$
only the term proportional to $f_{abc}$ is non-vanishing at the order $g^3\mu^2$. 

The structure of effective action for gluons and NG bosons is similar to the 2SC phase 
\cite{Son:1999cm, Casalbuoni:1999wu,  Manuel:2000wm,Zarembo:2000pj}.
Thus we find that all the three-boson vertices are proportional to $f_{abc}$ at leading order, and all the boson propagators are 
proportional to $\delta_{ab}$. Therefore the tadpole diagram in Fig. \ref{fig2} vanishes also in the CFL phase.

\section{Some remarks on the specific heat (outlook)}

We would like to conclude this chapter with some remarks on the specific heat of color superconducting phases.
First let us consider the contribution of a fermionic mode to the specific heat.
We assume that the imaginary part of the fermion self energy is negligible.
In a similar way as in Eq. (\ref{x2}) one finds then the following contribution to the entropy density,
\begin{equation}
  \cS_\mathrm{ferm.}=-2\int{d^3q\over(2\pi)^3}\int_{-\infty}^\infty{dq_0\over2\pi}{\partial n_f(q_0)\over\partial T}
  \imag\log S(q_0+i\eta,q)^{-1}, \label{spc1}
\end{equation}
where $S$ is the propagator of the corresponding fermionic mode\footnote{Compared to the notation of chapter \ref{cspecific}, 
we have shifted the energy variable by $\mu$, such that now the propagator instead of the distribution
function depends on $\mu$. This notation is much more convenient in the context of superconductivity.}.
 Since we neglect the imaginary part of the fermion
self energy, the imaginary part of the logarithm in Eq. (\ref{spc1}) gives essentially a step function 
$\Theta(q_0-\omega(q))$, where $\omega(q)$ is the solution of the dispersion relation $S^{-1}(\omega(q),q)=0$.
For the specific heat we find then [see Eq. (\ref{x1})]
\begin{eqnarray}
  &&\!\!\!\!\!\!\!\!\!\!\!\!\!\!
  \cC_{v,\mathrm{ferm.}}\simeq T\left(\partial \cS\over\partial T\right)_\mu
  =T\int{d^3q\over(2\pi)^3}\int_{\omega(q)}^\infty dq_0\, {\partial^2 n_f(q_0)\over\partial T^2}\nonumber\\
  &&\qquad=\int{d^3q\over(2\pi)^3}\omega(q){\partial n_f\left(\omega(q)\right)\over\partial T}\equiv
  \int{d^3q\over(2\pi)^3}\epsilon(q){\partial n_f\left(\epsilon(q)\right)\over\partial T},
\end{eqnarray}
with $\epsilon(q)=|\omega(q)|$.
For a gapped fermionic mode we have $\epsilon(q)=\!\sqrt{(q-\mu)^2+\phi^2}$. Then one finds that the corresponding
contribution to the specific heat is exponentially suppressed, $\propto\exp(-\phi/T)$. For an ungapped mode one
finds the usual Fermi liquid result, $\cC_v\propto\mu^2T$, provided that $\omega(q)$ is analytic in $q-\mu$.

For a bosonic mode one finds in a similar way (again neglecting the imaginary part of the bosonic self energy)
\begin{equation}
   \cC_{v,\mathrm{bos.}}\simeq \int{d^3q\over(2\pi)^3}\tilde\omega(q){\partial n_b\left(\tilde\omega(q)\right)\over\partial T},
\end{equation}
where $\tilde\omega(q)$ is the solution of the respective bosonic dispersion law. For massless bosons one finds $\cC_{v}\propto T^3$, 
whereas for massive bosons ($m\gg T$) the contribution is exponentially suppressed, $\propto\exp(-m/T)$.

In the CFL phase all fermionic quasiparticles are gapped. Therefore the specific heat is is at low temperature dominated by
light bosonic excitations \cite{Jaikumar:2002vg},
leading to a specific heat proportional to $T^3$. Other color superconducting phases
discussed in the literature  have also ungapped fermionic quasiparticles, leading to a
specific heat proportional to $\mu^2T$ \cite{Schafer:2004jp}. E.g., in the 2SC phase the quarks of color $3$ are ungapped. 
Since they have only couplings with massive gluons \cite{Rischke:2000qz}, one expects that there will be no 
anomalous $\mu^2T\log T$ contribution to the specific heat \cite{Schafer:2004jp}.
 Also for the LOFF phase the specific heat is linear in $T$, at least in the NJL model approach \cite{Casalbuoni:2003sa}.
Only for the gapless CFL phase  the specific heat 
is of the order $\mu^2\sqrt{\phi T}$ as a consequence of an (almost) quadratic dispersion relation of one 
of the fermionic quasiparticles \cite{Alford:2004zr}.

These results for the specific heat will receive corrections from the energy-momentum dependence and from the
non-vanishing imaginary parts of the self energies and the gap. These issues certainly deserve further studies,
since the specific heat is of central importance for the cooling behavior of compact stars.

\chapter{Conclusions and outlook\label{cconc}}

In this thesis we have investigated properties of cold dense quark matter.
The initial motivation was the fact that in the core of (some) neutron stars the density may be high enough for
deconfinement of quarks at comparatively small temperatures.

In chapters \ref{csigma} and \ref{cspecific} we have computed the quark self energy and the specific heat in ultradegenerate QCD.
At high density and zero temperature the chromomagnetic screening mass vanishes, and
quasistatic chromomagnetic fields are only dynamically screened. These long-range interactions lead to a non-Fermi-liquid behavior 
at high density and small temperature. In the quark self energy the leading result at zero temperature is proportional to
$\gf^2(E-\mu)\log(\gf\mu/(E-\mu))$ [see Eq. (\ref{sigma0})]. Correspondingly, the leading term in the interaction part of 
the specific heat is of the order $\gf^2\mu^2T\log(\gf\mu/T)$ [see Eq. (\ref{cv1})]. 
We have corrected an error in a recent paper \cite{Boyanovsky:2000zj}, in which it was claimed that the leading term 
in the interaction part of the specific heat should instead be of the order $g^2T^3\log(\gf\mu/T)$. 
Furthermore we have performed a perturbative expansion of the quark self energy and the specific heat in powers of $T/(\gf\mu)$. 
These expansions contain fractional powers which come from the dynamical screening scale $q\sim (\gf^2\mu^2q_0)^{1/3}$. 
An important application of the results for the specific heat and the quark self energy is neutrino emission from
ungapped quark matter in neutron stars \cite{Schafer:2004jp}.

Chapter \ref{ccsc} of the thesis has been devoted to color superconductivity. Using a general gauge dependence identity we 
have given a formal proof of gauge independence for the fermionic quasiparticle dispersion relations in 
a color superconductor. As long as the gauge dependence of the quark self energy can be neglected,
this implies gauge independence of the gap function on the quasiparticle mass shell. 
The application of gauge dependence identities
for gauge theories with spontaneous symmetry breaking at finite temperature has also been demonstrated in chapter \ref{csym}
for the Abelian and a non-Abelian Higgs model.
We found furthermore that the spontaneous breaking of global color symmetry induces a non-vanishing expectation
value for the gluon field in a color superconductor. This expectation value acts as an effective chemical potential
for the color charge. As shown explicitly in \cite{Dietrich:2003nu} this mechanism ensures color neutrality of
the color superconducting system.
The expectation value of the gluon field can be computed from the gluon tadpole diagram.
We have computed the leading order tadpole diagrams, both for the 2SC phase and for the CFL phase, where we have also included
a small, but non-vanishing strange quark mass in the latter case. We have also shown that the expectation value
of the gluon field is at leading order not modified by one-loop tadpole diagrams with resummed gluon or Nambu-Goldstone
boson propagators.

There are several natural continuations of the present work, some of which I would like to sketch briefly.
As mentioned already in Sec. \ref{sneutrino}
one could use the results of chapters \ref{csigma} and  \ref{cspecific} to compute the cooling behavior
of a neutron star with a normal quark matter component beyond the leading logarithmic accuracy of Ref. \cite{Schafer:2004jp}.
Still a lot of work has to be done in the field of color superconductivity.
We have computed the gluon tadpole diagrams only for the 2SC and CFL phases.
A straightforward exercise would be the computation of gluon tadpoles for other color superconducting phases, and
at finite temperature. It would certainly be more challenging to compute higher order corrections to the tadpole diagram.
Probably such a computation will only be possible after higher order corrections to the gap have been determined
consistently. Another important subject is the specific heat of color superconducting phases, which we briefly
discussed at the end of chapter \ref{ccsc}. Also the inhomogeneous color superconducting (LOFF) phases deserve further
studies. In particular it would be interesting to compute 
the neutrino emissivity of neutron stars which contain quark matter in the LOFF phase  \cite{Schafer:2004jp}.

It is fair to say that at present our understanding of cold dense quark matter is still very incomplete. 
As discussed in the Introduction, the major obstacle is the fact that lattice simulations at large chemical potential are not 
possible so far.  In view of this situation the (semi-)perturbative
methods which we have used in this thesis provide an extremely valuable tool for an (at least) approximate description of
the properties of cold dense quark matter.

\appendix

\chapter{Matsubara sums \label{appa}}

\begin{figure}
  \begin{center}
    \includegraphics[width=8cm]{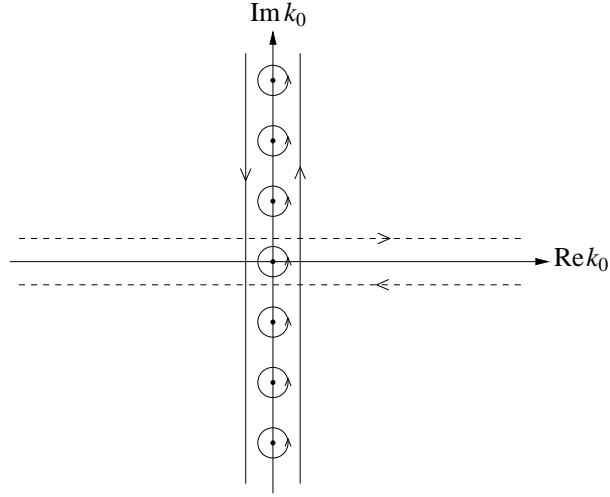}
  \end{center}
  \vspace{-0.5cm}
  {\it \caption{Integration contours for bosonic Matsubara sums. \label{figB1}}}
\end{figure}

In this appendix we shall briefly review some standard techniques for the
evaluation of Matsubara sums.

For the computation of Feynman diagrams at finite temperature
in the imaginary time formalism \cite{LeB:TFT,Kapusta:TFT} we need the following Matsubara sum in the bosonic case,
\begin{equation}
  J_b:=T\sum_{n\in\mathbbm Z} f( k_0=2\pi inT). \label {b1}
\end{equation}
One can rewrite this sum as a contour integral using the residue theorem.
Assuming that $f( k_0)$ is regular along the imaginary axis, one gets
\begin{equation}
  J_b={T\over 2\pi i}\oint_{\mathcal C}d k_0\,f( k_0) {\beta\over2}
  \coth{\beta  k_0\over2}, \label{b2}
\end{equation}
where we have used the relation
\begin{equation}
  \Res{\!\!\!\!k_0=2\pi inT}\left[{\beta\over2}\coth{\beta k_0\over2}
  \right]=1.
\end{equation} 
The contour ${\mathcal C}$ consists of the small circles shown
in Fig. \ref{figB1},
which enclose the singularities of the $\coth$ at $k_0=2\pi inT$. The 
contour can be deformed 
to two vertical lines enclosing the imaginary axis,
see Fig. \ref{figB1}. In this way we arrive at
\begin{equation}
  J_b={1\over 2\pi i}\int_{-i\infty}^{i\infty}dk_0\,f(k_0)
  +{1\over 2\pi i}\int_{-i\infty+\epsilon}^{i\infty+\epsilon}dk_0
  \left(f(k_0)+f(-k_0)\right)n_b(k_0). \label{b4}
\end{equation}
Here the first term on the right hand side is the vacuum contribution, 
and the second term is the thermal contribution involving the Bose-Einstein
distribution,
\begin{equation}
  n_b(k_0)={1\over e^{k_0/T}-1}. \label{nb}
\end{equation}

The sum (\ref{b1}) can be evaluated in a slightly different way by 
considering instead of (\ref{b2}) the object
\begin{equation}
  J_b(\tau):={1\over 2\pi i}\oint_{\mathcal C}dk_0\,f(k_0)
  e^{k_0\tau}n_b(k_0),
\end{equation}
which fulfills
$
  \lim_{\tau\to0} J_b(\tau)= J_b.
$
Since we have for $0<\tau<\beta$
\begin{equation}
  \lim_{k_0\to\pm\infty} e^{k_0\tau}n_b(k_0) =0,
\end{equation}
we can add arcs at infinity (loosely speaking) which
give no contribution, assuming that $f(k_0)$ decays sufficiently fast at 
infinity. In this way we obtain a new integration contour
that pinches the real axis, as shown by the dashed lines in Fig. \ref{figB1}.
Taking the limit $\tau\to0+$ and assuming that
\begin{equation}
  f(k_0-i\eta)=f^*(k_0+i\eta)
\end{equation}
(Schwarz reflection principle), we find\footnote{The point $k_0=0$ should be
excluded from the integration path.}
\begin{equation}
  J_b={1\over\pi}\int_{-\infty}^\infty dk_0\, n_b(k_0)\imag f(k_0+i\eta).
  \label{b8}
\end{equation}
If the function $f(k_0)$ fulfills $f(-k_0)=f(k_0)$, we can rewrite Eq. (\ref{b8})
as
\begin{equation}
  J_b={1\over\pi}\int_{0}^\infty dk_0\, (1+2n_b(k_0)) \label{b9}
  \imag f(k_0+i\eta).
\end{equation}
We refer to the part of Eq. (\ref{b9}) which contains $n_b$ as the
``$n_b$-part'', and to the other part as the ``non-$n_b$-part''.

For fermions one finds in place of (\ref{b4})
\begin{eqnarray}
  &&\!\!\!\!\!\!\!\!\!\!\!\!
  J_f:=T\sum_{n\in\mathbbm{Z}} f\left(k_0=2\pi i(n+{\textstyle{1\over2}})T\right)
  \nonumber\\
  &&\!\!={1\over 2\pi i}\int_{-i\infty}^{i\infty}dk_0\,f(k_0)
  -{1\over 2\pi i}\int_{-i\infty+\epsilon}^{i\infty+\epsilon}dk_0
  \big(f(k_0)+f(-k_0)\big)n_f(k_0),\qquad \label{b10}
\end{eqnarray}
and in place of (\ref{b8})
\begin{equation}
  J_f=-{1\over\pi}\int_{-\infty}^\infty dk_0\, n_f(k_0)\imag f(k_0+i\eta),
  \label{b11}
\end{equation}
where $n_f(k_0)$ is the Fermi-Dirac distribution,
\begin{equation}
  n_f(k_0)={1\over e^{k_0/T}+1}. \label{nf}
\end{equation}
At finite chemical potential one has to replace $n_f(k_0)$ with $n_f(k_0-\mu)$.

\chapter{HDL/HTL gluon spectral densities \label{appc}}
In this appendix we review the spectral densities of the gluon propagator in the
HDL/ HTL approximation, as given in \cite{LeB:TFT}.
From Eq. (\ref{g37}) one finds for the transverse spectral density
\begin{equation}
  {1\over2\pi}\rho_T(q_0,q)=Z_T(q)\left[\delta(q_0-\omega_T(q))-\delta(q_0+\omega_T(q))\right]
  +\beta_T(q_0,q).
\end{equation}
Here the dispersion law $\omega_T(q)$ is determined from the pole of the propagator,
\begin{equation}
  \omega_T(q)^2-q^2-\real\Pi_T^{\rm HDL/HTL}(\omega_T(q),q)=0, \label{wt}
\end{equation} 
and the residue $Z_T(q)$ is given by
\begin{equation}
  Z_T(q)={\omega_T(\omega_T^2-q^2)\over3\omega_p^2\omega_T^2-(\omega_T^2-q^2)^2}, \label{zt}
\end{equation}
with the plasma frequency 
\begin{equation}
  \omega_p=\sqrt{2/3}m, \label{wp}
\end{equation} 
where $m$ is defined in Eq. (\ref{g32}).
For the cut contribution $\beta_T(q_0,q)$ one finds ¢
\begin{eqnarray}
  &&\!\!\!\!\!\!\!\!\!\!\!\!\!\!\!\!\!\!\!\!\!\!\!\!\!\!\!\!\!
  \beta_T(q_0,q)=m^2x\big(1-x^2\big)\theta\big(1-x^2\big)/2\nonumber\\
  &&\times\Bigg[\left(q^2\big(x^2-1\big)-m^2\left(x^2+{x\big(1-x^2\big)\over2}
  \log\bigg|{x+1\over x-1}\bigg|\right)\right)^2\nonumber\\
  &&\quad+\pi^2m^4x^2{\big(1-x^2\big)^2\over4}\Bigg]^{-1}, \label{bt}
\end{eqnarray}
where $x=q_0/q$. For the longitudinal spectral density one finds from Eq. (\ref{g38}) 
\begin{equation}
  {1\over2\pi}\rho_L(q_0,q)=Z_L(q)\left[\delta(q_0-\omega_L(q))-\delta(q_0+\omega_L(q))\right]
  +\beta_L(q_0,q).
\end{equation}
Again the dispersion law $\omega_L(q)$ is determined from the pole of the propagator [see 
Eqs. (\ref{g10}), (\ref{g38})],
\begin{equation}
  q^2+\real\Pi_H^{\rm HDL/HTL}(\omega_L(q),q)=0, \label{wl}
\end{equation} 
and the residue $Z_L(q)$ is given by
\begin{equation}
  Z_L(q)={\omega_L(\omega_L^2-q^2)\over q^2\left(q^2+3\omega_p^2-\omega_T^2\right)}. \label{zl}
\end{equation}
For the cut contribution $\beta_L(q_0,q)$ one finds
\begin{eqnarray}
  &&\!\!\!\!\!\!\!\!\!\!\!\!\!\!\!\!\!\!\!\!\!\!\!\!\!\!\!\!\!
  \beta_L(q_0,q)=m^2x\,\theta\big(1-x^2\big)\nonumber\\
  &&\times\Bigg[\left(q^2+2m^2\left(1-{x\over2}
  \log\bigg|{x+1\over x-1}\bigg|\right)\right)^2+\pi^2m^4x^2\Bigg]^{-1}, \label{bl}
\end{eqnarray}
with $x=q_0/q$.

\begin{figure}
  \begin{center}
    \includegraphics[width=7cm]{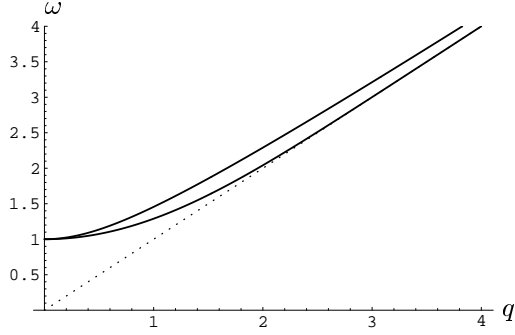}
  \end{center}
  \vspace{-0.5cm}
  {\it \caption{The dispersion laws $\omega_T(q)$ (upper curve) and $\omega_L(q)$ (lower curve) for $\omega_p=1$. 
  (Dotted line: light cone).\label{figH1}}}
\end{figure}
\begin{figure}
  \begin{center}
    \includegraphics[width=6cm]{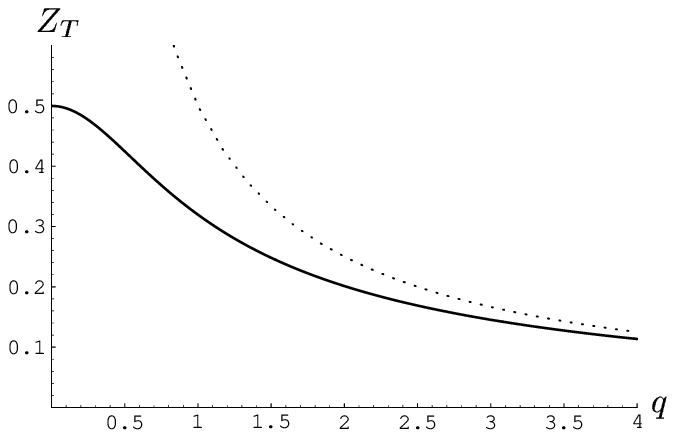} \includegraphics[width=6cm]{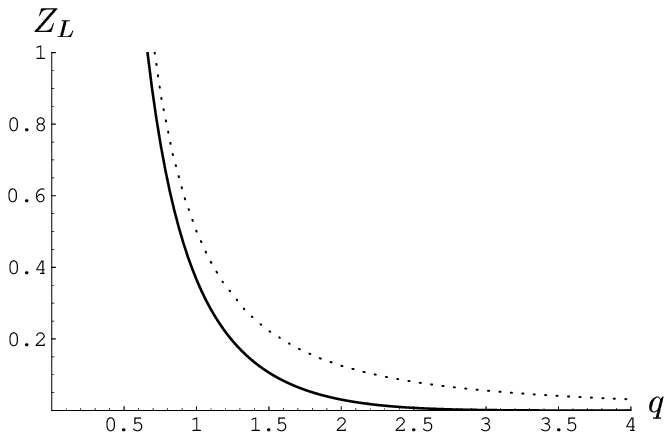}
  \end{center}
  \vspace{-0.5cm}
  {\it \caption{The residues $Z_T(q)$ and $Z_L(q)$ for $\omega_p=1$. (Dotted line in the left panel: ${1\over2q}$, 
  dotted line in the right panel: ${\omega_p\over2q^2}$).\label{figH2}}}
\end{figure}
Fig. \ref{figH1} shows the dispersion laws $\omega_T(q)$ and $\omega_L(q)$, as determined from Eqs. (\ref{wt})
and (\ref{wl}). Fig. \ref{figH2} shows the residues $Z_T(q)$ and $Z_L(q)$ which are given by
Eqs. (\ref{zt}) and (\ref{zl}).

\chapter{Matrix identities}
In order to be complete, we give here the proof for two matrix identities
which are needed in the main text.

{\bf Proposition 1:} For the variation of the determinant of an invertible matrix $\mathcal M$ one has
\begin{equation}
  \delta\det\bM=(\det\bM)\,{\rm Tr}[\delta\log\bM]=(\det\bM)\,{\rm Tr}[\bM^{-1}\delta\bM]. \label{c1}
\end{equation}
{\it Proof}: Using
$$
  \det\bM=\exp\left({\rm Tr}\log\bM\right)
$$
we find
$$
  \delta\det\bM=(\det\bM)\,\delta\left({\rm Tr}\log\bM\right)=(\det\bM)\,{\rm Tr}[\delta\log\bM].
$$
$\log\bM$ can be expanded in a series, for instance about the unit matrix. Under the trace
cyclic permutations of a product of matrices are allowed. Therefore one has
$$
  \qquad\qquad\qquad\qquad\qquad
  {\rm Tr}[\delta\log\bM]={\rm Tr}[\bM^{-1}\delta\bM].  
  \qquad\qquad\qquad\qquad\qquad\raisebox{3pt}\fbox{}
$$

{\bf Proposition 2:} For any $n\times n$ matrices $\bA$, $\bB$, $\bC$, 
$\bD$ (with $\bD$ invertible) one has
\begin{equation}
  \det\left(
    \begin{array}{cc}\bA&\bB\\ \bC&\bD
    \end{array}
  \right)=\det(\bD\bA-\bD\bB\bD^{-1}\bC). \label{c2}
\end{equation}
{\it Proof}: With the definition
$$
  \bE:=(\bA-\bB\bD^{-1}\bC)^{-1}
$$
we can write the inverse matrix as
$$
  \left(
    \begin{array}{cc}\bA&\bB\\ \bC&\bD
    \end{array}
  \right)^{-1}=
  \left(
    \begin{array}{cc}\bE&\-\bE\bB\bD^{-1}\\ ­\bD^{-1}\bC\bE&\bB^{-1}\bA\bE\bB\bD^{-1}
    \end{array}
  \right).
$$
A short calculation using this relation shows that
$$
  \det\left[\left(
    \begin{array}{cc}\bD\bE^{-1}&{\bf 0}\\ \bD^{-1}\bC&{\bf 1}
    \end{array}
  \right)\cdot
  \left(
    \begin{array}{cc}\bA&\bB\\ \bC&\bD
    \end{array}
  \right)^{-1}\right]=
  \det\left(
    \begin{array}{cc}\bD&-\bD\bB\bD^{-1}\\ {\bf 0}&\bD^{-1}
    \end{array}
  \right)=1,
$$
where $\bf 0$ and $\bf 1$ denote the $n\times n$ zero and unit matrices, respectively.
Therefore we have
$$
  \qquad\qquad\qquad
   \det\left(
    \begin{array}{cc}\bA&\bB\\ \bC&\bD
    \end{array}
  \right)= 
   \det\left(
    \begin{array}{cc}\bD\bE^{-1}&{\bf 0}\\ \bD^{-1}\bC&{\bf 1}
    \end{array}
  \right)=
  \det\left(\bD\bE^{-1}\right).
 \qquad\quad\ \ \raisebox{3pt}\fbox{}
$$

\end{document}